\documentclass[twocolumn,epsfig, graphics,floatfix]{revtex4-1}

\usepackage{amsmath,amsfonts,amssymb,graphics,graphicx,epsfig,color,times,bbm}

\usepackage{mathtools}
\usepackage{amsthm}
\usepackage{psfrag}
\usepackage{braket}
\usepackage[caption=false]{subfig}
\usepackage{graphicx}
\usepackage{color,soul}
\AtBeginDocument{\usepackage{booktabs}}
\usepackage[hidelinks]{hyperref}
\hypersetup{colorlinks=false}

\usepackage{float}
\usepackage{enumitem}

\newcommand{\zz}{\mathbbm{Z}}
\newcommand{\nn}{\mathbbm{N}}

\newcommand{\id}{\mathbbm{1}}

\usepackage{array}
\newcolumntype{L}{>{\arraybackslash}m{6.5cm}}

\usepackage[framemethod=tikz]{mdframed}
\usepackage{hyperref}

\theoremstyle{plain}

\theoremstyle{plain}

\theoremstyle{definition}
\newtheorem{defn}{Definition}

\theoremstyle{remark}

\usepackage{xcolor}

\usepackage[normalem]{ulem}
\usepackage{dsfont}
\newcommand{\resstab}{\mathcal{R}}
\newcommand{\resstabsix}{\mathcal{R}_6}
\newcommand{\checkgp}{\mathcal{C}}
\newcommand{\measgp}{\mathcal{M}}
\newcommand{\survstab}{\mathcal{S}}

\newcommand{\trivundterror}{T}

\newcommand{\spiderx}[1]{\hat{\boldsymbol{x}}_{#1}}
\newcommand{\spiderz}[1]{\hat{\boldsymbol{z}}_{#1}}

\usepackage{multirow}

\begin{document}

\title{Logical blocks for fault-tolerant topological quantum computation}

\newcommand*\leadauthor{\thanks{Lead authors:  \\ \mbox{sroberts@psiquantum.com} \\  \mbox{ryan.mishmash@gmail.com} }}

\author{H\'ector Bomb\'in}
\author{Chris Dawson}
\author{Ryan V. Mishmash} \leadauthor
\author{Naomi Nickerson}
\author{Fernando Pastawski}
\author{Sam Roberts}\leadauthor

\affiliation{PsiQuantum Corp., Palo Alto}
\date\today

\begin{abstract}

Logical gates constitute the building blocks of fault-tolerant quantum computation. 
While quantum error-corrected \emph{memories} have been extensively studied in the literature, explicit constructions and detailed analyses of thresholds and resource overheads of universal logical gate sets have so far been limited. 
In this paper, we present a comprehensive framework for universal fault-tolerant logic motivated by the combined need for (\textit{i}) platform-independent logical gate definitions, (\textit{ii}) flexible and scalable tools for numerical analysis, and (\textit{iii}) exploration of novel schemes for universal logic that improve resource overheads. 
Central to our framework is the description of logical gates holistically in a way which treats space and time on a similar footing. 
Focusing on quantum instruments based on surface codes, we introduce explicit, but platform-independent representations of topological logic gates---called logical blocks---and generate new, overhead-efficient methods for universal quantum computation. 
As a specific example, we propose fault-tolerant schemes based on surface codes concatenated with more general low-density parity check (LDPC) codes, suggesting an alternative path toward LDPC-based quantum computation.
The logical blocks framework enables a convenient software-based mapping from an abstract description of the logical gate to a precise set of physical instructions for executing both circuit-based and fusion-based quantum computation (FBQC). Using this, we numerically simulate a surface-code-based universal gate set implemented with FBQC, and verify that the threshold for fault-tolerant gates is consistent with the bulk threshold for memory. 
We find, however, that boundaries, defects, and twists can significantly impact the logical error rate scaling, with periodic boundary conditions potentially halving   resource requirements. 
Motivated by the favorable logical error rate suppression for boundaryless computation, we introduce a novel computational scheme based on the teleportation of twists that may offer further resource reductions.

\end{abstract}

\maketitle

\section{Introduction} 

Quantum fault tolerance will form the basis of large-scale universal quantum computation. 
The surface code~\cite{kitaev2003fault, kitaev1997quantum,dennis2002topological} and related topological approaches~\cite{raussendorf2007fault, raussendorf2007topological, bolt2016foliated, nickerson2018measurement, brown2020universal} are among the most appealing methods for near term fault-tolerant quantum computing (FTQC), primarily due to their high thresholds and amenability to planar architectures with nearest-neighbor interactions. 
In recent years there have been numerous studies exploring the surface code memory threshold, which is the error rate below which encoded information can be protected arbitrarily well in the limit of large code size~\cite{dennis2002topological,wang2003confinement, stace2009thresholds, duclos2010fast, bombin2012strong, fowler2012surface, watson2014logical, bravyi2014efficient, darmawan2017tensor}. 
However, to understand the thresholds and overhead for universal fault-tolerant quantum \emph{computation}, it is necessary to study the behavior of fault-tolerant \emph{logical gates}. 
In topological codes these gates can be implemented using methods that draw inspiration from condensed matter, where encoded operations are achieved by manipulating topological features such as boundaries, defects, and twists~\cite{raussendorf2007topological, raussendorf2007fault, bombin2009quantum, bombin2010topological, landahl2011fault, horsman2012surface, fowler2012time, barkeshli2013classification, barkeshli2013twist, hastings2014reduced, yoshida2015topological, terhal2015quantum, yoder2017surface, brown2017poking, yoshida2017gapped, roberts2017symmetry, bombin2018transversal, bombin20182d, lavasani2018low, lavasani2019universal, webster2020fault, hanks2020effective, roberts20203, webster2020fault, zhu2021topological,chamberland2021universal,landahl2021logical}. 

To date, there has been no taxonomic analysis that validates the effect on the error threshold and below-threshold scaling in the presence of topological features (see Refs.~\cite{beverland2021cost, chamberland2021universal} for developments in this direction).
It is known that the introduction of modified boundary conditions can have a significant impact on the error suppression of the code~\cite{fowler2013accurate,beverland2019role}, and therefore, as technology moves closer to implementing large-scale fault-tolerant quantum computations~\cite{doi:10.1126/science.abi8378,egan2021fault,ryan2021realization,postler2021demonstration}, it is critical to fully understand the behavior not only of a quantum memory, but of a universal set of logical gates.

\begin{figure}[h]
	\centering
	\includegraphics[width=0.99\columnwidth]{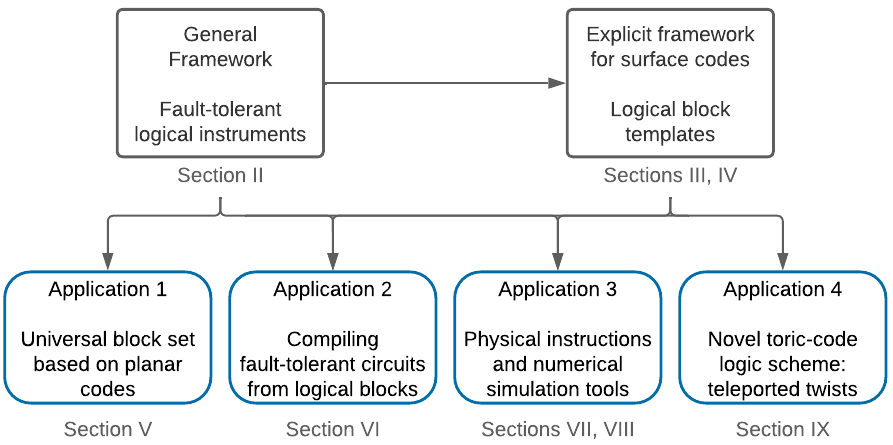} 
	\caption{Outline of the paper.}
	\label{LogicBlocksOutline}
\end{figure}

In this paper we comprehensively study universal logical gates for fault-tolerant schemes based on \textit{surface codes}. 
Our approach is centered around a framework for defining and analyzing logical instruments as three-dimensional (3D) objects called fault-tolerant logical instruments, allowing for a fully topological interpretation of logical gates. 
Focusing on fault-tolerant instruments directly---rather than starting with an error-correcting code and considering operations thereon---is beneficial for several reasons. 
Firstly, it provides a holistic approach to logical gate optimization, allowing us to explore options that are not particularly natural from a code-centric perspective. 
Secondly, it provides a way to define explicit logical instruments from physical instruments in a way that is applicable across different physical settings and models; in our case, we provide explicit instructions to compile these fault-tolerant gates to both circuit-based quantum computation (CBQC) based on planar arrays of static qubits and to fusion-based quantum computing (FBQC)~\cite{bartolucci2021fusion}. 
Thirdly, it enables a unified definition of the fault-distance of a protocol; while distance of a code is straightforwardly defined, logical failures in a protocol can occur in ways that cannot be associated with any single time-step of the protocol, (for instance, timelike chains of measurement errors in a topological code). 
The fault distance of a protocol provides a go-to proxy for fault tolerance, which avoids the computational overhead of full numerical simulations. 
Using this framework, we introduce several new approaches to topological quantum computation with surface codes (including both planar and toric), and numerically investigate their performance. An outline of the paper is displayed in Fig.~\ref{LogicBlocksOutline}.

\textbf{A framework to define fault-tolerant instruments.} 
Our first contribution is to define the framework of \textit{fault-tolerant logical instruments} to describe quantum computation based on stabilizer codes holistically as instruments in space-time rather than as operations on a specific code. 
This framework, defined in Sec.~\ref{secFTInstruments}, builds upon concepts first introduced in topological measurement-based quantum computation (MBQC)~\cite{raussendorf2007topological,raussendorf2007fault} and extended in related approaches~\cite{fujii2015quantum,brown2017poking,brown2020universal,hanks2020effective, bombin2021interleaving}. 
Within this framework, introduce a surface-code-specific construction called a \textit{logical block template}, which allows one to explicitly specify (in a platform-independent way) a fault-tolerant surface code instrument in terms of (2{+}1)D space-time topological features. 
The ingredients of a logical block template---the \textit{topological features}---consist of boundaries (of which there are two types), corners, symmetry defects, and twists of the surface code, as defined in Sec.~\ref{secElements}. In Sec.~\ref{sec3DElements}, we define logical block templates and how to compile them into physical instructions (for either CBQC or FBQC), with the resulting instrument being referred to as a \textit{logical block}. 
This framework highlights similarities between different approaches to fault-tolerant gates, for example between transversal gates, code deformations, and lattice surgeries, as well as between different models of quantum computation.

\textbf{Logic blocks for universal quantum computation.}
Our first application of the fault-tolerant instrument framework is to define a universal gate set based on planar codes~\cite{bravyi1998quantum}. Some of these logical blocks offer reduced overhead compared to previous protocols. For example, we show how to perform a phase gate on the distance $d$ rotated planar code~\cite{nussinov2009symmetry,bombin2007optimal,beverland2019role}, using a space-time volume of $4d^3$. This implementation requires no distillation of Pauli-$Y$ eigenstates, and thus we expect it to perform better than conventional techniques. In addition to its reduced overhead, the phase gate we present can be implemented in a static 2D planar (square) lattice of qubits using only the standard 4-qubit stabilizer measurements, without needing higher-weight stabilizer measurements~\cite{bombin2010topological,litinski2019game} or modified code geometries~\cite{yoder2017surface,brown2020universal} that are typically required for braiding twists. 
These logical blocks can be composed together to produce fault-tolerant circuits, which we illustrate by proposing an avenue for fault-tolerant quantum computation based on concatenating surface codes with more general quantum low-density parity check (LDPC) codes~\cite{tillich2014quantum,gottesman2013fault,fawzi2018constant,fawzi2018efficient,breuckmann2021ldpc,hastings2021fiber,breuckmann2021balanced,panteleev2021asymptotically}. 
Such concatenated code schemes may offer the advantages of both the high thresholds of surface codes, with the reduced overheads of constant-rate LDPC codes---an attractive prospect for future generations of quantum computers. 

\textbf{Fusion-based quantum computation---physical operations, decoding, and simulation.}
Fusion-based quantum computation is a new paradigm of quantum computation, where the computation proceeds by preparing many copies of a constant-sized (i.e., independent of the algorithm size) entangled resource state, and performing entangling measurements between pairs (or more) resource states. This model is motivated by photonic architectures, where such resource states can be created with high fidelity, and then destructively measured using \textit{fusion} measurements~\cite{bartolucci2021fusion}.

In Sec.~\ref{sec:topological_fbqc}, we review FBQC and show how logical block templates can be compiled to physical FBQC instructions. In Sec.~\ref{secNumerics}, we introduce tools to decode and simulate such blocks, and numerically investigate the performance of a complete set of logical operations in FBQC (these operations are complete in that they are universal when supplemented with noisy magic states). 

Firstly, we verify that the thresholds for these logical operations all agree with the bulk memory threshold. Secondly, we uncover the significant role that boundary conditions have in the resources required to achieve a target logical error rate. Namely, we see that qubits encoded with periodic boundary conditions offer more favorable logical error rate scaling with code size than for qubits defined with boundaries (as has been previously observed in Ref.~\cite{fowler2013accurate}). For instance, at half threshold, nontrivial logic gates can require up to $25\%$ larger distance (about 2 times larger volume) than that estimated for a memory with periodic boundary conditions in all three directions (i.e., lattice on a 3-torus). 
Our results demonstrate that entropic contributions to the logical error rate can be significant, and should be contemplated in gate design and in overhead estimates for fault-tolerant quantum algorithms.

\textbf{Logical instruments by teleporting twists.}
Finally, motivated by the advantages in error suppression offered by periodic boundary conditions, we introduce a novel computational scheme in Sec.~\ref{secPortals}, where fault-tolerant gates are achieved by teleporting twists in time. In this scheme, qubits are encoded in twists of the surface code, and logical operations are performed using space-time defects known as portals. These portals require nonlocal operations to implement, and are naturally suited to, for instance, photonic fusion-based architectures~\cite{bartolucci2021fusion,bombin2021interleaving} for which we prescribe the physical operations required. To our knowledge, this is the first surface code scheme that does not require boundaries to achieve a universal set of gates and may offer even further resource reductions in the overhead of logical gates. This logic scheme is an important example of the power of the fault-tolerant instrument framework, as the operations are difficult to understand as sequences of operations on a 2D quantum code.

\section{Fault-tolerant instruments}\label{secFTInstruments}

In this section, we describe the notion of a stabilizer \emph{fault-tolerant logical instrument}, suitable for describing a wide class of logical operations. 
A fault-tolerant logical instrument takes some number $k_\text{in}$ of encoded quantum states as inputs (encoded in stabilizer codes~\cite{gottesman1997stabilizer}), performs an encoded operation, and outputs some number of encoded states $k_\text{out}$.

We use the term {\it logical port}, to refer to a group of physical qubits that together represent logical input or output qubits of the instrument.
Each port has a quantum error-correcting code with a fixed number of physical and logical qubits associated with it.
In this way, fault-tolerant instruments can be composed with each other only through a pair of compatible input and output ports.
Alternatively, {\it logical ports} can be seen as a specific collection of cuts that partition a complex logical quantum circuit into logical blocks, the elementary quantum instruments that are amenable to independent study and optimization.
In this way, quantum error-correcting codes continue to play a crucial role in defining modular interface structure through which to compose fault-tolerant algorithms from elementary logical blocks.

The main feature of a fault-tolerant instrument is the classical data they produce (as intermediate measurement outcomes). 
These classical data are used both to identify errors (by relying on measurement outcome redundancy in the form of checks) as well as to determine the Pauli frame required to interpret the logical mapping and logical measurement outcomes, as described below. 
Examples of fault-tolerant instruments that are included within this model are transversal gates, code deformations, gauge fixing, and lattice surgeries~~\cite{bombin2006topological, raussendorf2007topological, raussendorf2007fault, bombin2009quantum, bombin2010topological, horsman2012surface, paetznick2013universal, hastings2014reduced}. 
Such operations are commonly understood in terms of a series of gates and measurements on a fixed set of physical qubits, a perspective originating from matter-based qubits. 
Nevertheless, several recent works have introduced new approaches to fault-tolerant memories beyond the setting of static codes~\cite{nickerson2018measurement, newman2020generating, hastings2021dynamically}. 
Here we want to generalize and extend these concepts to reach a holistic perspective on logical operations rather than as operations on an underlying code.

In this section we describe properties of a fault-tolerant logical instrument, building upon the formalism introduced for the one-way measurement-based quantum computer by Raussendorf \textit{et al}.~\cite{raussendorf2007fault} and extended in Refs.~\cite{brown2020universal, bartolucci2021fusion}. 
In subsequent sections we specialize to logical instruments that are achieved by manipulating topological features of the surface code in (2{+}1)D space-time.

\begin{figure}[h]
	\centering
	\includegraphics[width=0.8\columnwidth]{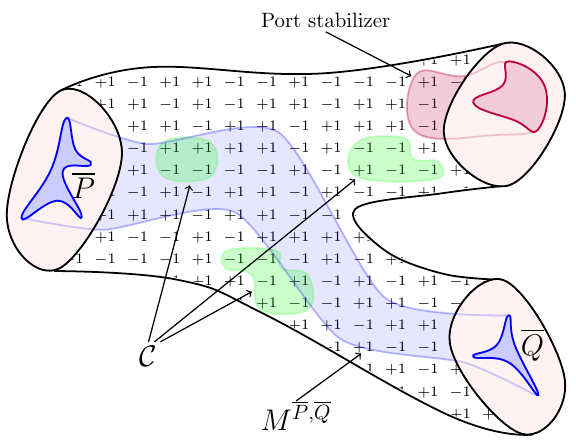} 
	\caption{
		Abstract schematic of a fault-tolerant instrument with one input port (left) and two output ports (right) (all shaded in pink). 
		In the stabilizer framework, outcomes correspond to measurement of Pauli product observables and checks take the form of joint parity constraints.
		The $+1$s and $-1$s represent the physical measurement outcomes in $\mathcal{O}$. 		
		While each measurement outcome may individually be random, {\it checks} identify subsets of outcomes (depicted here in green) for which the joint parity is fixed (in the absence of errors).
		Example of a logical correlator ${P} \otimes {Q} \in \mathcal{S}({U})$ represented by a \emph{logical membrane} $M^{\overline{P}, \overline{Q}}$. 
		A logical correlator ${P} \otimes {Q} \in \mathcal{S}(U)$ means that an input logical operator $\overline{P}$ is mapped to an output logical operator $\overline{Q}$ up to a sign depending on outcomes of measurements supported on a representative logical membrane $M^{\overline{P},\overline{Q}}$. Here, $\mathcal{S}(U)$ denotes the stabilizer group of the stabilizer channel $U$. \\		
		A concrete choice of such an instrument could be an encoding isometry for a 2-repetition code, mapping arbitrary states in the input space onto the subspace of the output space stabilized by $Z \otimes Z$. The stabilizer generators for this instrument are given by $\mathcal{S} = \langle Z_{\mathrm{in}_1} \otimes Z_{\mathrm{out}_1} I_{\mathrm{out}_2}, X_{\mathrm{in}_1} \otimes X_{\mathrm{out}_1}  X_{\mathrm{out}_2}, I_{\mathrm{in}_1}\otimes Z_{\mathrm{out}_1}  Z_{\mathrm{out}_2} \rangle$ (where the tensor product is only included to denote the input/output subsystems). For this choice, both $P$ and $Q$ correspond to the logical $Z$ operators of their corresponding port codes (see App.~\ref{secStabilizerOperatorsMapsAndInstruments} for more details).
}
\label{figAbstractChannel}
\end{figure}

\subsection{Quantum instrument networks}

Here we draw the curtain and present the stage: a general framework to think about FTQC.

\subsubsection{Quantum instruments}

Quantum instruments describe the most general process in which the input is a quantum system and the output is a combination of quantum and classical systems. 
We refer to classical outputs as {\it outcomes}, the collection of which is labeled by $\mathcal{O}$. 
Quantum instruments can model {\bf any} of the (idealized) physical devices that take part in a quantum computation (state preparation, unitaries and measurement).
A quantum instrument with a single value classical outcome may be used to represent quantum state preparations and quantum maps (also called channels).

A quantum instrument is specified as a collection
\begin{equation}\label{eqnQuantumInstrument}
\{\mathcal E_m\}
\end{equation}
of completely positive and trace nonincreasing linear maps, such that their sum is trace preserving. The maps are indexed by the outcome $m$: for an input state $\rho$, if the instruments outcome is $m$, the unnormalized final state is $\mathcal{E}_m(\rho)$ and the probability for the outcome $m$ to occur is the trace of this state.

\subsubsection{Networks}

A natural way to describe a fault-tolerant quantum computation is as a network (or circuit) of quantum instruments.

\begin{defn}
A {\it quantum instrument network} ({\bf QIN}) is a directed acyclic graph ({\bf DAG}) in which
\begin{itemize}
\item edges are interpreted as quantum systems,
\item vertices are interpreted as quantum instruments: their quantum input (output) is the tensor product of incoming (outgoing) edges.
\end{itemize}
\end{defn}

Recall that the vertices of a DAG can always be ordered so that all edges point towards the ``largest'' vertex (as per the ordering). 
Thus we can interpret a QIN as a process in which the quantum instruments are applied sequentially, each mapping a collection of subsystems to a new such collection. 
Since the specific ordering is immaterial, the DAG is enough to specify the process.

In such a process, each vertex of the DAG contributes an outcome. 
Classical beings as we are, the ultimate object of interest is the classical distribution of outcomes. 
The probability of an outcome configuration can be computed as a tensor network contraction: the tensor network has the same topology as the DAG and, given some choice of basis for each edge, the tensor at a given vertex is obtained from the corresponding (outcome-dependent) linear map.

\subsection{Stabilizer fault tolerance}

In order to make headway in the analysis of fault-tolerant logical blocks, we focus on {\it stabilizer quantum instruments} $\{\mathcal{E}_m\}$ and stabilizer QINs. 
In terms of domains, this restricts the input and output Hilbert spaces of each instrument to tensor products of qubits and classical outcomes to bit strings. 
For each classical outcome $m$, the quantum channel $\mathcal{E}_m$ is in fact required to be a stabilizer operator stabilized by $\mathcal{S}_m$.
We assume that all such operators share the same stabilizer group up to signs (i.e., $\langle -1, {\mathcal S}_m \rangle = \langle -1, {\mathcal S}_n \rangle$ for outcomes $m$, $n$).
Moreover, there is a linear transform (over $\mathbbm{Z}_2$) that relates outcome bits with signs of stabilizer generators (see App.~\ref{secStabilizerOperatorsMapsAndInstruments} for a full definition of the notion of stabilizer instruments).
Crucially, this property is preserved under composition (i.e., the quantum instruments resulting from composing elementary stabilizer quantum instruments in QINs will themselves be stabilizer quantum instruments).

This set of stabilizer quantum instruments includes full or partial Pauli product measurements as well as unitaries from the Clifford group and encoding isometries for stabilizer codes.
However, it does not include the full flexibility of {\it adaptivity}, i.e., conditionally applying distinct instruments depending on previously obtained outcomes.
While adaptivity is crucial to allow for universal quantum computation at the logical level, it is not required to achieve a restricted form of fault tolerance limited to logical stabilizer operations.

\subsubsection{Pauli frame}
The different signs for the resulting stabilizer group can be interpreted as being an outcome-dependent {\it Pauli frame correction}.
Thus, similarly to quantum teleportation, in the absence of noise, a specific Pauli correction can be directly ($\mathbbm{Z}_2$ linearly) inferred from a parity combination of the classical outcomes associated with the stabilizer instrument.

\subsubsection{Check generators}
Under a noise-free operation, not all outcome combinations are possible for a fault-tolerant logical block. 
In stabilizer fault tolerance, the set of possible outcomes is characterized by linear constraints (considering outcomes as a vector space over $\mathbbm{Z}_2$).
The generators for said set of constraints are called {\it check} generators.
In the case of topological fault tolerance, check generators can be chosen to be geometrically local (i.e., involving only outcomes in small  neighborhood with respect to the QIN graph).

It is the presence of checks that allows fault-tolerant protocols to reliably extract logical outcomes from noisy physical outcomes.
The presence of check violations, together with statistical understanding of the noise model, allows one to infer the correct way to interpret logical outcomes with an increasingly high degree of reliability.

\subsection{Properties of fault-tolerant instruments \\ (aka logical blocks)}\label{secPropeprtiesGeneralFTChannel}

A  fault-tolerant stabilizer instrument $\Phi$ is a stabilizer QIN on which the following specific structure has been identified.
\begin{itemize}[leftmargin=*]
	\item {\bf Check generators} are the basis of fault tolerance and jointly define a check group $\checkgp$ (a $\mathbbm{Z}_2$-affine subspace).
	The check operators are combinations of classical outcomes within the QIN that yield a predefined parity.
	We choose a minimal set of low-weight, geometrically local {\it check generators} whenever possible.
	\item External {\bf logical ports} are a partitioning of the {\it physical} input and output ports of the QIN into {\it logical} input and output ports. 
	Each logical port is thus a collection of physical ports in the QIN that have not been attached within it.
	\item The outputs and inputs of any given port are related by {\bf port stabilizers} whose signs generally  depend on the values of outcomes within the QIN.
	Port stabilizers jointly give the port the structure of a quantum error-correcting code (up to an outcome-dependent Pauli frame).
	Each port stabilizer is a Pauli stabilizer on the corresponding port qubits together with a collection of QIN outcomes that jointly determine its sign.
	\item {\bf Logical correlators} $\survstab$ in fault-tolerant stabilizer instruments are encoded through logical stabilizer operators in one (or more) ports together with an outcome mask $m$.
	Drawing from topological codes, the outcome masks will be commonly referred to as {\it logical membranes} as this is the shape these have in the topological protocols we focus on.
	There is no mathematical difference between logical correlators and port stabilizers, but rather in their intent. 
	In fact, logical correlators are defined up to multiplication by check generators and/or port stabilizers (a form of gauge freedom). We denote by $\mathcal{S}(U)$ the stabilizer group of the channel $U$. See App.~\ref{secStabilizerOperatorsMapsAndInstruments} for more details. 
\end{itemize}
An abstract depiction of these properties is shown in Fig.~\ref{figAbstractChannel}. 
The concepts will be further clarified in the context of surface codes in Secs.~\ref{sec3DElements}~and~\ref{secLogicBlocks}. 

\begin{figure*}[t]
	\centering
	\includegraphics[width=0.85\linewidth]{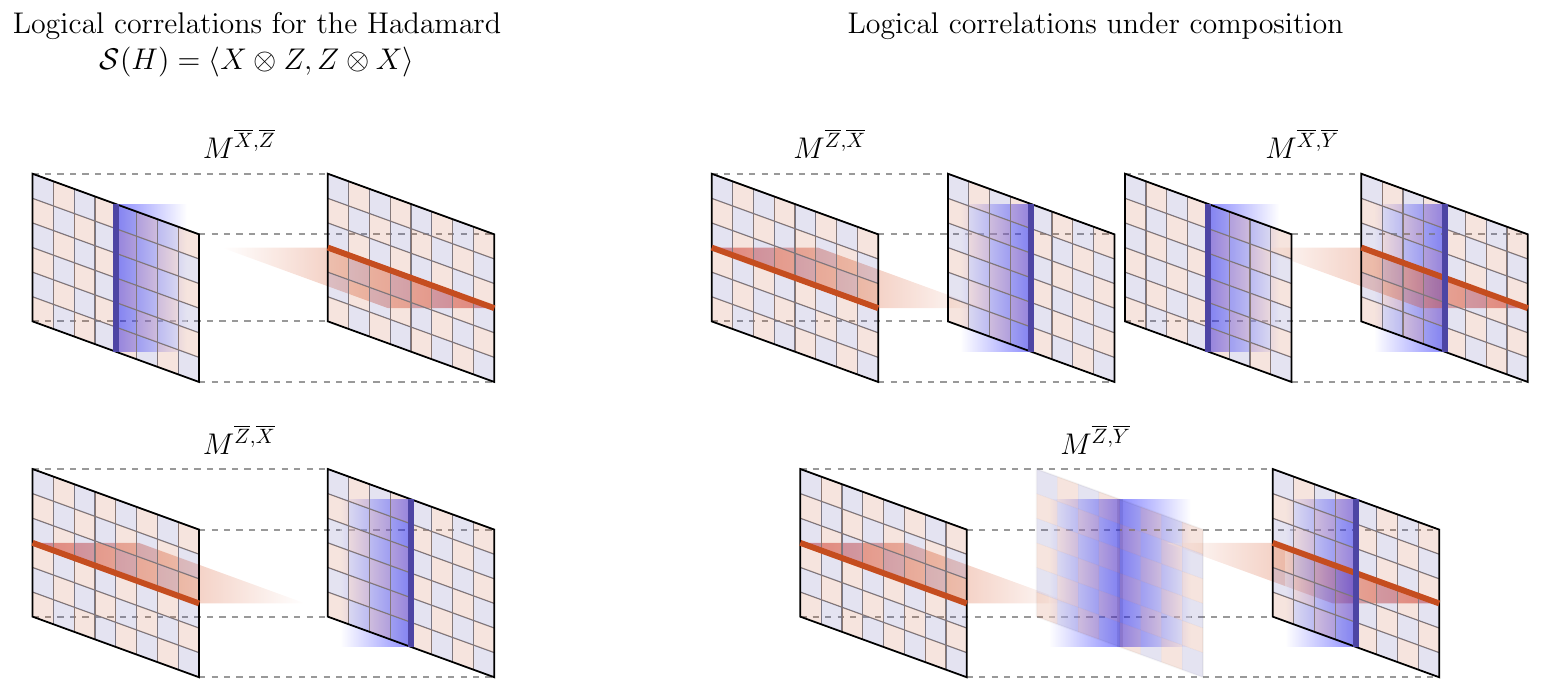} 
	\caption{
		Schematic representation of ports, logical correlator (membranes) for surface code computations. 
		The internal structure for the fault-tolerant quantum instrument  is omitted, leaving only the ports depicted, with the input [output] being on the left [right] of each instrument. 
		(left) correlator representation of the logical Hadamard. 
		The first (second) tensor factor corresponds to the input (output) of the instrument. 
		A fault-tolerant instrument realizing this operation has logical membranes for each stabilizer in $\mathcal{S}(\overline{H})$, which can be understood as mapping logical observables between input and output ports. 
		(right) Fault-tolerant instruments can be composed along a pair of input and output ports that share a common code, leading to new composite logical correlators.}
	\label{figLogicalBlockConcatenation}
\end{figure*}

In CBQC, we can understand logical correlators and membranes as follows. 
Logical operators at an input port are transformed by every constituent instrument of the QIN.
These transformations can be tracked stepwise, following the temporal order of application in the circuit model.
At each intermediate step, the logical operator admits an instantaneous representation.
This {\it instantaneous representation} is supported on {\it intermediate qubits}, which correspond to contracted quantum inputs and outputs of the constituent quantum  instruments.
In the case of a noiseless CBQC fault tolerantly representing unitary instruments, the logical correlators need not become correlated with any of the classical outcomes $\mathcal{O}$ of the QIN.
Elements of $\mathcal{S}$ correspond to combinations of logical operators on input and output ports.
However, this lack of correlation does not persist in other models of computation such as MBQC and FBQC;
even the logical correlators for CBQC  must be sign corrected in the presence of noise.
Noisy operation requires identifying and tracking the most likely error class consistent with visible outcomes, with such errors possibly changing the sign of the logical correlator.
This sign correction is referred to as (logical) Pauli frame tracking, as it amounts to applying a logical Pauli operator to the quantum ports of the fault-tolerant instrument.

In certain situations, there may be elements of $\mathcal{S}$ corresponding to state preparation isometries, which are supported exclusively on output ports. 
Similarly, for measurements and partial projections, there will be elements of $\mathcal{S}$ supported exclusively on input ports.
In this latter case, in order for the corresponding fault-tolerant instruments to be trace preserving the corresponding logical stabilizers must be correlated with the classical outcome of the corresponding QIN.
An example of this is given by the $X$ or $Z$ measurement instruments in Fig.~\ref{figAllLogicalBlocks} below, wherein the logical block has no quantum output and is closed off by a layer of single-qubit measurements revealing the value of a corresponding logical operator.

In general, outcome masks $m\subset \mathcal{O}$ are required to determine the Pauli frame correction; 
the product of their outcomes determines the sign of the logical correlator, and hence the Pauli correction operator that is additionally applied. 
Check generators in $\mathcal{C}$ as well as local port stabilizers, can be understood as trivial logical membranes (i.e., equivalent to logical identity), and thus, logical membranes form equivalence classes under multiplication by checks.
In this way, \emph{fault tolerance} can be observed at the level of the logical membranes: a fault-tolerant instrument should have many equivalent logical membranes for a given logical correlator.
Whereas in the absence of noise all these membranes lead to a consistent Pauli frame correction, in the presence of noise, it is the job of the decoder algorithm to identify the most likely fault equivalence class and corresponding correction.

\textbf{Logical faults}. 
For the instrument to be fault tolerant, it is essential that a small number of elementary faults retain the intended noiseless logical interpretation. 
We define the fault distance as the minimal number of elementary faults yielding trivial (i.e., correct) check outcomes and an incorrect logical outcome.
The choice of what to interpret as elementary faults is model specific and should be motivated by the physics of the device(s) being described.
It is common to choose arbitrary single-qubit Pauli error on any of the underlying physical input or output qubits of the constituent instruments of the QIN.
These generally include outcome or measurement errors that have a purely classical interpretation as an outcome being flipped. 
The \textit{weight} of a fault combination is defined as the number of elementary faults that compose it. 
A fault combination is called \textit{undetectable} if it leaves all checks invariant. 
An undetectable fault combination is a \textit{logical fault} if it leads to an incorrect logical Pauli frame (i.e., it flips the sign of a logical correlator with respect to that of a noiseless scenario). 
The \textit{fault distance} of a quantum protocol, is defined as the smallest number of \textit{elementary faults} that combine to form a logical fault.

In practice, to make progress,  we focus on the fault distance of individual logical blocks, wherein local port stabilizers are assumed to be perfectly measured.
This approach is more pragmatic than focusing on quantum circuits, which are composed of a number of {\it logical blocks} as large as demanded by their intended function rather than by fault-tolerance considerations.
The approach we use to isolate individual logical blocks can be interpreted as attaching an idealized decoding partial projection at output ports and an ideal encoding isometry at input ports and, as such, is slightly optimistic.
Further discussion on the port boundary conditions and block decoding simulation can be found in App.~\ref{appPortBoundaryConditionsAndDecodingSims}.

\textbf{Example: quantum memory}.
We consider a simple circuit-based example of a fault-tolerant instrument performing a logical identity operation (or any single-qubit Pauli operation for that matter) on a surface-code-encoded qubit.
We follow the usual circuit model assumption wherein the set of data qubits is fixed throughout the computation.
Thus, the only two logical ports ${\mathcal{P}} = \{\text{in}, \text{out}\}$ can be seen as using a common set of labels for the physical qubits.
The quantum instrument network is composed of elementary circuit elements repeatedly performing {\it stabilizer measurements}. 
For the circuit model these measurements are typically described in terms of (i) auxiliary qubit initialization, (ii) two-qubit entangling gates such as controlled-NOT ({\texttt{CX}}) and controlled-phase ({\texttt{CZ}}) gates that map stabilizer information onto auxiliary qubits and (iii) single-qubit measurement applied to auxiliary qubits.
The auxiliary qubits are {\it recycled} between measurement and initialization, allowing a local realization through geometrically bound quantum information carriers such as superconducting qubits~\cite{fowler2012surface}.
Of these, only (iii) yields a classical outcome, and the circuits are designed such that they yield the measurement value for a code stabilizer generator.
These measurements are repeatedly performed throughout the code for a certain number $T$ of {\it code cycles}.

We identify a check generator of $\mathcal{C}$ for every pair of consecutive measurements of the same stabilizer.
In practice, the {\it consecutive} qualifier is important as it leads to relatively compact check generators that allow directly revealing information on low-weight faults; 
the underlying assumption is that the outcome of these check generators can only be affected (i.e., flipped) by a small set of fault generators between the two consecutive stabilizer measurements.
For each stabilizer generator of the underlying code, the first (final) measurement round for it leads to a corresponding localized input (output) port stabilizer with sign correlated to the measurement outcome.
Finally, as the logical operators of the code are preserved by the stabilizer measurements, the logical operators can be seen as propagating as unperturbed logical membranes $M^{\overline{X},\overline{X}}$, $M^{\overline{Z},\overline{Z}}$ through the (2{+}1)D structure of the QIN from input to output.
In the usual circuit model, the logical Pauli frame---in the absence of errors---is trivial (i.e., the identity), as logical operators map deterministically from input to output. This can change when performing logic gates through code deformation, for example.
An additional example is shown in Fig.~\ref{figLogicalBlockConcatenation} for a block implementing a logical Hadamard on a surface code. 

\section{Elements of topological computation}\label{secElements}

We are mostly interested in constructing \textit{topological} fault-tolerant instruments; 
that is, for instruments whose inputs and output ports are encoded using topological codes, a useful subset of logical correlations can be generated by  manipulating topological \emph{features} such as boundaries, domain walls, and twists.
These section is devoted to the detailed description of these features. 

We now introduce the ingredients that make up a topological fault-tolerant instrument network. 
In topological quantum computation, fault tolerance is achieved by creating a fault-tolerant \emph{bulk}, generally with a periodic repeating structure that contains parity checks (stabilizers) to enable error correction. An example of this would be the repeated measurement of stabilizer operators on a toric code, or a three-dimensional fusion network in a FBQC setting. The resulting bulk allows logically encoded quantum information to be stored. However, in order to use this system to perform a nontrivial quantum \emph{computation}, the homogeneous nature of the bulk must be broken. One way of achieving this is to introduce \emph{topological features} that can be manipulated to perform logical gates. The simplest example of a topological feature is a boundary, which terminates the bulk in a certain location~\cite{kitaev2006anyons,preskill1999lecture, raussendorf2007topological, raussendorf2007fault}.
In surface codes one can also introduce further topological features known as domain walls and twists~\cite{bombin2009quantum}. By introducing these features in an appropriate configuration, logical information can not only be stored, but also manipulated to perform all Clifford operations~\cite{bombin2010topological, barkeshli2013twist, levin2013protected, liu2017quon, yoder2017surface, brown2017poking, barkeshli2019symmetry}.

In this section, we introduce the surface code and its topological features, beginning with explicit examples in two-dimensions. We then interpret these features and their relationships in (2{+}1)D space-time. In particular we describe the symmetries of the code and how they relate to the behavior of anyonic excitations of the code---a tool we use throughout this paper to define and describe the behavior of topological features. These features form the anatomy of a general topological fault-tolerant logical operation, which we explore in the following section.

\subsection{The surface code, anyons, and their symmetries}
\label{secSurfaceCode}

The simplest example of the surface code~\cite{kitaev2003fault,wen2003quantum} with no topological features consists of qubits positioned on the vertices of a square lattice with periodic boundary conditions. (Note that there are many variations of the surface code, originally introduced by Kitaev~\cite{kitaev2003fault}. We utilize the symmetric, \textit{rotated} version due to Wen~\cite{wen2003quantum}, because of its better encoding rate~\cite{bombin2006topologicalencoding,bombin2007optimal,tomita2014low}, and comment on the relationship between the two in App.~\ref{appKitaevToWen}.)

\begin{figure*}
	\centering
  	\includegraphics[width=0.90\linewidth]{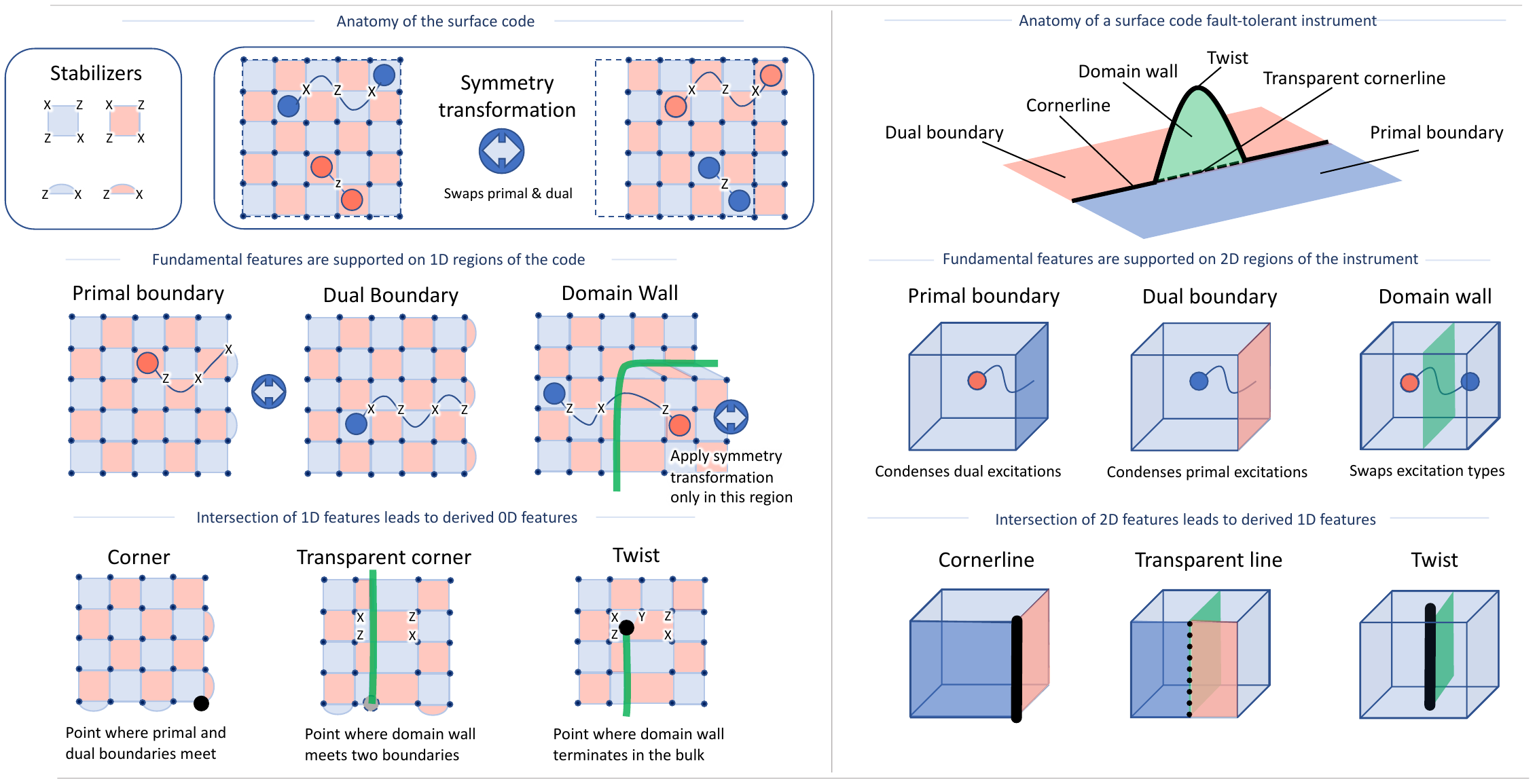}
	\caption{
	(left) Topological features in 2D and their relationship to the $\zz_2$ primal-dual symmetry transformation. Primal (dual) anyonic excitations are created on primal (dual) plaquettes at the end of open dual (primal) string operators. The symmetry swaps primal and dual stabilizer checks, and anyonic excitations. Primal boundaries absorb dual-type excitations, dual boundaries absorb primal-type excitations. Primal and dual anyons are swapped upon crossing a domain wall. Corners and transparent corners appear on the interface between primal and dual boundaries in the absence and presence of domain walls, respectively. Twists appear on the boundary of domain walls (with corresponding stabilizer shown). 
	(right) The (2{+}1) depiction of topological features and their relationships. 
	Twists and corners are two different manifestations of the same object.
	Their location is physical and has observable consequences.
	In contrast, transparent domain walls and corners can be relocated and simply correspond to a book-keeping gauge choice. 
	}
	\label{figToricCode}
\end{figure*}

\textbf{Stabilizers and logical operators.} The surface code is a stabilizer code, and for each plaquette (face) of the lattice, there is a stabilizer generator $s_{(i,j)} = Z_{(i,j)} X_{(i+1,j)} Z_{(i+1,j+1)} X_{(i,j+1)}$, where $(i,j)$ labels the vertices of the lattice, as described in Fig.~\ref{figToricCode}. We bicolor the faces of the lattice in a checkerboard pattern, as depicted in Fig.~\ref{figToricCode}, and call stabilizers on blue (red) plaquettes primal (dual) stabilizers. This primal and dual coloring is simply a gauge choice. The logical Pauli operators of the encoded qubits are associated with noncontractible cycles on the lattice, and so the number of logical qubits encoded in a surface code depends on the boundary conditions. For example, the surface code on a torus encodes two logical qubits. The surface code can also be defined for many different lattice geometries by associating qubits with the edges of an arbitrary 2D cell complex~\cite{kitaev2003fault}, but for simplicity, we restrict our discussion to the square lattice. The descriptions of topological features that follow apply to arbitrary lattice geometries, provided one first finds the corresponding symmetry representation, which, for a general surface code, is given by a constant-depth circuit.

\textbf{Errors.} 
When low-weight Pauli errors act on the toric code, they anticommute with some subset of the stabilizers, such that if measured, these stabilizers would produce ``${-}1$'' outcomes. 
The measurement outcomes for stabilizers form the classical \emph{syndrome} that can then be decoded to identify a suitable correction. 
Stabilizers with associated ${-}1$ outcome are said to be flipped and are also viewed as ``excitations" of the codespace that behave as anyonic quasiparticles. 
The behavior of these anyons has been widely studied~\cite{kitaev2006anyons,preskill1999lecture,barkeshli2019symmetry}, but for our purposes, they will be a useful tool for characterizing the behavior of topological features, and how they affect encoded logical information. 
As illustrated in Fig.~\ref{figToricCode}, anyons can be thought of as residing on any plaquette of the lattice and are created at the endpoints of open strings of Pauli operators. 
We refer to anyons residing on primal (blue) plaquettes as ``primal'' anyons, and those on dual (red) plaquettes as ``dual'' anyons \footnote{The two types of anyons of the surface code are also often also referred to as $e$-type and $m$-type, or alternatively $X$-type and $Z$-type.}. 
Primal and dual anyons are topologically distinct, in that (in the absence of topological features) there is no local operation than can change one into the other. 
We refer to the string operators that create primal (dual) anyons as dual (primal) string operators. We can understand primal (dual) stabilizers as being given by primal (dual) string operators supported on closed, topologically trivial loops. 
Similarly, Pauli-$X$ and Pauli-$Z$ logical operators consist of primal and dual string operators (respectively) supported on nontrivial loops of the lattice.

\textbf{Symmetries.} Before defining features of the surface code, we first examine its symmetries. The symmetries of the surface code allow us to explicitly construct topological features, as well as to discuss the relationships between them. We refer to any locality-preserving unitary operation (a unitary that maps local operators to local operators) that leaves the bulk codespace invariant as a \emph{symmetry} of the code. 
Transversal logical gates acting between one or more copies of a code are examples of symmetries. Symmetry operations can also be understood as an operation that, when applied to the code(s), generates a permutation on the anyon labels that leaves the behavior of the anyons (i.e., their braiding and fusion rules) unchanged (see, e.g., Refs.~\cite{beverland2016protected, yoshida2017gapped}). 
For a single surface code, there is only one nontrivial symmetry generator, which for the surface code in Fig.~\ref{figToricCode}, is realized by shifting the checkerboard pattern by one unit in either the horizontal or vertical direction. 
This symmetry permutes the primal and dual anyons, as shown in Fig.~\ref{figToricCode} (bottom left), and we refer to it as the \emph{primal-dual symmetry} or $\zz_2$ (translation) symmetry.

\subsection{Topological features in the surface code}\label{sec:topological_features_surface_code}

We now present three types of topological features that can be introduced in the surface code and how they relate to the primal-dual symmetry. The features are termed boundaries, domain walls and twist defects, and each breaks the translational invariance of the bulk in a different way. 
Explicit static examples are provided for the square lattice surface code in Fig.~\ref{figToricCode}. 
As the topological features in a code are dynamically changed over time, they trace out world lines and world sheets in space-time and are naturally interpreted as (2{+}1)D topological objects. We provide a schematic representation of these features in Fig.~\ref{figToricCode} and give a more precise meaning of such (2{+}1)D features in Sec.~\ref{sec3DElements}. 
We focus on the codimension of a feature (as opposed to its dimension), as it applies to both the 2D code and (2{+}1)D space-time instrument network. Here a codimension-$k$ feature is a ($2{-}k$)D object in a 2D code, or a ($3{-}k$)D object in a space-time instrument network.

\textbf{Primal and dual boundaries.} 
Boundaries of the surface code are codimension-1 objects that arise when the code is defined on a lattice with a boundary~\cite{bravyi1998quantum} (i.e., they form 1D boundaries in a 2D code, or 2D world sheets in a (2+1)D quantum instrument network). 
Specific anyonic excitations can be locally created or destroyed at a boundary~\cite{kitaev2012models, levin2012braiding, levin2013protected, barkeshli2013classification, lan2015gapped} (a process referred to as anyon condensation by parts of the physics community). 
In terms of the code, one can understand this in terms of error chains that can terminate on a boundary without flipping any boundary stabilizers, as shown in Fig.~\ref{figToricCode}.
There are two boundary types: primal boundaries condense dual anyons (i.e., they terminate primal error strings), while dual boundaries condense primal anyons (i.e., they terminate dual error strings). 
We can encode logical qubits using configurations of primal and/or dual boundaries; logical operators can be formed by string operators spanning between distinct boundaries. 
Note that in the 2D surface code these boundaries are often referred to as ``rough'' and ``smooth'' boundaries~\cite{bravyi1998quantum}. Examples of these two boundary types are shown in Fig.~\ref{figToricCode}.

\textbf{(Transparent) Domain walls.} 
A domain wall is a codimension-1 feature formed as the boundary between two bulk regions of the code, where the symmetry transformation has been applied to one of the regions\footnote{Note that generally, the term ``domain wall'' refers to any boundary between two topological phases. Here, we specifically use it to refer to the primal-dual swapping boundary.}, as shown in Fig.~\ref{figToricCode}. When an anyon crosses a domain wall it changes from primal to dual, or vice versa (which follows from the fact that the symmetry implements the anyon permutation). 
Microscopically this transformation can be induced by the interpretation of the checks straddling the domain wall---they are primal on one side and dual on the other (which we can view as a change of gauge).
Thus, the corresponding string operators must transform from primal to dual (and vice versa) in order to commute with the stabilizers in the neighborhood of the domain wall. 
In (2+1)D space-time, domain walls form world sheets that can be used to exchange $\overline{X}$ and $\overline{Z}$ logical operators, as well as primal and dual anyons upon crossing. We note that such domain walls are also referred to as ``transparent boundaries" in the literature \cite{kitaev2012models, lan2015gapped}.

\textbf{Corners.}
In the absence of any transparent domain walls, the codimension-2 region where primal and dual boundaries meet is called a corner [i.e., it is a 0-dimensional point in the 2D code, or a 1D line in the (2+1)D instrument]. 
Corners can condense arbitrary anyons as they straddle a primal boundary to one side and a dual boundary to the other and can be used to encode quantum information. 
For example, using surface codes with alternating segments of primal and dual boundaries, one can encode $n$ logical qubits within $2n{+}2$ corners~\cite{bravyi1998quantum,freedman2001projective}. 
In the (2+1)D context we also refer to corners as cornerlines, and their manipulation (e.g., braiding) can lead to encoded gates. 
They can be understood as twist defects (defined below) that have been moved into a boundary.

\textbf{Twists.} 
A twist is a codimension-2 object that arises when a transparent domain wall terminates in the bulk~\cite{bombin2010topological,barkeshli2013twist}. 
Similarly to corners, twist defects are topologically nontrivial objects and can carry anyonic charge. 
In particular, the composite primal-dual anyon can locally condense on a twist and one can use the charge of a twist to encode quantum information.
Indeed, twists and corners can be thought of as two variations of the same topological object; namely, a twist can be thought of as a corner that has been moved into the bulk, leaving behind a transparent corner (defined below), as is readily identified in the space-time picture as per Fig.~\ref{figToricCode}. Like corners, we can use $2n{+}2$ twists to encode $n$ logical qubits. 

\textbf{Transparent corners.}
In the presence of domain walls, primal and dual boundaries may meet in another way. We call the codimension-2 region at which a primal boundary, a dual boundary, and a domain wall meet a transparent corner. Unlike the previous corners, transparent corners carry no topological charge, cannot be used to encode logical information, and should be thought of as the region at which a primal and dual boundary are locally relabeled (i.e., a change of gauge). 

\vspace{1em}

For the purposes of defining logical block templates in the following section, we refer to the codimension-1 features (boundaries and domain walls) as \textit{fundamental features}, and the codimension-2 features (twists, cornerlines, and transparent cornerlines) as \textit{derived features}: the locations of derived features are uniquely determined by the locations of fundamental features. We remark that this terminology is a matter of convention and not a statement about the importance of a feature---indeed most encodings and logical gates can be understood from the perspective of twists and corners alone. This represents all possible features of one copy of the surface code. As we consider more copies of the surface code, the symmetry group becomes richer, and thereby corresponds to a much larger set of symmetry defects (domain walls and twists) that can be created between the codes\footnote{For example, the 2D color code (which is locally equivalent to two copies of the surface code code~\cite{bombin2012universal}) has a symmetry group containing 72 elements~\cite{yoshida2015topological,scruby2020hierarchy}, compared to the $\zz_2$ symmetry of a single surface code.}. 
Sec.~\ref{secPortals}, for example, introduces a particularly interesting defect known as a \emph{portal} that arises when we create defects between copies of the same code.

\section{Fault-tolerant instruments for the surface code in (2{+}1)D}\label{sec3DElements}

Having defined the topological features of the surface code, we now introduce a framework for fault-tolerant logical instruments that are achieved by manipulating these topological features in (2{+}1)D space-time. 
The central object we define is called a \textit{logical block template}, which is a platform-independent set of instructions for (2{+}1)D surface code fault-tolerant instruments. 
The template provides an explicit description of the space-time location of topological features (boundaries, domain walls, twists, cornerlines, and transparent cornerlines), allowing for flexibility in the design of logical operations. Templates provide a direct way of identifying checks and logical membranes of the logical instrument, as well as a method of verifying that it is fault tolerant.
They can be directly compiled to physical instructions of a QIN, prescribing the qubits, ports and instruments to implement a fault-tolerant instrument in a physical architecture for CBQC and FBQC. 

\subsection{Logical block templates: diagrammatic abstraction for (2{+}1)D topological computation}\label{secLogicalTemplate}

\begin{figure*}
	\centering
  	\includegraphics[width=0.95\linewidth]{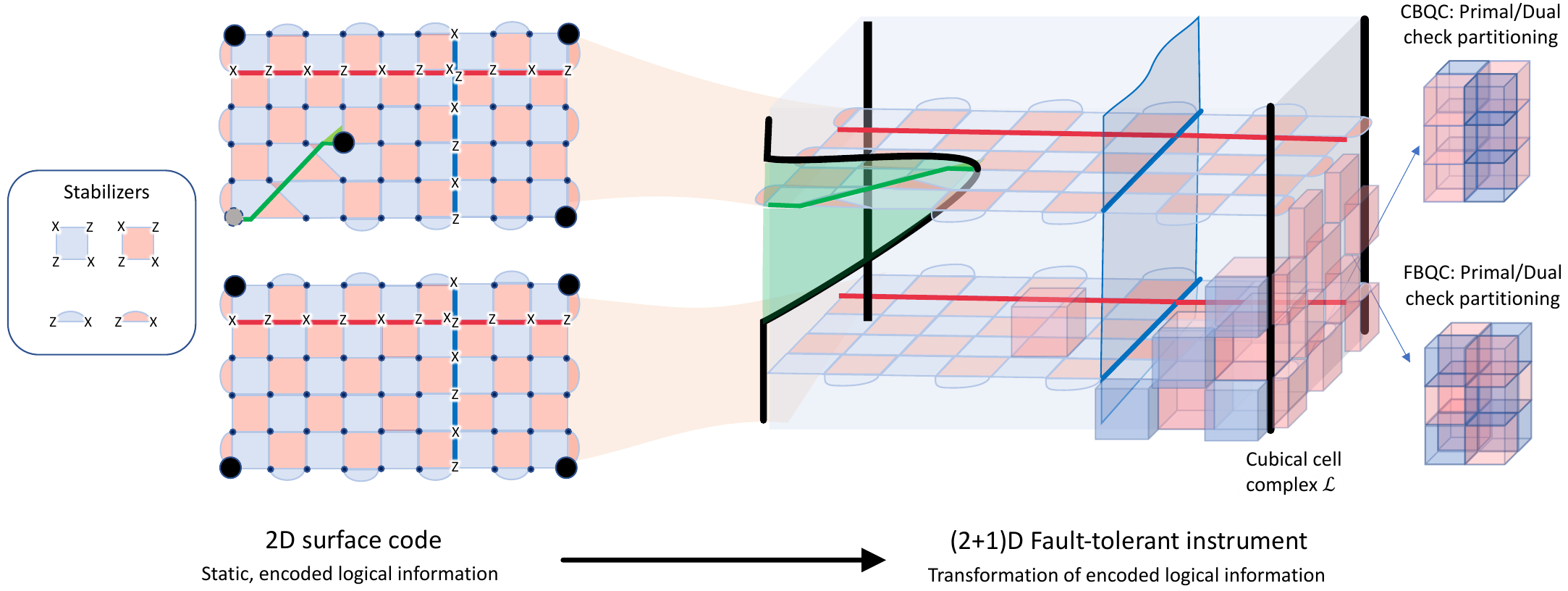}
	\caption{(left) The surface code consists of stabilizer check operators supported on primal [blue] and dual [red] plaquettes. Stabilizers on the primal (dual) boundary consist of truncated primal (dual) stabilizer operators. Logical Pauli operators for the surface code consist of operators supported on nontrivial cycles of the underlying surface, or between distinct boundaries.
	(right) Representation of a topological instrument network; if implemented in CBQC, each slice of the network describes stabilizer measurement instructions to perform on a plane of qubits. If implemented in FBQC, the network describes globally what fusions and measurements one should perform between and on resource states. 
	}
	\label{figTemplateInterpretation}
\end{figure*}

Each logical block template is given by a 3D cubical cell complex $\mathcal{L}$ and an accompanying set of cell labels. 
Here, by cubical cell complex, we mean that $\mathcal{L}$ is a cubic lattice consisting of sets of vertices $\mathcal{L}_0$, edges $\mathcal{L}_1$, faces $\mathcal{L}_2$, and volumes $\mathcal{L}_3$ with appropriate incidence relations\footnote{This cell complex is distinct from the cell complex commonly used in the context of fault-tolerant topological MBQC~\cite{raussendorf2007fault,raussendorf2007topological, nickerson2018measurement} in which the checks correspond to 3-cells and 0-cells of the complex.}.
The labels are to determine regions of the cell complex that support a primal boundary, dual boundary, domain wall, or a port. 
We denote the set of features by 
\begin{align}
\mathcal{F} = \{&  \text{PrimalBoundary}, \text{DualBoundary}, \text{DomainWall}, \nonumber \\ 
& \text{Port}_i, \emptyset \},
\end{align}
where $i$ indexes the distinct ports of the instrument network (e.g., $1, 2, 3$ for the network depicted in Fig.~\ref{figAbstractChannel}). The label $\emptyset$ is used as a convenience to denote the absence of any feature.

We now formally define a logical block template as follows.
\begin{defn} 
	A logical block template is a pair $(\mathcal{L}, F)$ where $\mathcal{L}$ is a cubical cell complex and $F: \mathcal{L}_2 \rightarrow \mathcal{F}$ is a labeling of the 2-cells $\mathcal{L}_2.$
\end{defn}

To define a logical block template, we have only specified the location of the ports and \textit{fundamental}, codimension-1 features (the boundaries and domain walls). 
The \textit{derived}, codimension-2 features (twists, cornerlines, and transparent cornerlines) are all inferrable from them. 
Recall that twists reside at the boundary of a domain wall in the absence of any boundaries, while cornerlines (transparent cornerlines) reside on the interface between primal and dual boundaries in the absence (presence) of a domain wall.

In order to simplify the construction of derived features, it is convenient to decompose the label for topological features into two indicator functions $B$ and $T$ on $\mathcal{L}_2$.
The first one,  $B$, indicates whether the 2-cell is a boundary or not (this may simply be derived as $\partial \mathcal{L}_3$).
The second, $T$, identifies transparent domain walls as well as distinguishing dual boundaries ($B \wedge T$) from primal boundaries ($B \wedge \neg T$). 
The union of {\it twist defects} and {\it cornerlines} can then be identified as $\partial T$, of which only $(\partial T) \cap B$ are identified as {\it cornerlines}.
Finally, elements of $\partial (T \cap B)$ that are not in $\partial T$ are considered {\it transparent cornerlines}.
In this way, cornerlines, transparent cornerlines, and twists can be described as labels on 1-cells of the template and one can extend the domain of $F$ to include 1-cells accordingly (see App.~\ref{secLogicalBlockTemplateExtension} for the explicit definition of the extension). 

\textbf{Remarks.}
The logical block templates do not make any explicit reference to a causal or temporal direction.
As such, they provide a natural starting point to describe pictures of topological fault tolerance that do not explicitly present a local temporal ordering such as FBQC and MBQC.
In order to ascribe a circuit model interpretation, it is necessary to extend the template with a causal (i.e., temporal) order compatible with the input and output status of logical ports.
Contrary to the static 2D case, boundaries, domain walls, twists, cornerlines and transparent cornerlines may all exist along planes normal to the time direction. 
The physical operations generating such features will be explained in the following sections. 

Finally, we remark that the cubical complex is natural for blocks based on the square lattice surface code (in CBQC) and the six-ring fusion network (in FBQC). 
For other surface code geometries or fusion networks one can generalize the logical block template to other cell complexes.
This case remains important for the 3-cells in the bulk to remain 2-colorable as they will continue to represent primal and dual checks and locations where these are not locally two colorable will correspond to twist defects (and cornerlines if one considers the exterior as a color).
Twist defects in $\mathcal{L}_1$ are associated with an odd number of incident faces from $\mathcal{L}_2$.

\subsection{Fault-tolerant instruments from logical block templates}

Logical templates define fault-tolerant logical instruments without reference to the computational model, but can be directly compiled into physical instructions for different models of computation. We now explain how the different features of the template correspond to measurement instructions, checks and logical membranes.

\textbf{Compiling templates to physical instructions.}
Logical block templates can be directly compiled into a network of quantum instruments realizing surface-code-style fault tolerance for CBQC, and FBQC.
We provide an overview of this mapping here, leaving the explicit mapping from templates to CBQC instructions in App.~\ref{secTemplateToCBQC}, and to FBQC instructions in Sec.~\ref{sec:topological_fbqc}. 
One can also obtain MBQC instructions on cluster states using the framework of Ref.~\cite{brown2020universal} applied to the CBQC instructions.

To compile a template into physical instructions for CBQC we must choose a coordinate direction as the temporal direction or otherwise equip the template with a causal structure. 
From this, each time slice defines a 2D subcomplex of the template, each vertex of which corresponds to a qubit, and each bulk 2-cell of which corresponds to a bulk surface code stabilizer measurement. The feature labels on the subcomplex result in modifications to the measurement pattern, as per Fig.~\ref{figToricCode} and as shown in the example in Fig.~\ref{figTemplateInterpretation}.

In FBQC, the symmetry between space and time is maintained, and measurements may be performed in any order. 
The flavor of FBQC presented in \cite{bartolucci2021fusion}, which uses 6-qubit ring graph states as resource states, is naturally adapted to the logical block template. 
To each vertex of the template we place a resource state, while each edge of the template corresponds to a fusion measurement (a two-qubit projective measurement) between resource states as determined by the feature labels.

\textbf{Checks.}
There is a close connection between the elements used to describe topological codes with those required to characterize fault-tolerant instruments.
For instance, stabilizer operators of the code give rise to check operators for the QIN as the stabilizers are repeatedly measured. 
Similarly, the logical operators of the code give rise to logical correlators, which track how the corresponding degree of freedom map between ports.
In topological fault tolerance, going from code to protocol involves increasing the geometric dimension by one, which in the circuit model is naturally interpreted as the temporal direction.
In particular, checks correspond to parity constraints on the outcomes of operators supported on closed (i.e., without boundary), homologically trivial surfaces of codimension 1 (i.e., two dimensional) in the template complex. 
This is analogous to how 2D surface code stabilizers consist of Pauli operators supported on closed, homologically trivial loops.
In particular, the surface of every bulk 3-cell of the template corresponds to a check, in the following way. 
In CBQC, a surface code stabilizer measurement repeated between two subsequent timesteps gives rise to a check, and this check can be identified with the 3-cell whose two faces the measurements are supported on. 
In FBQC, fusion measurements between resource states supported on the vertices of a 3-cell constitute a resource state stabilizer, and thus a check (this will be carefully validated in Sec.~\ref{sec:topological_fbqc}).

Much like the two-dimensional case, bulk check generators can be partitioned into two disjoint sets, either primal or dual, as depicted in Fig.~\ref{figTemplateInterpretation}. 
For CBQC, this partition consists of the 2D checkerboard pattern extended in time, while for FBQC, the primal and dual checks follow a 3D checkerboard pattern. Thus, we may label a bulk 3-cell (and its surface) by either Primal or Dual, depending on what subset it belongs to. These checks can be viewed in terms of Gauss's law---they detect the boundaries of chains of errors\footnote{In condensed matter language, this check operator group can be understood as a $\zz_2\times \zz_2$ 1-form symmetry~\cite{gaiotto2015generalized,kapustin2017higher,roberts2020symmetry}.}. 
The presence of features modifies the check operator group: primal and dual boundaries lead to checks supported on truncated 3-cells, while defects and twists lead to checks supported on the surfaces of pairs of 3-cells sharing a defect 2-cell or twist 1-cell. 
We discuss the check operator group and how it is modified by features in much more detail in Sec.~\ref{sec:topological_fbqc} for FBQC and App.~\ref{secTemplateToCBQC} for CBQC.

\textbf{Logical correlators and membranes}.
Logical membranes determine how logical information is mapped between input and output ports of the instrument. 
In the template, logical membranes---which can be thought of as the world sheets of logical operators---are supported on closed, homologically nontrivial surfaces of codimension 1. 
Logical membranes can be obtained by finding relatively closed surfaces $M \subseteq \mathcal{L}_2$, each with their 2-cells taking labels from $\mathcal{F}_M = \{\text{Primal}, \text{Dual}, \text{PrimalDual}\}$ satisfying certain requirements. 
Here, by relatively closed, we mean that the surface is allowed to have a boundary on the boundary of the template cell complex. 
The labels must satisfy the following conditions: (i) only faces with Primal (Dual) labels can terminate on primal (dual) boundaries, (ii) upon crossing a domain wall, the Primal and Dual labels of the membrane are exchanged (and the PrimalDual label is left invariant). 

Any such surface corresponds to a logical membrane in the following ways. 
In CBQC, the membrane can be projected into a given time slice where it corresponds to a primal, dual, or composite logical string operator. The components of a membrane in a plane of constant time correspond to stabilizer measurements that must be multiplied to give the equivalent representative on that slice (and thus their outcome is used to determine the Pauli frame). In FBQC, the membrane corresponds to a stabilizer of the resource state, whose bulk consists of a set of fusion and measurement operators used to determine the Pauli frame. In both cases, when projected onto a port, these membranes correspond to a primal-type ($X$-type), dual-type ($Z$-type), or a composite primal-dual-type ($Y$-type) string logical operator. Checks can be considered ``trivial'' logical membranes, with logical membranes forming equivalence classes up to multiplication by them (i.e., by local deformations of the membrane surfaces).

\textbf{Logical errors}. 
Errors can be understood at the level of the template. 
Elementary errors---Pauli or measurement errors---are categorized as either primal or dual, according to whether they flip dual or primal checks, respectively. 
Undetectable chains of elementary errors comprise processes involving creation of primal or dual excitations, propagating them through the logical instrument, transforming them through domain walls, and absorbing them into boundaries or twists. 
Specifically, primal (dual) excitations can condense on dual (primal) boundaries, composite primal-dual excitations can condense on twists, and primal and dual excitations are swapped upon crossing a transparent domain wall. 
The primal and dual components of a logical membrane can be thought of as measuring the flux of primal and dual excitations, respectively. 
If an undetectable error results in an odd number of primal or dual excitations having passed through the primal and dual components of a logical membrane, then a logical error has occurred. 
The fault distance of the logical instrument is the weight of the smallest weight logical error.

\textbf{Graphical conventions.}
Throughout the rest of the paper, as per Figs.~\ref{figLogicalBlockConcatenation},~\ref{figToricCode}, and ~\ref{figTemplateInterpretation}, we pictorially represent primal boundaries and logical membranes in blue, while dual boundaries and dual logical membranes will be represented in red. Domain walls are represented in green.

\section{Universal block-sets for topological quantum computation based on planar codes}\label{secLogicBlocks}

In this section we apply the framework of logic block templates to construct a universal set of fault-tolerant instruments based on planar codes~\cite{bravyi1998quantum}. 
The gates we design are based on fault-tolerant Clifford operations combined with noisy magic state preparation, which together are sufficient for universal fault-tolerant logic (via magic state distillation~\cite{bravyi2005universal,bravyi2012magic,litinski2019game}). To the best of our knowledge, some of the Clifford operations we present---in particular the phase gate and controlled-NOT gate---are the most efficient versions in the literature in terms of volume (defined as the volume of the template cell complex). These Clifford operations form the backbone of the quantum computer, and set for instance the cost of and rate at which magic states can be distilled and consumed (the latter of which can become quite expensive for applications with large numbers of logical qubits~\cite{litinski2019magic,kim2021faulttolerant}). Later, in Sec.~\ref{secPortals}, we show another way of performing Clifford operations---namely, Pauli product measurements---for twist-encoded qubits, using a space-time feature known as a portal. These portals require long-range operations in general, and will be discussed in the context of FBQC.

\begin{figure*}[t]
    \centering
	\includegraphics[width=0.86\linewidth]{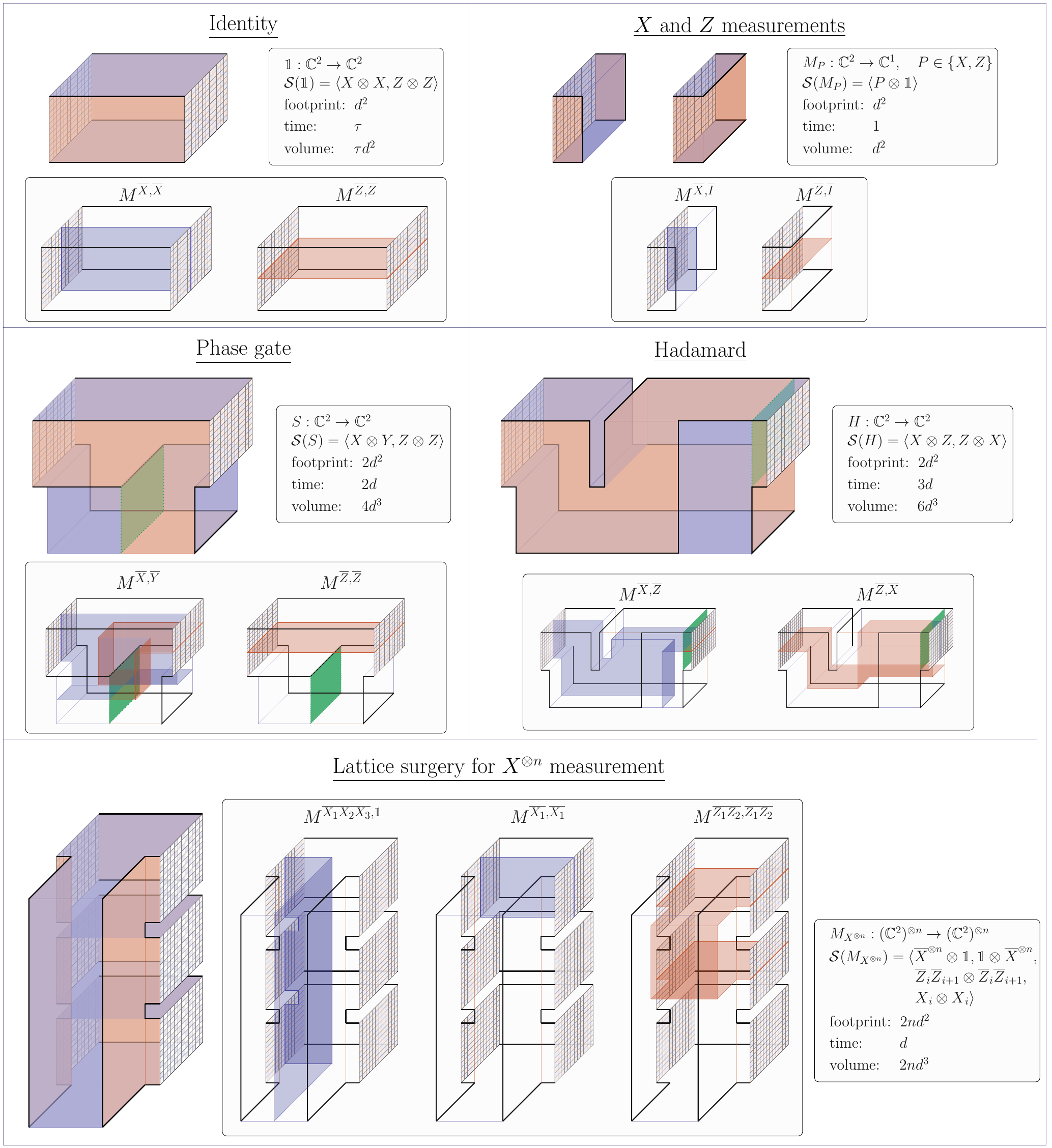} 
	\caption{Logical templates for Clifford operations. Primal and dual boundaries are depicted in blue and red, domain walls in green, cornerlines and twists in bold black. For each operation $U$, a membrane $M^{\overline{P}, \overline{Q}}$ is shown for each generating logical correlator $\overline{P}\otimes\overline{Q}\in \mathcal{S}(U)$. Time and footprint overheads are based on traversing the block from left to right in 2D slices, however we emphasize that networks of such operations may be traversed in any direction. \textbf{The identity gate} for time $\tau$. 
    \textbf{Measurements of Pauli $X$ and $Z$} amount to placing a primal or dual boundary and can be done in constant time; the measurement outcome is inferred from the logical membrane terminating on the boundary. Preparations are given by the respective time-reversed template. 
    \textbf{The phase gate} can be realised by the braiding of two cornerlines. One of the cornerlines is propagated as a twist through the bulk, while the other moves around the exterior of the block.
    \textbf{The Hadamard} can be realised by a clockwise or counterclockwise rotation of the cornerlines. Note that the "trench" created by the dual boundary may be as thin as a width-0 plane (meaning the boundaries on either side coincide). 
    \textbf{Lattice surgery} to measure $X^{\otimes{n}}$ can be realised by fusing together cornerlines from different logical qubits. The outcome of the logical measurement is inferred from the $M^{\overline{X}^{\otimes n}, \id}$ membrane terminating on the boundary.
	}
	\label{figAllLogicalBlocks}
\end{figure*}

\subsection{Planar code logical block templates for Clifford operations}\label{secCliffordOperators}

We begin by defining logical block templates for a generating set of the Clifford operations on planar codes~\cite{bravyi1998quantum}: the Hadamard, the phase gate, the lattice surgery for measuring a Pauli-$X$ string, along with Pauli preparations and measurements.
Recall that a fault-tolerant instrument realizes an encoded version of the Clifford operator $U\in \mathcal{C}_{l,m}$ if it has $l$ distinct input ports, $m$ distinct output ports, and an equivalence class of logical membranes $M^{\overline{P},\overline{Q}}$ for every logical correlator $P\otimes Q \in \mathcal{S}(U)$. 
Each of the block templates are depicted in Fig.~\ref{figAllLogicalBlocks} along with a generating set of membranes, showing the mapping of logical operators between the input and output ports. To complete a universal block set, we present the template for noisy magic state preparation in Fig.~\ref{figFBQCMagicStatePreparation} and App.~\ref{secMagicStatePreparation}. We reemphasize that the instruments we discuss are only true up to a random but known Pauli operator---known as the Pauli frame. The logical membranes determine the measurement outcomes that are be used to infer this Pauli frame.

\textbf{Logical block criteria.}
The logical blocks we present are designed for planar code qubits with the following criteria in mind:
(i) They each have a fault distance of $d$ for both CBQC and FBQC under independent and identically distributed (IID) Pauli and measurement errors, where $d$ is the (tunable) distance of the planar codes on the ports.
(ii) They are composable by transversal concatenation and that composition preserves distance. This means that for a given distance, we can compose two blocks together by identifying the input port(s) of one with the output(s) of another. In particular, this requires that inputs and output ports are a fixed planar code geometry.  
(iii) They admit a simple implementation in 2D hardware: after choosing any Cartesian direction as the flow of physical time, the 2D time slices can be realized in a 2D rectangular layout. 

These criteria allow the logical blocks to be used in a broad range of contexts, however, it is possible to find more efficient representations of networks of logical operations by relaxing them. For instance, we need not require the input and output codes to be the same, as we can classically keep track of rotations on the planar code and compensate accordingly. Secondly, networks of such operations can be compiled into more efficient versions with the same distance. In particular, one may also find noncubical versions of these blocks with reduced volumes.

\textbf{Hadamard and phase gates.}
To the best of our knowledge, the phase gate presented in Fig.~\ref{figAllLogicalBlocks} is the most volume-efficient representation in the literature for planar code qubits using the so-called rotated form~\cite{nussinov2009symmetry,bombin2007optimal,beverland2019role}. Moreover, the scheme we present can be implemented in CBQC with a static 2D planar lattice using at most four-qubit stabilizer measurements on neighboring qubits on a square lattice. In particular, the physical operations for the phase gate can be ordered in a way that does not require the usual five-qubit stabilizer measurements~\cite{bombin2010topological,litinski2019game} or modified code geometry~\cite{yoder2017surface,brown2020universal} that are typically required for twists. We explicitly show how to implement this phase gate in CBQC in Fig.~\ref{figPhaseGateCBQC} in App.~\ref{AppPhaseGateCBQC}. We remark that with access to nonlocal gates or geometric operations called ``folds'', one can find an even more efficient phase gate based on the equivalence of the toric and color codes~\cite{kubica2015unfolding,moussa2016transversal}. The Hadamard gate has the same volume as the ``plane rotation" from Ref.~\cite{litinski2019game}.

\textbf{Lattice surgery and Pauli product measurements}.
To complete the Clifford operations, we consider the nondestructive measurement of an arbitrary $n$-qubit Pauli operator, known as a Pauli product measurement (PPM)~\cite{elliott2009graphical} (a general Clifford computation can be performed using a sequence of PPMs alone~\cite{litinski2019game}). By nondestructive, we mean that only the specified Pauli is measured, and not, for example, its constituent tensor factors. A general PPM $M_{P}: (\mathbb{C}^2)^{\otimes n} \rightarrow (\mathbb{C}^2)^{\otimes n}$,  $P \in \mathcal{P}_n$, has a stabilizer given by
\begin{align}
\mathcal{S}(M_{P}) = 
\langle P \otimes I, I \otimes P, Q\otimes Q^{\dagger} ~|~ Q\in \mathcal{Z}_{\mathcal{P}_n}(P) \rangle.
\end{align}
These PPMs can be performed using lattice surgery~\cite{horsman2012surface, litinski2019game}. With access to single-qubit Clifford unitaries, an arbitrary PPM can be generated using lattice surgery in a fixed basis, such as the $X$ basis as depicted in Fig.~\ref{figAllLogicalBlocks}.

To efficiently perform a general PPM using lattice surgery~\cite{horsman2012surface}, one may utilize planar codes with six corners, as described in Ref.~\cite{litinski2019game}. Each six-corner planar code encodes two logical qubits and supports representatives of all logical operators $\overline{X}_i$, $\overline{Y}_i$, and $\overline{Z}_i$ of each qubit $i\in \{1,2\}$ on a boundary. This enables us to measure arbitrary $n$-qubit Pauli operators in a more efficient way as no single-qubit Clifford unitaries or code rotations are required between successive lattice surgery operations. The price to pay is that single-qubit Pauli measurements and preparations can no longer be done in constant time. As a further improvement, by utilizing periodic boundary conditions, we can compactly measure any logical Pauli operator on $k$ logical qubits using a block with at most $2kd^3$ volume, as shown in Ref.~\cite{bombin2021interleaving, kim2021faulttolerant}.

\begin{figure}
	\centering
	\includegraphics[width=0.47\linewidth]{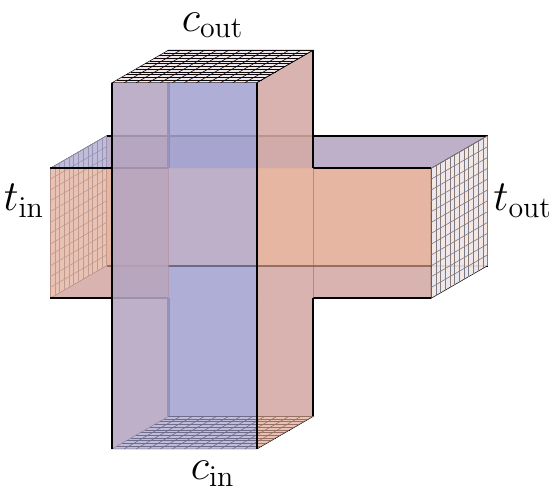} \quad
	\includegraphics[width=0.47\linewidth]{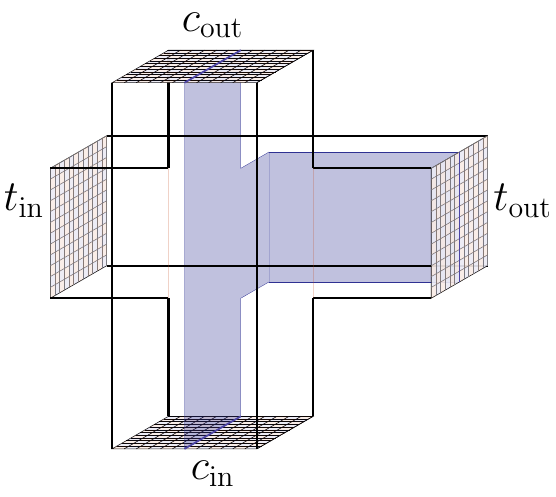} 	
	\caption{ (left) A $CX$ (CNot) gate with control qubit $c$ and target qubit $t$ has two ports which are treated as inputs ($c_\text{in}$ and $t_\text{in}$) and two ports which are treated as outputs  ($c_\text{out}$ and $t_\text{out}$). 
	The control qubit $c$ is presented as progressing from bottom to top, whereas the target qubit is presented as progressing from left to right.
	 (right) The support of one instance of the logical correlator membrane  is presented.
	 In this case, the support of the membrane at the ports fully determines which stabilizer it represents at the level of the encoded input and output qubits $X_{c_\text{in}}\otimes X_{c_\text{out}}X_{t_{\text{out}}} \in \mathcal{S}(CX)$. 
	 Other correlator membranes and stabilizer generators may be similarly determined}
	\label{figBlockCNOT}
\end{figure}

\textbf{Qubits in space and time: from lattice surgery to controlled-Pauli operations}.
The logical flow (the order in which the logical block maps inputs to outputs) and physical flow (the order in which physical operations are implemented to realize the fault-tolerant instrument) for a logical block do not need to be aligned. One can take advantage of this in the design of logic operations. In particular, each qubit participating in the lattice surgery may be regarded as undergoing a controlled-$X$, controlled-$Y$, or controlled-$Z$ gate with a spacelike ancilla qubit (i.e., a qubit propagating in a spatial direction). For example, the $X$-type lattice surgery of Fig.~\ref{figAllLogicalBlocks} can be understood as preparing an ancilla in the  $\ket{+}$ state, performing controlled-$X$ gates between the ancilla and target logical qubits, then measuring the ancilla in the $X$ basis. 

One can use this to define a logical block template for the controlled-$X$ gate, as depicted in Fig.~\ref{figBlockCNOT}. Therein, one can verify that the block induces the correct action by finding membranes representing stabilizers of $CX: (\mathbb{C}^2)^{\otimes 2}\rightarrow (\mathbb{C}^2)^{\otimes 2}$
\begin{align}
\mathcal{S}(CX) = \langle X_1\otimes X_1X_2, X_2 \otimes X_2,~ \nonumber\\
 Z_1 \otimes  Z_1,  Z_2 \otimes Z_1Z_2 \rangle.
\end{align}
One can similarly define $CZ$ and $CY$ gates by appropriately including domain walls in the template. The stabilizers for the $CZ$ and $CY$ can be obtained by applying a Hadamard or phase to the second qubit after the $CX$. 

In the next section, we expand on this concept by designing a logic scheme using ``toric code spiders'' that can be considered an alternative approach to lattice surgery.

\section{Assembling blocks into circuits: Concatenating LDPC codes with surface codes}\label{secConcatenation}

We now introduce a logic scheme that takes advantage of the space-time flexibility inherent to surface code logical blocks and show how to generate larger circuits using these building blocks. An application of this scheme is the construction of fault-tolerant schemes based on surface codes concatenated with more general LDPC codes.

While the surface code (and topological codes) are advantageous due to their very high thresholds, they are somewhat disadvantaged by their asymptotically zero encoding rate (i.e., the ratio of encoded logical qubits to physical qubits vanishes as the code size goes to infinity). Fortunately, there are families of LDPC codes that have nonzero rates~\cite{tillich2014quantum,gottesman2013fault,fawzi2018constant,fawzi2018efficient,breuckmann2021ldpc,hastings2021fiber,breuckmann2021balanced,panteleev2021asymptotically}, meaning that the number of encoded logical qubits increases with the number of physical qubits. Such codes may offer ways of greatly reducing the overhead for fault-tolerant quantum computation~\cite{gottesman2013fault}.

Code concatenation allows us to take advantage of the high threshold of the surface code and the high rates of LDPC codes. The space-time language of the previous sections provides a natural setting for the construction and analysis of the resulting codes. The building blocks for these constructions consist of certain topological projections that we refer to as ``toric code spiders''---these correspond to encoded versions of the $Z$ and $X$ spiders of the ZX-calculus~\cite{coecke2008interacting,van2020zx,de2020zx}. Spiders are encoded Greenberger-Horne-Zeilinger (GHZ) basis projections for surface code qubits. Our protocol is intended to be illustrative, and we emphasize that further investigation into the performance of such concatenated codes is an interesting open problem.

\subsection{Toric code spiders}
We now define spiders and toric code spiders, the building blocks for the concatenated codes we consider. There are two types of spiders, which we label by $\spiderx{k}$ and $\spiderz{k}$, where $k\in \nn$ labels the number of input and output ports. We do not distinguish between input and output ports here, and, as such, we write the stabilizer groups for $k>1$ as 
\begin{align}
    \mathcal{S}(\spiderx{k}) =& \langle Z^{\otimes k}, X_{i}X_{i+1} ~|~ i = 1,\ldots, k-1 \rangle,  \\
    \mathcal{S}(\spiderz{k}) =& \langle X^{\otimes k}, Z_{i}Z_{i+1} ~|~ i = 1,\ldots, k-1 \rangle,
\end{align}
with $\mathcal{S}(\spiderx{1}) = \langle Z \rangle$ and $\mathcal{S}(\spiderz{1}) = \langle X \rangle$. If all ports are considered outputs, then spiders can regarded as preparing GHZ states (up to normalization)
\begin{align}
    \ket{\spiderx{k}} &= \ket{+\ldots +} + \ket{-\ldots -}, \label{eqSpiderxState} \\
    \ket{\spiderz{k}} &= \ket{0\ldots 0} + \ket{1\ldots 1}. \label{eqSpiderzState} 
\end{align}
Similarly, if all ports are considered inputs, then the spider performs a GHZ basis measurement. The flexibility arises by considering networks of spiders where each spider may have both input and output ports, i.e., where each type of spider $\spiderx{k}$ and $\spiderz{k}$ is a map $(\mathbb{C}^2)^{\otimes k-a} \rightarrow (\mathbb{C}^2)^{\otimes a}$ for any choice of $a\in \{0,1,\ldots, k\}$, and can be obtained by turning some kets to bras in Eqs.~(\ref{eqSpiderxState}), (\ref{eqSpiderzState}). We note that in this language, Pauli-$X$ (Pauli-$Z$) measurements and preparations may be regarded as 1-port spiders $\spiderz{1}$ ($\spiderx{1}$), while the identity gate may be regarded as 2-port spiders of either type.

\begin{figure}
	\centering
	\includegraphics[width=0.9\linewidth]{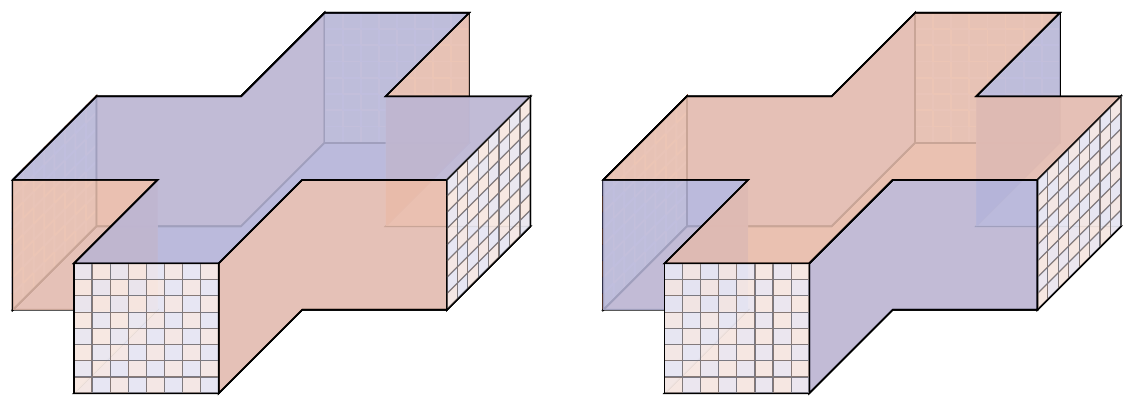} 
	\includegraphics[width=0.9\linewidth]{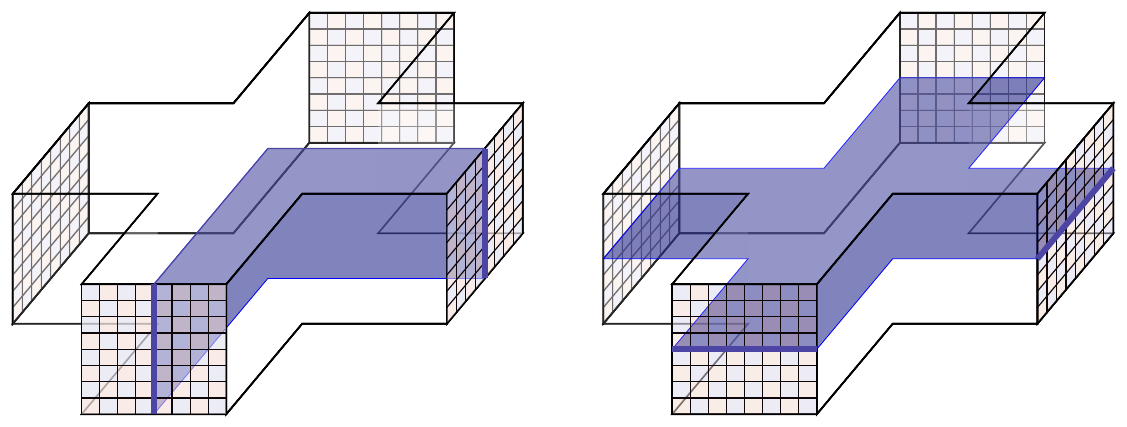} 
	\caption{Logical block templates for toric code spiders. On the left (right) we have the toric code spider for the $\spiderx{4}$ ($\spiderz{4}$) operation. Lengths of the legs are exaggerated for illustration purposes. On the bottom row we depict two logical membranes; labelling the ports 1 to 4 clockwise from the left, we have example membranes $X_3X_4 \in \mathcal{S}(\spiderx{4})$ and $X_1X_2X_3X_4 \in \mathcal{S}(\spiderz{4})$. The length of the legs of the spider is exaggerated for illustration purposes.}
	\label{figToricCodeSpiders}
\end{figure}

Toric code spiders are logical blocks representing encoded versions of these spiders such that each input and output is a qubit encoded in a surface code.
We depict example toric code spiders corresponding to $\spiderx{4}$ and $\spiderz{4}$ in Fig.~\ref{figToricCodeSpiders}.

\textbf{Stabilizer measurements.}
By composing many toric code spiders in a network along with single-qubit Hadamard and phase gates, we can perform any Clifford circuit. Such circuits can be used to measure the stabilizers of any stabilizer code, and in the following we show how networks of spiders alone are sufficient to measure the stabilizers of any Calderbank-Shor-Steane (CSS) code~\cite{calderbank1996good,steane1996multiple}, thus giving us a recipe to measure the stabilizers of the concatenated codes of interest.

\begin{figure*}
	\centering
	\includegraphics[width=0.27\linewidth]{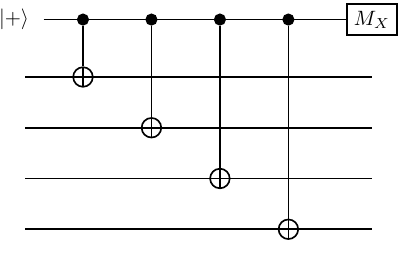} \qquad 
	\includegraphics[width=0.27\linewidth]{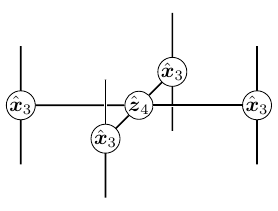} \qquad
	\includegraphics[width=0.27\linewidth]{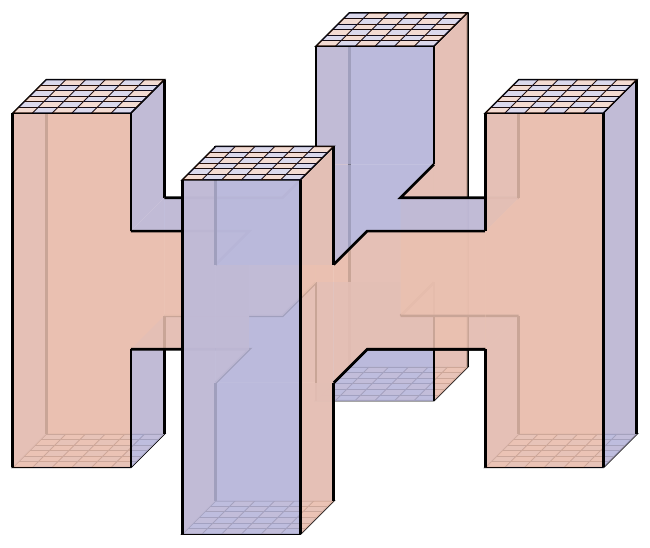}
	\caption{Arranging toric code spider networks for encoded Clifford circuits. (left) a Clifford circuit for measuring the Pauli operator $X^{\otimes 4}$ using an extra ancilla. Time moves from left to right. (middle) Converting the Clifford circuit into a network of $\spiderx{k}$ and $\spiderz{k}$ operations. Time moves from bottom to top. (right) The resulting space-time network. In App.~\ref{appSpiderNetworks} we demonstrate how the spider network is obtained, along with a larger network measuring encoded surface code stabilizers. Time moves from bottom to top. The length of the legs of the toric code spiders along with their spacing is exaggerated for illustration purposes.}
	\label{figToricCodeSpiderNetwork}
\end{figure*}

As a simple example, consider the Clifford circuit depicted in Fig.~\ref{figToricCodeSpiderNetwork} that uses one ancilla to measure the Pauli operator $X^{\otimes 4}$ on four qubits. Such a circuit may be regarded as the syndrome measurement of a surface code stabilizer. One can rewrite the circuit as a network of operations consisting of $\spiderx{k}$ and $\spiderx{k}$ (also depicted in the figure), where each spider is represented by a $k$-legged tensor that is composed along ports. One can verify this using standard stabilizer techniques or using the ZX-calculus~\cite{coecke2008interacting,van2020zx} (see also App.~\ref{appSpiderNetworks} for more details). We can arrange a space-time network as depicted in the figure. To measure $Z^{\otimes 4}$, one simply swaps the roles of $\spiderx{k}$ and $\spiderz{k}$.

\textbf{Concatenation with toric codes.} 
The previous example of performing stabilizer measurements can be generalized to the stabilizers of arbitrary concatenated CSS codes. Namely, denote the concatenation of an \textit{inner code} $C_{\text{in}}$ and an \textit{outer code} $C_{\text{out}}$ by $C_{\text{out}}\circ C_{\text{in}}$. It is the result of encoding each logical qubit of $C_{\text{out}}$ into $C_{\text{in}}$. We consider using a surface code as the inner code, and a general CSS LDPC code as the outer code, (The high-threshold surface code is used to suppress the error rate to below the threshold of the LDPC outer code.)

We can construct toric code spider networks to measure the stabilizers of the outer code $C_{\text{out}}$ as follows. 
For each round of $X$-type ($Z$-type) measurements of $C_{\text{out}}$, we place 
\begin{enumerate}
    \item a $\spiderx{\delta + 2}$ ($\spiderz{\delta + 2}$) spider for each code qubit, with $\delta$ equal to the number of $X$ ($Z$) stabilizers the code qubit is being jointly measured with 
    (the two additional ports can be considered as an input and an output port for the code qubit), 
    \item a $\spiderz{k}$ ($\spiderx{k}$) spider for each stabilizer measurement, connected to the corresponding code qubit spiders.
\end{enumerate}
Whenever an $X$ or $Z$ spider with more than four legs arises, it can be decomposed into a sequence of connected spiders with degree at most 4, and thus implementable with toric code spiders. Such a protocol allows us to measure stabilizers of the outer code. We emphasize that in constructing the instrument network to realize the stabilizer measurements, only the graph topology matters, and one may order the instruments in many different ways. 

As a simple example, we depict a surface code concatenated with itself in Fig.~\ref{figConvertingCircuitToSpider} in App.~\ref{appSpiderNetworks}. Note that for simplicity, the outer surface code is the Kitaev version that consists of independent $X$-type and $Z$-type generators.

In general, the stabilizers of a LDPC code cannot be made local in a planar layout, and long-range interactions will be required (necessarily so if the code has an nonzero rate and nonconstant distance~\cite{bravyi2010tradeoffs}). Such long-range connections can be facilitated by toric code spiders with long legs (topologically these ``long legs'' look like long identity gates), each of which comprises local operations, thus dispensing with the need for long-range connections. Alternatively, one may connect distant toric code spiders using portals (as we describe in Sec.~\ref{secPortals}). 

For example, embedding the qubits of an $[N,K,D]$ quantum LDPC code in a finite-dimensional Euclidean space will in general require connections between qubits of range $r = \mathrm{poly}(N)$. If the error rate on qubits is proportional to their separation, one can use the surface code concatenation scheme to reduce the error rate experienced by the qubits of the (top-level) LDPC code to a constant rate (independent of range). In particular, this can be achieved using surface code spiders with a distance of $d = \mathcal{O}(\log(r_{\mathrm{max}})) = \mathrm{polylog}(N)$ for each connection (where $r_{\mathrm{max}}$ is the largest qubit separation). This leads to a protocol with rate $\frac{K}{N \mathrm{polylog}(N)}$, which despite being asymptotically zero, may still be larger than a pure topological encoding, providing one starts with a LDPC code with good rate~\cite{hastings2021fiber,breuckmann2021balanced,panteleev2021asymptotically}.

\textbf{Logical operations.} 
One can perform fault-tolerant logical gates on such codes using networks of toric code spiders along with magic state injection.
In particular, gates are performed by measuring a target logical Pauli operator (a PPM) jointly with an ancilla magic state~\cite{litinski2019game}. The logical Pauli operator is measured either directly or through a code deformation approach~\cite{gottesman2013fault,krishna2021fault,cohen2021low}, using a toric code spider network as described above. The ancilla magic states can be obtained by magic state distillation at the inner (surface) code level. The initial state preparations in the $\overline{X}$ and $\overline{Z}$ bases can be achieved by preparing the inner code qubits in either the $X$ or $Z$ eigenbasis (through an appropriate choice of primal or dual boundaries on the input) followed by a round of measurement of the outer code stabilizers\footnote{More generally, the outer code encoding circuit for an arbitrary state is Clifford for a general CSS code, but it  may not be constant depth.}. This approach may be viewed as a spider network approach to LDPC lattice surgery.

\section{Implementing logical blocks in topological fusion-based quantum computation}
\label{sec:topological_fbqc}

\begin{figure*}
	\centering
	\includegraphics[width=0.95\linewidth]{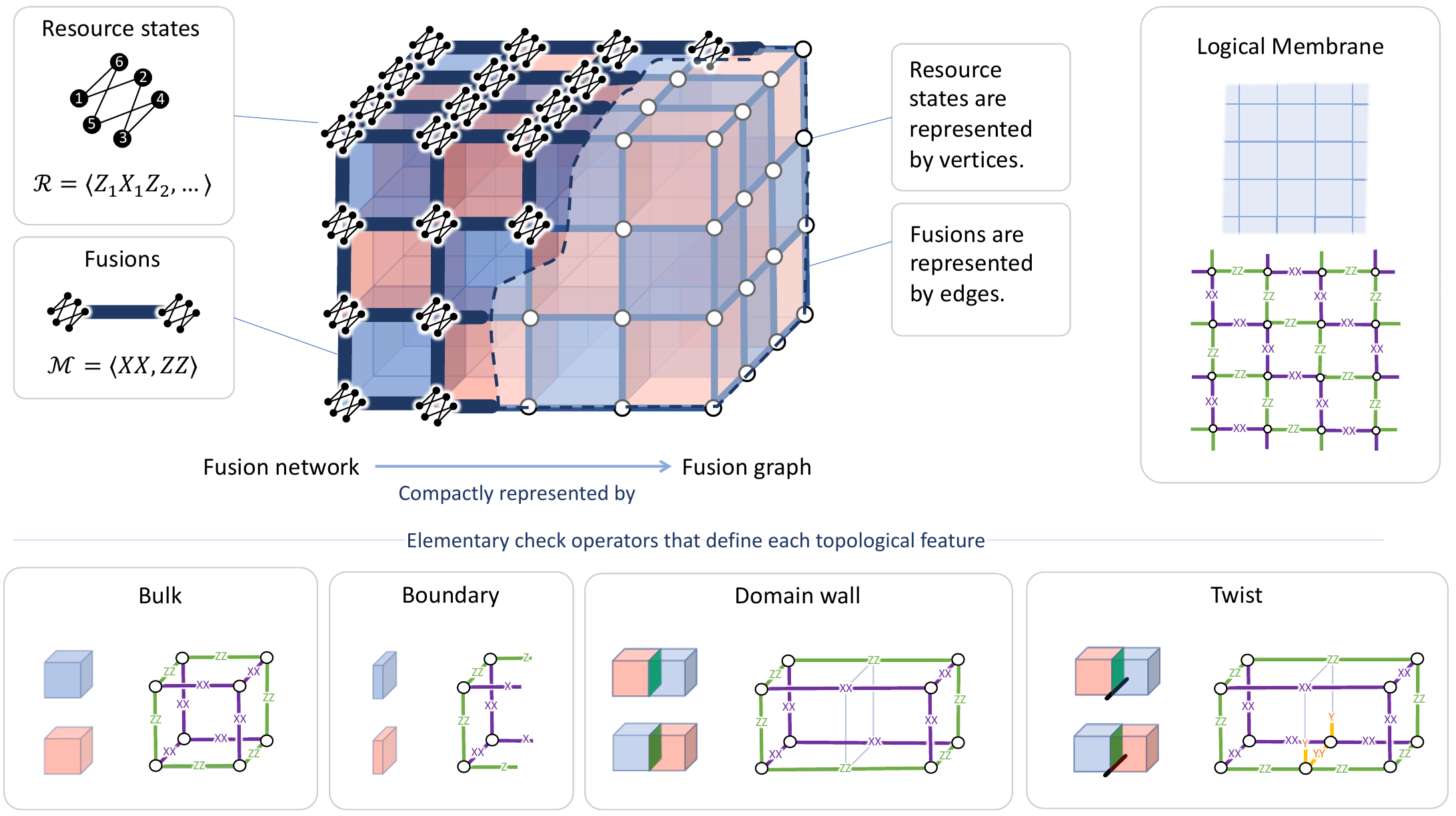} 
	\caption{The $6$-ring fusion network. The fusion network can be compactly represented by a fusion graph---a cubic graph whose vertices represent resource states ($6$-ring cluster states) and whose edges represent fusions. Logical membranes and check operators for various features are depicted.}
	\label{figFBQCFeatures}
\end{figure*}

This section describes the use of logical block templates to implement fault-tolerant instruments in FBQC~\cite{bartolucci2021fusion}. FBQC is a universal model of quantum computation in which entangling \emph{fusion} measurements are performed on \emph{resource states}. It is a natural physical model for many architectures, including photonic architectures where fusions are native operations~\cite{browne2005resource,gimeno2015three,bartolucci2021fusion}. In FBQC a topological computation can be expressed as a sequence of fusions between a number of resource states arranged in (2+1)D space-time. This gives rise to a construction termed a \emph{fusion network}. For concreteness, we consider an example implementation based on the six-ring fusion network introduced in Ref.~\cite{bartolucci2021fusion}, and with it construct fusion networks that realize the templates of Sec.~\ref{sec3DElements}.

\subsection{Resource state and measurement groups}
\label{sec:resource_and_measurement_groups}

To begin, we consider the cubical cell complex of a logical template introduced in Subsection~\ref{secLogicalTemplate}. Recall that this complex is referred to as the lattice $\mathcal{L}$. In a FBQC implementation of the template we define a fusion network from the vertices and edges of $\mathcal{L}$. Each vertex specifies the location of a \emph{resource state}, while the edges represent fusions (multiqubit projective measurements) and single-qubit measurements on its qubits. 

The six-ring resource state is a graph state~\cite{hein2004graph} with stabilizer generators
\begin{equation}
\label{eqClusterRingStabilizers}
\resstabsix = \langle K_i ~|~  K_i = Z_{i-1} X_i Z_{i+1}, i \in \{0,\ldots 5\} \rangle.
\end{equation}
The indexing is not arbitrary and will be described in Sec.~\ref{sec:surviving_stabilizers} below. 

The \emph{resource state group} $\mathcal{R}$ is defined as the tensor product
\begin{equation}
\resstab = {\otimes}_{v \in \mathcal{L}_0} \mathcal{R}_{6},
\end{equation}
where $\mathcal{L}_0$ is the set of all vertices in $\mathcal{L}$. 
Each edge $e$ of $\mathcal{L}_1$ thereby corresponds to a pair of qubits from distinct resource states. In the featureless bulk of the lattice we perform fusion measurements on each such pair in the basis
\begin{equation}\label{eqFusion}
M_{e} = \langle XX, ZZ\rangle.
\end{equation}

For features and boundaries, the template labels in $\mathcal{F}$ may specify alternative fusion bases or single-qubit measurements. In the latter case we continue to identify the measurement with an edge $e$; however, the template will specify one vertex as vacant so only a single qubit is involved. The \emph{measurement group} $\mathcal{M}$ is then defined as the group generated by
\begin{equation}
\measgp = \langle  M_e ~|~ e \in \mathcal{L}_1 \rangle
\end{equation}
where $\mathcal{L}_1$ is the set of all edges in $\mathcal{L}$.\footnote{In Ref.~\cite{bartolucci2021fusion} this group is termed the \emph{fusion group} and denoted $F$. It is renamed the measurement group $\measgp$ here to note the inclusion  of single-qubit measurements when required.} Importantly, in FBQC, the measurements to be performed are all commuting, and thus may be performed in any order (for example, layer by layer, or sequentially~\cite{bombin2021interleaving}). 

Elements in the resource state group that commute with the measurement group play an important role in FBQC, as we will see in the next section.

\subsection{Surviving stabilizers: checks and membranes}
\label{sec:surviving_stabilizers}

Both the error-correcting capabilities and logical instruments of FBQC may be understood using the \emph{surviving stabilizers} formalism \cite{raussendorf2007topological, brown2020universal, bartolucci2021fusion}. The surviving stabilizer subgroup $\mathcal{S}$ is defined as those stabilizers $r \in \mathcal{R}$ that commute with all elements of the measurement group
\begin{equation}
\mathcal{S} = \mathcal{Z}_\mathcal{R}\left(\mathcal{M}\right) = \left\{ r \in \mathcal{R} ~|~ rm = mr  \textrm{ for all } m \in \mathcal{M} \right\}.
\end{equation}
Elements of $\mathcal{S}$ are termed surviving stabilizers. The lattice structure allows us to determine this centralizer relatively easily as will be shown below. 

One important subgroup of $\mathcal{S}$ is the intersection $\mathcal{C} = \mathcal{R} \cap \mathcal{M}$, whose elements provide the check operators of the FBQC instrument network. Qubits of each six-ring resource state are arrayed on the edges in such a way that each generator of Eq.~(\ref{eqClusterRingStabilizers}) can be associated with a corner of a $3$-cell; one suitable indexing on Cartesian axes is $\{z_+, x_+, y_+, z_-, x_-, y_-\}$.

\begin{figure}[ht]
\centering
\subfloat[Stabilizers in $\mathcal{R}$\label{sfig:check_from_r}]{%
 \includegraphics[width=0.35\columnwidth]{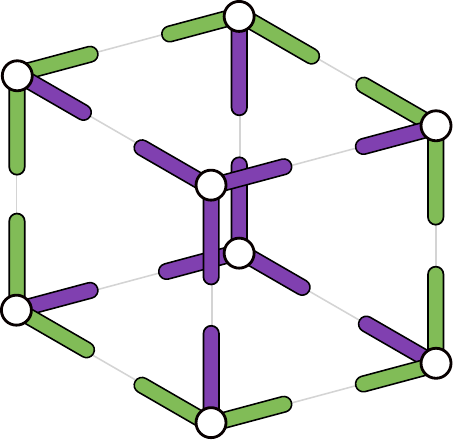}%
}\hspace{0.08\columnwidth}%
\subfloat[Whose product is in $\mathcal{M}$\label{sfig:check_from_m}]{%
 \includegraphics[width=0.35\columnwidth]{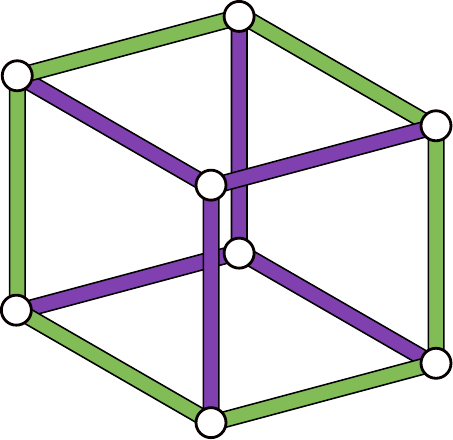}%
}
\caption{ A surviving stabilizer in the intersection $\mathcal{R} \cap \mathcal{M}$. On the left are stabilizers of the $6$-ring resource states arrayed on each vertex of a lattice cell. These are of the form $ZXZ$ (green-purple-green) or $XXX$ (purple-purple-purple). The product can be rewritten in terms of $XX$ (purple) and $ZZ$ (green) and is therefore also in the measurement group. This type of surviving stabilizer is a check operator of the FBQC instrument network.\label{fig:surviving_check}}
\end{figure}

Consider now a 3-cell in the bulk of the lattice as shown in Fig.~\ref{fig:surviving_check}. For six of its vertices we may choose a $ZXZ$ generator of Eq.~(\ref{eqClusterRingStabilizers}) on a corner of the cell as shown. For the remaining two corners we take a product of three six-ring stabilizer generators to obtain an $XXX$ operator on the corner. The product of all eight corner stabilizers can be rewritten as a product of elements $XX$ and $ZZ$ from $\mathcal{M}$, and is therefore contained in $\mathcal{C}$. 

These check stabilizers corresponding to 3-cells of the lattice are analogous to the plaquette operators of the surface code. Note that each cell edge shows only one outcome of the $\langle XX, ZZ\rangle$ fusion, the other outcome is included in a neighboring check operator. This partitions the cells into a 3D checkerboard of primal and dual check operators, as shown in Fig.~\ref{figFBQCFeatures}. Errors on fusion outcomes flip the value of checks, which can be viewed as anyonic excitations as before. 

When all qubits are measured or fused, the surviving stabilizer group is given by the intersection $\mathcal{S} = \mathcal{R} \cap \mathcal{M}$. We now turn to the case where some qubits remain unmeasured, and construct additional elements of $\mathcal{S}$ that will be important in understanding the FBQC implementations of logical instruments.

\begin{figure}[ht]
	\centering
	\includegraphics[width=0.7\linewidth]{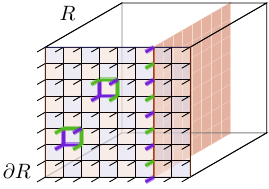}
	\caption{Topological boundary states for the six-ring fusion network. By fusing resource states in a region $R$, we are left with a series of unmeasured, outer qubits on the boundary $\partial R$ (represented by short lines coming out of the page). These outer qubits are in a code state of the surface code (up to signs depending on measurement outcomes), as can be inferred from the surviving stabilizers $XZZX \otimes s_\mathrm{inner}$ depicted. Therein, short purple (green) lines correspond to Pauli-$X$ (Pauli-$Z$) operators, while long purple (green) lines correspond to $XX$ ($ZZ$) operators. Another type of surviving stabilizer $\overline{Z}\otimes s_\mathrm{inner}$ is depicted, where $\overline{Z}$ is a logical operator for the surface code on $\partial R$ and $s_\mathrm{inner}$ is supported on a two-dimensional membrane.}
	\label{fig:surviving_on_boundary}
\end{figure}

\textbf{Topological boundary states.}
Consider a boxlike region $R$ within a bulk fusion network, and suppose that fusions are performed only inside this box. Qubits involved in these fusions will be referred to as \emph{inner} qubits, and the remaining unmeasured qubits referred to as \emph{outer} qubits. As the measurement group has full (stabilizer) rank on the inner qubits all surviving stabilizers can be written as $s = s_\mathrm{outer} \otimes s_\mathrm{inner}$. In the interior of the region we can define check operators as above where $s_\mathrm{outer} = \id$. However, in the presence of unmeasured qubits, we will find surviving stabilizers that reveal topological boundary states on the boundary of the measured region $\partial R$. In particular, if we consider a planarlike boundary of the bulk fusion network, we find surviving stabilizers as shown in Fig.~\ref{fig:surviving_on_boundary}. For this boundary geometry, the first type are of the form $XZZX \otimes s_\mathrm{inner}$, where $s_\mathrm{inner}$ consists of a product of fusion measurements. We recognize this first type from Sec.~\ref{secSurfaceCode} as a surface code check operator acting on qubits on the region's boundary. These \emph{boundary checks} partition into primal and dual types, as shown in Fig.~\ref{fig:surviving_on_boundary}.
The second type of surviving stabilizers are two-dimensional sheets, some of which, as we will see in the next section, define membranes of the logical instrument implemented by the fusion network. Note that membranes may be deformed by multiplying by the cubic and boundary check operators.

\subsection{Topological instrument networks in FBQC}
We now turn to the implementation of the logical instruments described in Sec.~\ref{secCliffordOperators} and the topological features by which they are implemented. Additionally, using an example, we describe the second type of surviving stabilizer; the logical membrane. In Sec.~\ref{secElements} we understood features in terms of their properties with respect to anyonic excitations, which in turn were derived from the structure of the stabilizer check operators of the code. The FBQC approach is similar in nature. We introduce local modifications to the fusion and measurement patterns of the basic six-ring fusion network to create topological features, and show that this generates a surviving stabilizer group that includes the appropriate checks and stabilizers. 

\subsubsection{Logical membranes and the identity gate}
\label{sec:fbqc_identity_channel}

In this subsection we describe the FBQC implementation of the identity gate, the simplest logical operation shown in Fig.~\ref{figAllLogicalBlocks}, and in particular the implementation of its ports and boundaries. The fusion graph is cubic with dimensions $d \times d \times d$, and fusions on qubits of neighboring resource are performed as before. Qubits on the primal and dual boundaries are measured to ensure that only the appropriate boundary checks remain as members of $\mathcal{S}$ (defined in the following). The remaining boundary checks on each port then exactly define a surface code, with membrane stabilizers given by 
\begin{equation}\label{eqMembraneFBQC}
\begin{array}{c}
    \overline{X}_\mathrm{in} \otimes \overline{X}_\mathrm{out} \otimes m^{X,X}, \\
    \overline{Z}_\mathrm{in} \otimes
    \overline{Z}_\mathrm{out} \otimes m^{Z,Z}.
\end{array}
\end{equation}

\begin{figure}[h]
\centering
\subfloat[Stabilizers in $\mathcal{R}$\label{sfig:membrane_from_r}]{%
 \includegraphics[width=0.4\columnwidth]{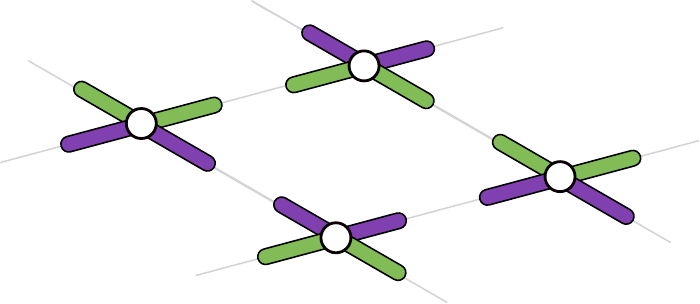}%
}\hspace{0.06\columnwidth}%
\subfloat[Whose product is in $\mathcal{M}$\label{sfig:membrane_from_m}]{%
 \includegraphics[width=0.4\columnwidth]{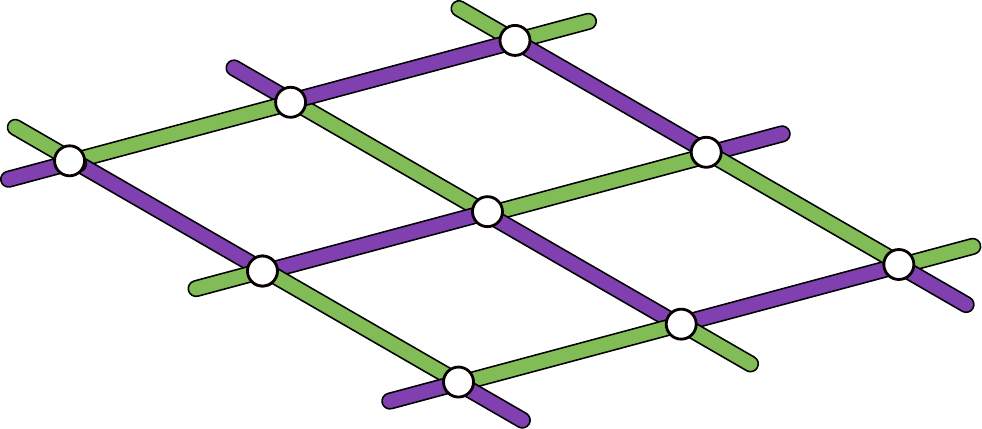}%
}
\caption{The bulk of the logical membranes. Both $m^{X, X}$ and $m^{Z,Z}$ from Eq.~(\ref{eqMembraneFBQC}) are elements of the surviving stabilizer $\mathcal{R} \cap \mathcal{M}$. They can be constructed from stabilizers of the six-ring resource states in such a way that they also clearly belong to the measurement group. Purple and green full (half) edges represent $XX$ ($X$) and $ZZ$ ($Z$) operators.
\label{fig:fbqc_membranes}}
\end{figure}

The membranes $m^{X, X}, m^{Z,Z}$ represent the world sheet of each logical operator through the fusion network, and they define the action of the logical instrument. They are depicted in Figs.~\ref{fig:surviving_on_boundary},~\ref{fig:fbqc_membranes}. Undetectable errors strings are those that cross from one primal (dual) boundary to the other, thereby introducing an error on the membrane.

\subsubsection{Ports}
While not a topological feature, ports are an important component of the logical block framework that allow us to define logical blocks. In terms of the fusion network, a port identifies a set of resource states that have unpaired (and thus unfused) qubits. These unpaired qubits remain unmeasured and form the set of outer qubits, each connected component of which forms a surface code state (postmeasurement) encoding the input or output of a block (as depicted in Fig.~\ref{fig:surviving_on_boundary}).

\subsubsection{Boundaries}\label{sec:fbqc_boundaries}
Recall from Section~\ref{secElements} that boundaries come in two types according to the type of excitation that may condense on them. The primal boundary absorbs only dual-type excitations, and the dual boundary absorbs only primal-type excitations. These boundaries can be created using single-qubit measurements on a two-dimensional sublattice.

To see how they are created, consider a region $R$ with a boundary $\partial R$. As we have seen, by fusing the resource states within $R$, we are left with a topological surface code state on $\partial R$. There are surviving stabilizers $r\in \survstab$ that describe this boundary state. We consider in particular the surviving stabilizers $r\in \survstab\setminus \checkgp$ with support on $\partial R$. These surviving stabilizers admit a generating set in terms of primal and dual stabilizers (recall they can be viewed as truncated bulk check operators).

To create a primal or dual boundary on $\partial R$, we perform a set of single-qubit measurements that commute with either the primal or dual surviving stabilizers. In other words, we perform measurements such that only the primal or dual surviving stabilizers $r\in \survstab\setminus \checkgp$ remain (but not both). For the planar geometry of Fig.~\ref{fig:surviving_on_boundary}, the measurement patterns to create a primal boundary and a dual boundary differ only by a translation. Namely, they consist of an alternating pattern of $X$ and $Z$ single-qubit measurements, as shown in the example of Fig.~\ref{figFBQCMagicStatePreparation} below. The boundary checks are shown in Fig.~\ref{figFBQCFeatures}. Other geometries can be found similarly by implementing the single-qubit measurement pattern that completes either the primal or dual check operators (which are obtained by restricting bulk checks to the region with boundary). We remark that it is often convenient to describe the fusion graph on the dual of the template complex, such that resource states belong to 3-cells, and the measurement basis for each qubit is uniquely determined by the feature label on the 2-cell on which it resides.

\subsubsection{Domain walls}
\label{secK6SymmetriesAndDefects}

\begin{figure}
\centering
\includegraphics[width=0.36\columnwidth]{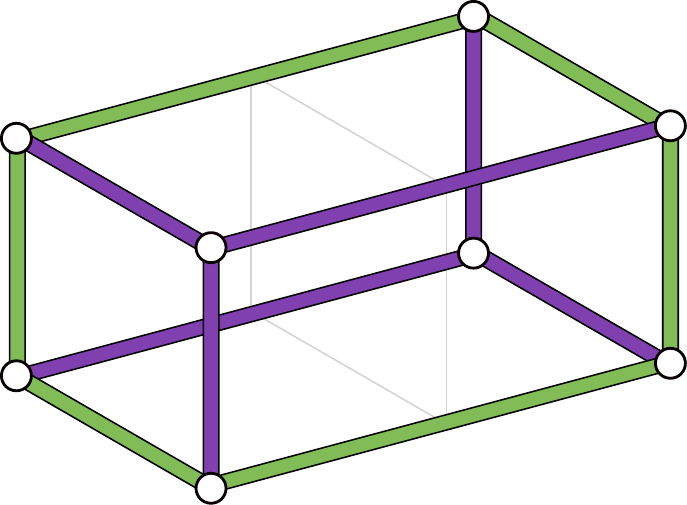}%
\caption{The measurement pattern and check operator of a FBQC domain wall defect. Purple and green edges represent $XX$ and $ZZ$ operators. The extended edges are the measurements from next-to-nearest-neighbor fusions.\label{fig:domain_wall_check}}
\end{figure}

As is the case for the surface code, nontrivial logic operations can be implemented by introducing defects using the underlying $\zz_2$ symmetry. The domain wall defect was described in Sec.~\ref{sec:topological_features_surface_code} as a codimension-1 feature formed when the symmetry transformation, i.e., the exchange of primal and dual checks, is applied to a subregion of the lattice. In the logical block template this domain wall is denoted by labeled 2-cells identifying this region. The fusion pattern is modified such that the \emph{next-to-nearest} resource states on opposite sides of the domain wall are fused together directly. Resource states assigned to vertices within the domain wall plane do not participate and may be discarded. The local check operators along the 2D domain wall plane are supported on the two 3-cells intersecting on the domain wall, as shown in Fig.~\ref{fig:domain_wall_check}. This check structure results in the exchange of primal and dual excitations upon crossing.

\subsubsection{Twist defects}

\begin{figure}[ht]
\centering
\includegraphics[width=0.36\columnwidth]{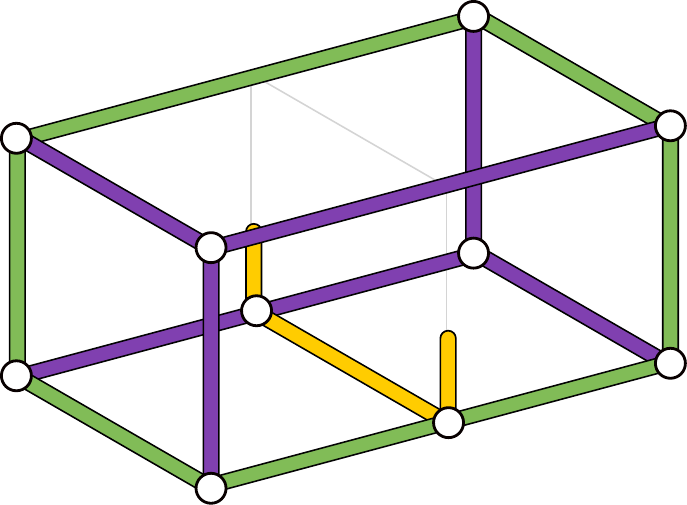}%
\caption{The measurement pattern and check operator of a FBQC twist defect. Purple and green edges represent $XX$ and $ZZ$ operators. Yellow full and half edges represent $YY$ and $Y$ operators. The $YY$ operator is obtained as the product of $XX$ and $ZZ$ measurements of the given fusion.}
\label{fig:twist_check}
\end{figure}

Twist defect lines occur along the 1D boundaries of a domain wall. In the fusion network, the line of resource states associated with vertices on the twist line each have one qubit that does not partake in a bulk fusion or domain wall fusion. These qubits are associated with the edge directed into the domain wall, and are measured in the $Y$ basis producing the measurement pattern and check operators shown in  Fig.~\ref{fig:twist_check}. These operators have overlap with both primal and dual checks, and it can be verified that composite primal-dual excitations may condense on twist lines.

\subsubsection{Cornerlines and transparent cornerlines}

The last feature we consider in the FBQC implementation of logical templates is that of \emph{cornerlines}. Cornerlines and transparent cornerlines naturally arise when the two distinct boundaries meet, depending on whether a domain wall is present or not (i.e., no further modifications to the measurement pattern are required). In particular, one performs the appropriate single-qubit measurements on either side of the (transparent) cornerline, according to whether that qubit belongs to the primal or dual boundary type. 
We depict an example of the measurement pattern that gives rise to cornerlines in Fig.~\ref{figFBQCMagicStatePreparation}.

\subsubsection{Fusion-based magic state preparation}
\begin{figure}[ht]
	\centering
	\includegraphics[width=0.48\linewidth]{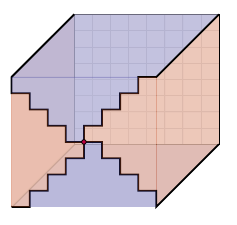}
	\includegraphics[width=0.48\linewidth]{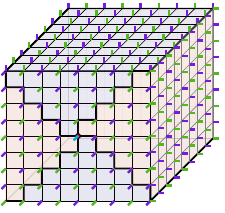} 
	\caption{Noisy magic state preparation block. (left) Noisy magic state preparation can be regarded as a fault-tolerant instrument taking a distance-$1$ code state (i.e., an unencoded state), to a distance-$d$ code state. The block produces (on the rear port) an encoded version of an arbitrary input qubit (on the front center). The depth is exaggerated for the purpose of illustration. (right) The $6$-ring fusion network measurement pattern for the block (defined here by considering the fusion graph as the 1-skeleton of the dual of the block template). Individual qubits belonging to resource states on the boundary are measured in the $X$ ($Z$) basis as accordingly depicted by purple (green) edges. To prepare an encoded $\ket{T} = \frac{1}{\sqrt{2}}(\ket{0} \pm e^{\frac{i \pi}{4}} \ket{1})$ on the output, the qubit belonging to the central resource state is measured in the $\frac{1}{2}(X + Y)$ basis; the $\pm$ sign is determined by the measurement outcomes.
	}
	\label{figFBQCMagicStatePreparation}
\end{figure}

For completeness, we depict in Fig.~\ref{figFBQCMagicStatePreparation} the fusion network and precise measurement pattern that can be used to prepare a noisy magic state. This can be viewed as the fusion-based analogue of the protocol described in Refs.~\cite{lodyga2015simple,brown2020universal}.

\section{Simulating logical blocks}\label{secNumerics}

In this section, we introduce tools to simulate the error rates of logical blocks, and present numerical results for the thresholds and logical error rates of the logical blocks depicted in Fig.~\ref{figAllLogicalBlocks}. We begin by describing the \textit{syndrome graph} representation of a fault-tolerant instrument, which will be a convenient data structure to decode and simulate logical blocks.  Moreover, the logical block templates that we have defined can be implemented in software as a tool to automatically generate such syndrome graphs, thereby allowing for the simulation of complex logical block arrangements. Using this software framework, we perform numerical simulations for logical blocks based on the six-ring fusion network under a phenomenological noise model\footnote{Note that this is distinct from phenomenological noise model that is commonly used to model measurement and Pauli errors in a code based fault tolerance scheme. Our model is closer to a gate error model in that it accounts for large scale entanglement generation from finite sized resources.} known as the \textit{hardware-agnostic fusion error model} in Ref.~\cite{bartolucci2021fusion}. We demonstrate that all blocks have a threshold that agrees to within error with the bulk memory threshold. We also evaluate the logical error rate scaling of each block in the subthreshold regime and evaluate the resources required to achieve a target logical error rate for practical regions of interest. In doing so, we see the important role that boundary conditions have on the logical error rate, and in particular that periodic boundary conditions provide a favorable scaling. Finally, we observe that for sufficiently large distances and low error rates, the logical error rates for different logical qubits participating in lattice surgery behave approximately independently\footnote{A potentially related observation is found in Ref.~\cite{farrelly2020parallel} for a different family of codes, whereby different logical qubits can be decoded independently while remaining nearly globally optimal.}.

\subsection{Simulation details}
To perform simulations of complex logical block arrangements, we implement the template in software as a cubical complex with labeled 2-cells. Each template corresponds to a set of physical operations, and can be used together with a set of rules (building upon the description in Sec.~\ref{sec:topological_fbqc}) to automatically generate a set of syndrome graphs (as defined below), and a set of bit masks representing the logical membranes thereon. We can subsequently use these syndrome graphs to perform sampling and decoding of errors.

\subsubsection{Syndrome graphs}
To evaluate the thresholds and below threshold scaling behavior of logical blocks we rely on the \textit{syndrome graph} representation of errors and checks. This representation  can be used for flexible error sampling and decoding, sufficient for many decoders such as minimum-weight perfect matching (MWPM)~\cite{dennis2002topological,kolmogorov2009blossom} and union find (UF)~\cite{delfosse2017almost}. 
The syndrome graph representation can be defined for any fusion network where each measurement ($X\otimes X$, $Z \otimes Z$, or single-qubit measurement) belongs to precisely two local check operator generators, such as the six-ring network. We define the syndrome graph as follows.
\begin{defn}
	(Syndrome graph). Let $\checkgp_{\text{local}}$ be the set of local check generators for $\checkgp$ (depicted in Fig.~\ref{figFBQCFeatures} for the six-ring network). We define the syndrome graph $G_{\text{synd}} = (V_{\text{synd}},E_{\text{synd}})$ by placing a vertex $v\in V_{\text{synd}}$ for every local check generator $c\in \checkgp_{\text{local}}$, and an edge between two vertices if their corresponding checks share a measurement.
\end{defn}

Flipped measurement outcomes are represented by edges of the syndrome graph, and the syndrome can be obtained by taking their mod-2 boundary (in other words, a flipped check operator corresponds to a vertex with an odd number of flipped edges incident to it). 
Logical membranes are represented on the syndrome graph as a collection of edges corresponding to the fusions and measurements that it consists of. We refer to this subset of edges as a logical mask. 
A logical error corresponds to a set of edges whose mod-2 boundary is zero, and that has odd intersection with a logical mask.

\begin{figure}
	\centering
	\includegraphics[width=0.615\linewidth]{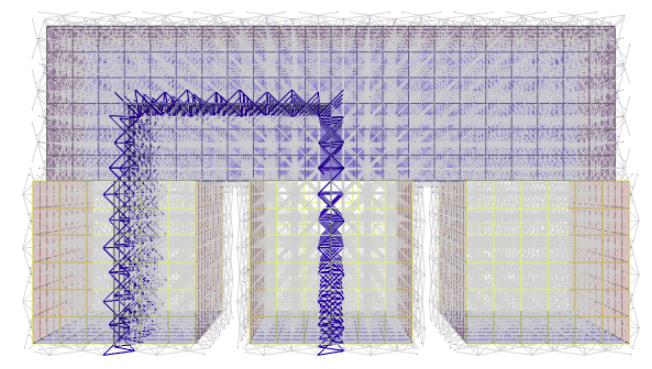}
	\includegraphics[width=0.375\linewidth]{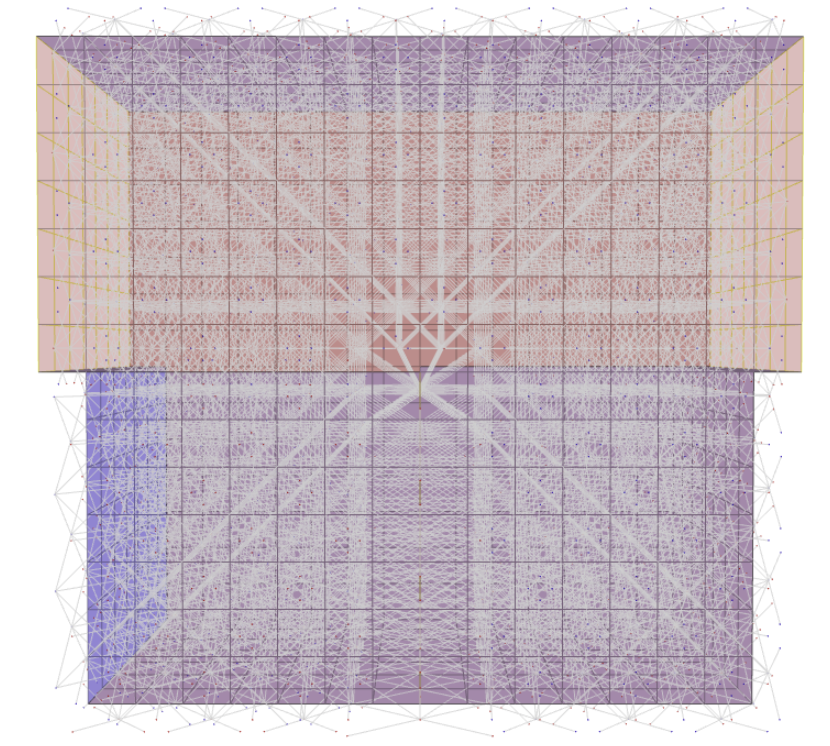}
	\caption{(left) Syndrome graph and a syndrome mask (i.e., a logical membrane projected onto the syndrome graph) for lattice surgery. The three input ports are displayed at the front along the bottom of the diagram. (right) Syndrome graph for the phase gate.}
	\label{figLatticeSurgerySyndromeGraph}
\end{figure}

For the bulk six-ring fusion network, the syndrome graph consists of two decoupled 12-valent graphs, which are referred to as primal and dual syndrome graphs.
Domain walls and twists may prevent the syndrome graph from decomposing into two disconnected components (as is the case in the phase gate for example), as, in particular, the two syndrome graphs are swapped across domain walls and fused together along twists.  Example syndrome graphs for a lattice surgery block and phase gate block are depicted in Fig.~\ref{figLatticeSurgerySyndromeGraph}. 

\subsubsection{Noise model and decoder}
Resource states are in general noisy, each subject to Pauli noise and erasures that can arise during preparation, propagation and measurement. 
The net effect of noise processes affecting both resource states and operations on them can be phenomenologically captured by modeling each measurement outcome (both fusion outcomes and single-qubit measurement outcomes) as being subject to an erasure error with probability $p_E$ or bit-flip error with probability $p_P$. This is known as the \textit{hardware-agnostic fusion error model} in Ref.~\cite{bartolucci2021fusion}. Specifically, we assign each measurement a probability of $p_E$ that the outcome is erased, a probability of $p_P(1-p_E)$ that the outcome is incorrect (i.e., bit flipped but not erased), and a probability of $(1-p_E)(1-p_P)$ that the measurement is correct. We refer to $p=(p_P, p_E)$ as the physical error rate. (Note that up to a reparametrization, the bit-flip errors are equivalent to an IID Pauli-$X$ and Pauli-$Z$ error channel acting on each qubit.)

To decode these errors, we utilize the union-find decoder of Ref.~\cite{delfosse2017almost} due to its optimal performance against erasures, high performance against bit-flip noise, and fast runtime. We remark that higher tolerance to bit-flip noise can be achieved with the minimum-weight perfect-matching decoder~\cite{dennis2002topological,kolmogorov2009blossom}. More details on the general decoding problem for (2+1)D logical blocks can be found in Sec.~\ref{secDecoding}.

\subsubsection{Simulated logical blocks}
We simulate six-ring fusion networks for the identity gate, the Hadamard, the phase gate, and the $Z$-type lattice surgery involving a variable number of logical qubits as described in Sec.~\ref{secLogicBlocks}. 
We also simulate the bulk fusion network on a cube with periodic boundary directions (i.e., a 3-torus), in order to have a bulk comparison. While this block contains no ports nor nontrivial logical correlators, for simulation purposes, we may define logical membranes on each of the nontrivial 2-cycles of the torus, such that failure is declared for any error spanning a nontrivial 1-cycle of the torus.

For each of these block families, we generate a family of syndrome graphs of varying distance to be used for logical error rate Monte Carlo simulations (involving error sampling and decoding). For the purposes of simulation, we assume certain fictitious boundary conditions for the ports where all qubits are perfectly read out (such that errors terminating on ports always generate syndromes). This allows logical operators to be noiselessly read out on each port.

\begin{figure*}
	\centering
	\includegraphics[width=0.82\linewidth]{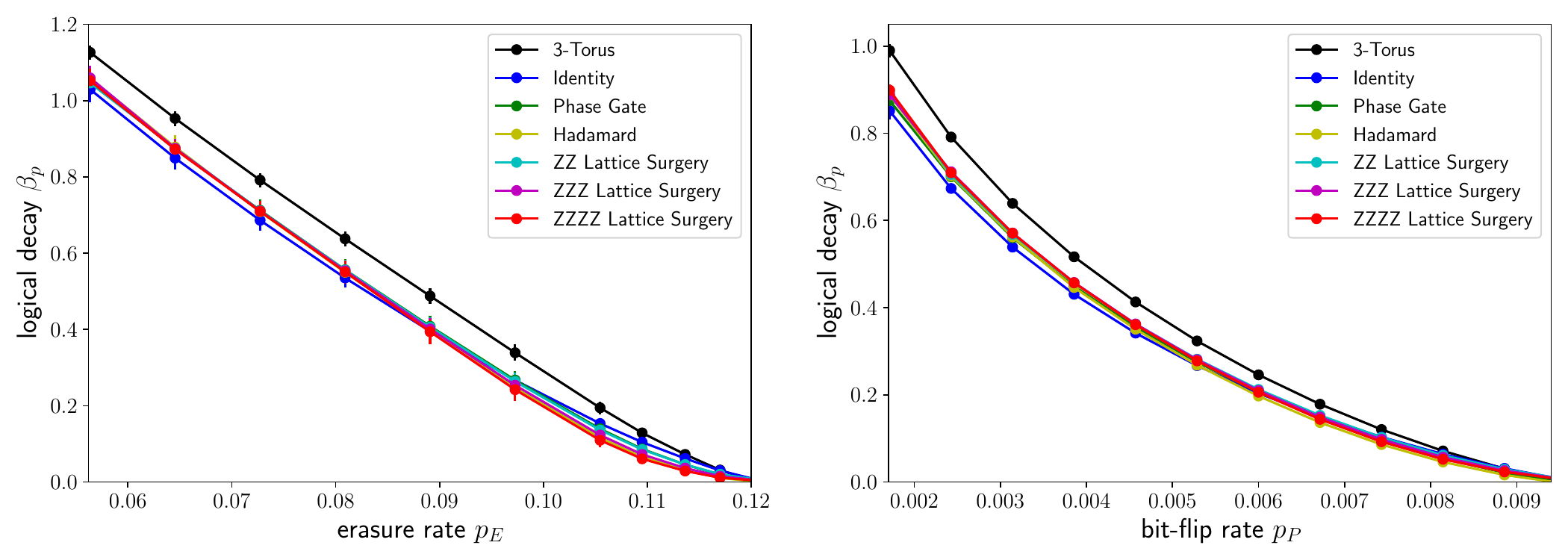} 
	\includegraphics[width=0.82\linewidth]{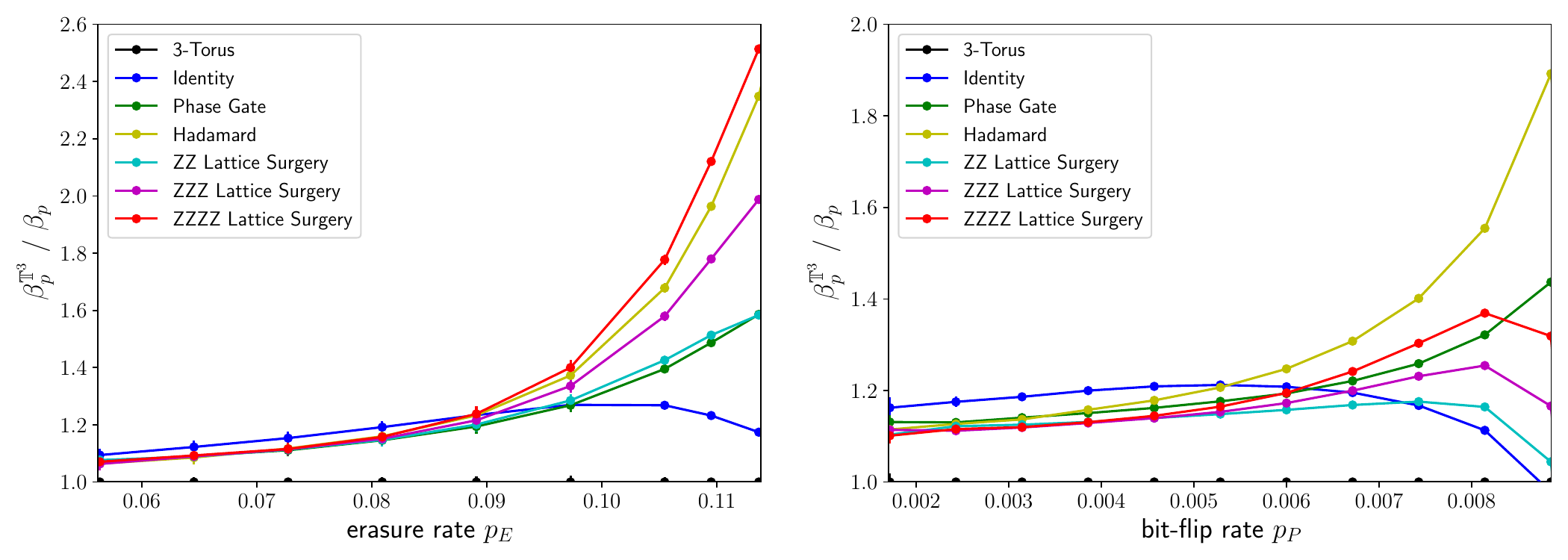} 
	\includegraphics[width=0.82\linewidth]{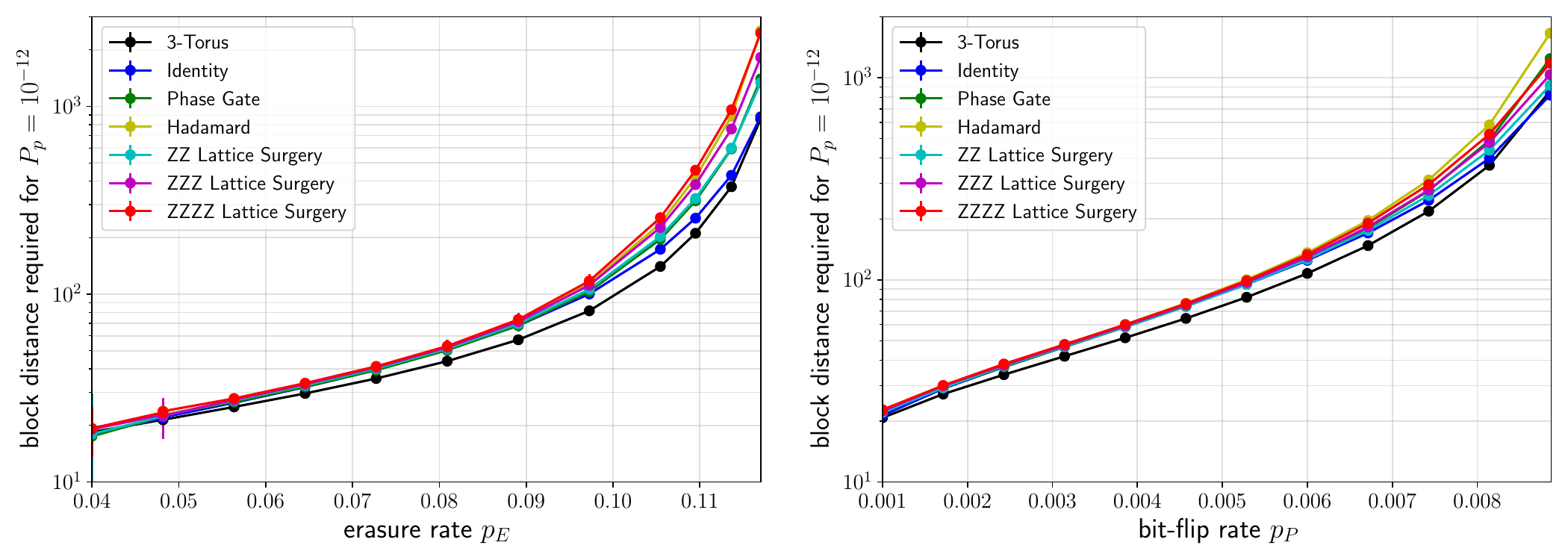} 
	\caption{
    (top) Fits of $\beta_p$ to the logical block error rate $P_{p}(d) = \alpha_p e^{-\beta_p d}$ under IID erasure and IID bit-flip noise models for the identity, 3-torus, phase, Hadamard, and multi-qubit lattice surgery blocks. Each data point is obtained by performing $10^{9}$ to $10^{10}$ decoder trials for each block distance and physical error rate, with block distances up to $14$. Error bars correspond to uncertainty in the fit, as described in App.~\ref{appLogicalPrefactor}.
    The 3-torus has the largest logical decay out of the simulated blocks including the identity block---meaning in the case of memory, the toric code has better error suppression than the planar code. Further, the logical decay $\beta_p$ for lattice surgery is independent of the number of logical qubits involved, for sufficiently low physical error rates. 
	(middle) Plots of $\beta_{p}^{\mathbb{T}^3}/\beta_p$ against physical error rate for different logical blocks, where $\beta_{p}^{\mathbb{T}^3}$ is the logical decay for the 3-torus. This ratio determines how much larger a given block distance must be, compared to the 3-torus, to achieve the same logical error rate scaling. 
	(bottom) The block distance $d$ required to achieve a target logical error rate of $P_p=10^{-12}$ per block as a function of physical error rate $p$ for both erasure and bit-flip physical errors. The distances required can become very large near threshold, and as such, it is important to be significantly below threshold to have reasonable fault-tolerance overheads.
	In all plots, lines are guides for the eye.
	}
	\label{figNumericsDecay}
\end{figure*}

\subsection{Numerical results}\label{secNumericalResults}

\subsubsection{Logical block thresholds} 
We provide numerical estimates of the noise threshold values for each logical block against both IID erasure noise with rate $p_E$ and IID outcome bit-flip noise with rate $p_P$. 

To evaluate the threshold value, we sweep along different error rates $p_E$ and $p_P$ and evaluate the logical error rate for different block sizes. The logical error rate at each physical error rate is obtained by performing $10^{7}$ to $10^{8}$ decoder trials for each block distance, up to $d = 26$. Each decoder trial consists of sampling an error configuration, running the decoder, and declaring success if and only if all of the correlation operators are successfully recovered.

The threshold values for each block are displayed in Table~\ref{tab:threshtable} and the threshold plots from which these values are obtained are displayed in Figs.~\ref{figThresholdsEr} and \ref{figThresholdsPa} below. For each error model, we estimate thresholds for each block that all agree to within error bars. This verifies the conventional intuition that the threshold should be determined by the bulk properties alone, and insensitive to the presence of codimension-1 and codimension-2 features. We remark that \textit{a priori} this may not have been true, due to the well-known error-correction stat-mech correspondence~\cite{dennis2002topological} and the fact that bulk phase properties can be driven by boundary conditions~\cite{henkel1994boundary}. See App.~\ref{secThreshAnalysis} for more details on how the thresholds are estimated.

\subsubsection{Overhead for target error rate and logical error rate fits}

The threshold sets an upper bound on the rate of errors that are tolerable for the scheme. However, to estimate the overhead for fault-tolerant quantum computation, it is important to estimate the block distances required for each logical operation to achieve a target logical error rate.

For a given physical error rate $p$ and block distance $d$, we perform $10^{9}$ to $10^{10}$ trials to obtain estimates of the logical error rate. Following Ref.~\cite{bravyi2013simulation}, we fit the logical error rate to an exponential decay as a function of distance according to 
\begin{equation}\label{eqLERfit}
P_{p}(d) = \alpha_p e^{-\beta_p d},
\end{equation}
where both $\alpha_p$ and $\beta_p > 0$ depend on the physical error rate $p$. We refer to $\beta_p$ as the logical decay and to $\alpha_p$ as the logical prefactor. In Fig.~\ref{figNumericsDecay}, we plot the logical decay $\beta_p$ as a function of the physical error rate. In App.~\ref{appLogicalPrefactor} we explain the fitting methodology and also plot the logical prefactor $\alpha_p$ as a function of the physical error rate.

One can invert Eq.~(\ref{eqLERfit}) to obtain the distance required to achieve a target logical block error rate at a given physical error rate. In Fig.~\ref{figNumericsDecay} we display estimates for the required distance based on numerical fits for $\alpha_p$ and $\beta_p$. For concreteness, we choose a target logical block error rate of $10^{-12}$. Such error rates are relevant for many existing algorithms in the fault-tolerant regime, for example for quantum chemistry applications~\cite{kivlichan2020improved, von2020quantum, kim2021faulttolerant, su2021faulttolerant}. The figure shows the importance of having error rates significantly below threshold, as otherwise the distance and fault-tolerance overheads required can become extremely large.

\begin{table}
\begin{tabular}{ p{1.5cm}||p{2.3cm}|p{2.5cm}  }
 \multicolumn{3}{c}{Threshold values} \\
 \hline
 Block & Erasure rate $p_E$ & Bit-flip rate $p_P$ \\
 \hline
 3-Torus & $(12.02\pm 0.04)\%$ & $(0.948\pm 0.005)\%$ \\
 Identity & $(12.04\pm 0.05)\%$ & $(0.955\pm 0.006)\%$ \\
 Phase gate & $(12.05\pm 0.06)\%$ & $(0.951\pm 0.005)\%$ \\
 Hadamard & $(12.02\pm 0.06)\%$ & $(0.948\pm 0.005)\%$ \\
 LS $Z^{\otimes 2}$ & $(12.06\pm 0.07)\%$ & $(0.956\pm 0.006)\%$ \\
 LS $Z^{\otimes 3}$ & $(12.07\pm 0.08)\%$ & $(0.957\pm 0.006)\%$ \\
 LS $Z^{\otimes 4}$ & $(12.08\pm 0.09)\%$ & $(0.958\pm 0.008)\%$ \\
 \hline
\end{tabular}
\caption{\label{tab:threshtable} Threshold values for various logical blocks. `LS' stands for lattice surgery.}
\end{table}

\subsubsection{Periodic vs. open boundary conditions}
Next, we observe the importance of boundary conditions for logical error rates. In Fig.~\ref{figNumericsDecay} we note that the distance required for a target logical error rate for the 3-torus block (having periodic boundary directions) is notably smaller than any of the other blocks and in particular the identity block. In other words, the logical error rates are more greatly suppressed with periodic boundary conditions as opposed to open boundary conditions, as has previously been observed in Ref.~\cite{fowler2013accurate}. In particular, we plot $\alpha_p^{\mathbb{T}^3} / \alpha_p$ for each block, where $\alpha_p^{\mathbb{T}^3}$ is the logical decay for the 3-torus block and $\alpha_p$ is the logical decay for a given block. This ratio is an approximate measure of the distance saving offered by periodic boundary conditions---it tells us the factor that each block distance must be increased by in order to have the same logical error rate scaling as the 3-torus.

This difference in performance may be explained by \textit{entropy}; for a given distance $d$, the identity block (i.e., with open boundary conditions) contains a larger number of logical errors of weight $k\geq d$ than the 3-torus block (i.e., with periodic boundary conditions). In addition to the favorable error rates, periodic boundary conditions can be used to encode more qubits, offering potentially further advantages at least when used for memory. These results demonstrate that entropy plays an important role in the design and performance of logical gates (as has previously been noted for quantum error-correcting codes in Ref.~\cite{beverland2019role}).
In particular, the scaling advantage motivates us to study schemes without boundaries, and we present one such proposal---based on the teleportation of twists---in Sec.~\ref{secPortals}.

\subsubsection{Stability of lattice surgery}
Finally, we observe that at low physical error rates, the logical decay for lattice surgery is insensitive to the number of logical qubits. This is expected at sufficiently large distances and low error rates, as each logical error behaves approximately independently on each qubit participating in the lattice surgery. More precisely, the logical error rate for $n$ independent planar codes undergoing memory is expected to behave like 
    $\overline{P}_n = 1 - (1 - \overline{P}_1)^n = n\overline{P}_1 + \mathcal{O}(\overline{P}_1^2)$,
where $\overline{P}_1$ is the logical error rate of a single planar code. Therefore, at sufficiently low error rates and to first order, we expect the logical decay to be invariant to the number of qubits undergoing lattice surgery, and the logical prefactor $\alpha_p$ to increase proportionally to the number of qubits $n$. This agrees well with the observed data in Fig.~\ref{figNumericsDecay} and Fig.~\ref{figLogicalPrefactor} below.

\section{Topological quantum computation without boundaries: Portals and teleported twists}\label{secPortals}

In this section, we introduce a new computational scheme for twist-encoded qubits. In this scheme, logical information is encoded in twist defects and fault-tolerant logical operations are achieved by introducing space-time defects known as portals, which teleport the twists forward in time. This scheme is motivated by the favorable logical error rate suppression observed for logical blocks that have no boundaries. To our knowledge, this is the first universal surface code scheme that does not require boundaries. 

The native gates implementable with twists and portals are PPMs (i.e., measurement of $n$-qubit Pauli operators), which are universal when supplemented with noisy magic states. In this scheme, a PPM is implemented by introducing a pair of portals that teleport a subset of twists to another space-time location. These portals generally require long-range operations to implement. For concreteness, we focus on portals and twists in a photonic FBQC architecture, where such long-range operations are conceivable~\cite{bartolucci2021fusion, bombin2021interleaving}.

\subsection{Encoding in twists}

\begin{figure}[t]
	\centering
	\includegraphics[width=0.95\linewidth]{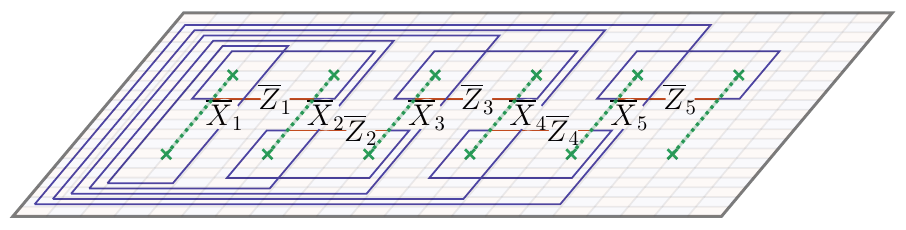} \vspace{3ex}
	\includegraphics[width=0.95\linewidth]{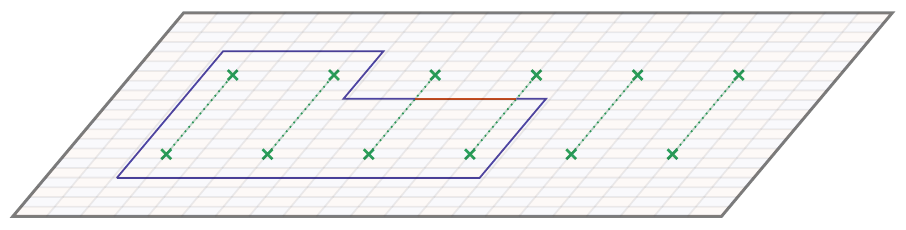} 
	\caption{
	Encoding $n$ qubits in $2n+2$ twists. (top) The generating set of logical operators is traceable. (bottom) Example of a product of Pauli generators that is also traceable. A general Pauli operator can be associated by the twists it encloses. In Fig.~\ref{figLogicalTraceability} we show that all Pauli operators are traceable.
	}
	\label{figTwistsEncoding}
\end{figure}

First consider a standard encoding whereby $n$ logical qubits can be encoded in $2n{+}2$ twists on a (topological) sphere (see, e.g., Refs.~\cite{bombin2010topological,barkeshli2013twist}). This can be understood using the Majorana fermion mapping~\cite{bombin2010topological,barkeshli2013twist}, whereby each twist defect can be expressed by a Majorana fermion operator $\gamma_i$, $i \in \{1,\ldots, 2n{+}2\}$ satisfying 
\begin{equation}\label{eqMFCommutation}
\gamma_j \gamma_k + \gamma_k \gamma_j = 2\delta_{jk},
\end{equation}
where $\delta_{jk}$ is the kronceker delta. In terms of these Majorana fermions, the logical operators may be expressed
\begin{align}\label{eqMFEncoding}
\overline{X}_k = i\prod_{1\leq i \leq 2k} \gamma_i, \quad  \overline{Z}_k = i\gamma_{2k} \gamma_{2k+1}.
\end{align}
One can verify that these logical operators satisfy the correct (anti)commutation relations using Eq.~(\ref{eqMFCommutation}). An arbitrary Pauli operator is thus represented by an even number of Majorana operators, $\mathcal{P}_{n} \cong \langle i , \gamma_j \gamma_k ~|~ j,k \rangle$. In Fig.~\ref{figTwistsEncoding} we depict surface code string operator representatives for the logical operators of Eq.~(\ref{eqMFEncoding}), which are realized as loops enclosing the corresponding twists.

\textbf{Operator traceability}. The enabling property for the teleported twist scheme is that all logical Pauli operators $\mathcal{P}_n$ in the twist encoding are \textit{traceable}. Traceability was introduced in Ref.~\cite{krishna2020topological}, and for our purposes, we say that a logical operator is traceable if it can be represented as a connected, non-self-intersecting string operator that is piecewise (i.e., locally) primal or dual. (Note that a traceable string operator can swap between primal and dual as it crosses a domain wall).\footnote{For instance, logical $\overline{X}$ and $\overline{Z}$ operators of a toric code or planar code are traceable, but the logical $\overline{Y}$ is not, due to the unavoidable self-intersection of string operators (see, for example, Fig.~\ref{figLogicalBlockConcatenation}).} That twist-encoded logical operators are traceable is shown in App.~\ref{appTwistTraceability}. As a consequence of traceability, every logical operator $\overline{P}\in \mathcal{P}_n$ can be identified by a subset of twists $\mathcal{T}_{\overline{P}}$ that it encloses. Furthermore, any logical operator $\overline{Q} \in \mathcal{P}_n$ that commutes with $\overline{P}$ can be generated by traceable loop operators both contained within $\mathcal{T}_{\overline{P}}$ or outside of $\mathcal{T}_{\overline{P}}$. Examples of commuting traceable logical operators for the twist encoding are shown in Fig.~\ref{figTwistsEncoding}.

\begin{figure*}[t]
	\centering
	\includegraphics[width=0.48\linewidth]{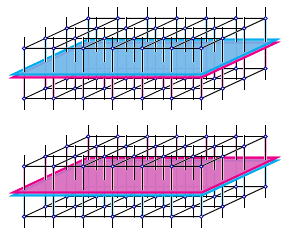}
	\includegraphics[width=0.48\linewidth]{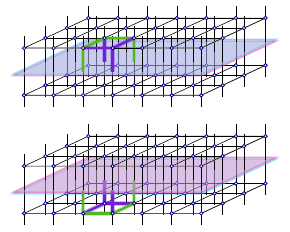}
	\caption{ Portals in the $6$-ring fusion network. (right) The modified fusion pattern to create a portal pair between two topological disks. The qubits (belonging to resource states) on either side of each portal are fused with the qubits on the corresponding side of the other portal (i.e., qubits of like-colors are fused; pink is fused with pink, and teal is fused with teal). (right) An example of a check operator spanning the portals.
	}
	\label{figPortalMicroscopics}
\end{figure*}

\subsection{Portals}
To perform logic on these twist-encoded qubits, we introduce the concept of a portal. Portals are two-sided, codimension-1 (i.e., two-dimensional) objects that can be thought of as a new type of geometric defect for space-time surface code instruments. They represent geometrically nonlocal correlations. We are specifically interested in portals that come in pairs, but note that self-portals (i.e., those defined on a single connected surface) are also possible and may have other interesting applications.
To be concrete, let us consider how to create portals in the six-ring fusion network. These portals are similar to the wormholes introduced in the 2D case in Ref.~\cite{krishna2020topological}.

To microscopically define a portal pair, we modify the bulk fusion pattern of the six-ring fusion network along two two-dimensional topological disks. Firstly, consider the dual of the fusion complex $\mathcal{L}^*$. (We obtain $\mathcal{L}^*$ from $\mathcal{L}$ by replacing vertices with volumes, edges by faces, and so on. In $\mathcal{L}^*$, volumes represent resource states and faces represent fusions between resource states.) Consider a topological disk $D$ consisting of a number of faces in $\mathcal{L}^*$. Let $D'$ be another topological disk obtained by translating (and potentially rotating) $D$. Disks $D$ and $D'$ specify a set of fusions. Each disk has two sides, separating qubits from resource states on either side. In the bulk, qubits on one side of a given disk are fused with qubits on the other side. To create a portal pair, we pair sides from $D$ and $D'$ and fuse qubits on each side of $D$ with qubits on the matching side of $D'$. This is depicted in the fusion graph in Fig.~\ref{figPortalMicroscopics}.

By changing the fusion group in this way, we obtain a new check operator group, $\mathcal{C}$, which contains check operators supported between the two disks, as depicted in Fig.~\ref{figPortalMicroscopics}. One can view the check operator entering one side of the disk as being mapped to the matching side of the other. Correspondingly, excitations can be mapped between disks by chains of errors entering one disk and emerging from the matching side of the other disk. We therefore refer to these disks as portals and their effect is to modify the connectivity geometry of the fusion network (leading to changes in topology and geometry).

\subsection{Logical operations by teleporting twists}

\begin{figure}[b]
	\centering
	\includegraphics[width=0.98\linewidth]{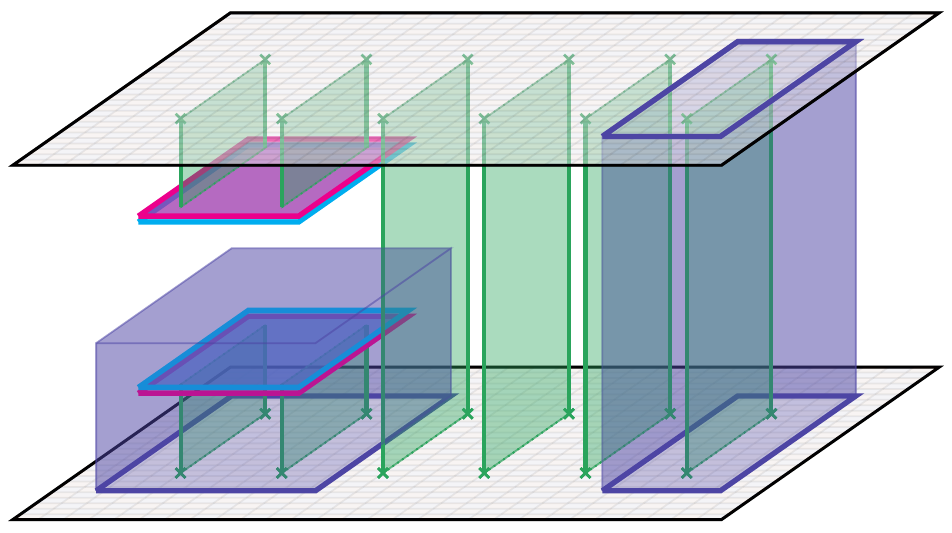}
	\caption{
	Teleporting the set of twists $\mathcal{T}_P$ to measure $\overline{X}_2\in \mathcal{P}_n$. The four twists enclosed by the $\overline{X}_2$ loop operator are teleported to a later timeslice. Depicted is a logical membrane $M^{\overline{X}_2,I}$ whose outcome gives the logical $\overline{X}_2$ measurement outcome. As shown by an example primal membrane on the right twist pair, Pauli operators that commute with the measured operator are unaffected.
	}
	\label{figTeleportedTwist}
\end{figure}

We now show how portals enable nondestructive measurements of logical operators for the twist encoding. As before, for any logical Pauli operator $\overline{P} \in \mathcal{P}_n$, we let $\mathcal{T}_{\overline{P}}$ be the set of twists enclosed by $\overline{P}$. To measure $\overline{P}$, construct a pair of portals to teleport the twists belonging to $\mathcal{T}_{\overline{P}}$ to a future temporal slice. Namely, define a topological disk $D_{\overline{P}}$ that encloses precisely the twists belonging to $\mathcal{T}_{\overline{P}}$ and another topological disk $D_{\overline{P}}'$ obtained by translating $D_{\overline{P}}$ by $d$ timesteps into the future (where $d$ is the fault distance). By matching the top face of one disk with the bottom face of the other (and vice versa), disks $D_{\overline{P}}$ and $D_{\overline{P}}'$ define a pair of portals. The the fusion pattern is modified such that twists and defects entering through the side of one portal are transmitted through the corresponding side of the other portal (one can verify that the check operator structure is valid, as we showed for portals in the bulk). We depict this in Fig.~\ref{figTeleportedTwist}. 

In general, one needs to resolve the locations of domain walls to ensure compatibility with the locations of twists that are traveling through the portals. Fortunately, the number of twists enclosed by a Pauli operator is always even, and therefore so is the number of twists entering a portal implementing its measurement. Therefore we can always find a compatible domain wall configuration, as exemplified in Fig.~\ref{figTeleportedTwist}.

We claim that this block and fusion pattern implements measurement of $\overline{P}$. In particular, we need to firstly check that the instrument network contains a logical membrane $M^{\overline{P},\id}$ corresponding to the logical correlation $P\otimes \id \in \mathcal{S}(M_{P})$. We verify this graphically in Fig.~\ref{figTeleportedTwist}, where the correlation surfaces of the fusion network corresponding to the input logical operator $\overline{P}$ is be ``capped off" and thus measured. Secondly, we need to check that any logical operator $\overline{Q} \in \mathcal{P}_n$ commuting with $\overline{P}$ is undisturbed, meaning that there are logical membranes $M^{\overline{Q},\overline{Q}}$ corresponding to the stabilizer $Q\otimes Q \in \mathcal{S}(M_{P})$. This is verified by the traceability property---every such commuting $\overline{Q}$ is generated by loop operators wholly within $D_{\overline{P}}$ or its complement, and the corresponding membranes propagate through the instrument network either through the portals or bulk, following Fig.~\ref{figTeleportedTwist}.

If the twists are separated by distance $d$ then it is sufficient to separate the portals by a distance $d$ to maintain an overall fault distance of $d$ for the protocol. We remark that many other portal and twist configurations are possible, including spacelike separated portals. Surprisingly, no boundaries need to be utilized in this construction. As we have seen in Sec.~\ref{secNumerics}, the lack of boundaries provides a favorable logical error rate scaling relative to the fault distance. Therefore, this scheme may provide an efficient approach to logical operations, as one can arrange twists in compact geometries. 

Beyond twist-based encodings, portals can also be used to save overhead for logical blocks based on planar encodings, as observed in Ref.~\cite{bombin2021interleaving}. For instance, portals can be used to compose the toric code spiders of Sec.~\ref{secConcatenation} that may be spatially separated. This is attractive in the context of measuring the stabilizer checks of a LDPC code.

\section{Conclusion}\label{secConclusion}

In this paper we have introduced a comprehensive framework for the analysis and design of universal fault-tolerant logic. The key components of this are the concept of fault-tolerant logical instruments, along with their specific application to surface-code-based fault tolerance with logical block templates. 

\textbf{Platform-independent logical gate definitions.} The framework of logical templates introduces a platform-independent method for defining universal logical operations in topological quantum computation based on surface codes. We have demonstrated how symmetry defects and boundaries can be used to encode and manipulate logical information, and have explicitly shown how these can be mapped onto fusion-based quantum computation as well as circuit-based models. As an application of our framework, we have presented volume-efficient logic templates, which, in addition to fusion-based quantum computation, can be utilized in any surface-code-based architecture. We hope that this can provide a valuable basis for a more unified approach to the study and design of fault-tolerant gates such that new techniques can map easily between different hardware platforms and models of computation.

\textbf{Flexible and scalable tools for numerical analysis.} The logical blocks framework enables a software-based mapping to physical instructions and a powerful tool for performing numerical analysis on complex logic blocks and their composition into small circuits. 
Using these tools, we have numerically investigated the performance of a set of Clifford logical operations, which when supplemented with noisy magic states (which can be distilled), are universal for fault-tolerant quantum computation. We have verified that the gate and memory thresholds are in agreement, and we have observed the important role that geometry and topology play in the fault-tolerance overhead---an important consideration when estimating resource costs for useful computations. The numerical results indicated that boundaryless computation appears to be a promising direction due to the further suppression of logical error rates for blocks without boundaries. 
As quantum technologies advance closer to implementing large-scale fault-tolerant computation~\cite{doi:10.1126/science.abi8378,egan2021fault,ryan2021realization,postler2021demonstration}, it is essential to have scalable software tools that allow analysis of complex logical operations. Our simulation framework based on logical block templates enables these advanced simulations by providing explicit definition, error sampling, and decoding of error prone logical operations on dozens of logical qubits with complex sets of topological features. 

\textbf{Exploration of novel logic schemes.} We have focused on designing logical gates directly---as fault-tolerant instruments---rather than as operations on a code, as this holistic view enables the construction of schemes that would be nonobvious from a code-centric view.
As a specific example of this, we introduced a new scheme for surface code computation based on the teleportation of twist defects, which was motivated by the improved performance of boundaryless computation. In this scheme, logical qubits are encoded in twists, and logical operations are performed by modifying the global space-time topology with nonlocal defects called portals. Such portals in general require the ability to perform long-range operations, such as those that are available in a photonic FBQC architecture. This scheme may offer further reductions in resource overheads.

\textbf{Future work.}
We have focused primarily on surface code fault-tolerant instruments and local operations, but the concepts we have introduced can be applied much more broadly. More general topological codes, in higher dimensions, with non-Euclidean geometry, or color codes may benefit from study in the framework of quantum instrument networks. Further study into the possible operations and resource reductions that can be offered by such codes is an important problem, as such codes can support significantly richer symmetry groups and domain walls. 
Further study into the power of nonlocal operations is also a promising avenue, with the teleported twist logic scheme we have introduced as one such example. 
For instance, transversal gates between multiple copies of the surface code may offer drastic resource reductions up to $\mathcal{O}(d)$ per logical operation, where $d$ is the block distance (i.e., the gate volume may reduce from $\mathcal{O}(d^3)$ to $\mathcal{O}(d^2)$). These transversal gates can be thought of in terms of symmetry domain walls between two or more copies of the surface code. 
It is advantageous to study the properties of transversal gates in the setting of fault-tolerant instruments, as they appear as codimension-1 defects on a constant time slice, where one can more easily reason about fault distances.
Furthermore, one may study nontransparent domain walls (e.g., to implement Pauli measurements) that, in contrast to the domain walls we have focused on, allow certain anyonic excitations to condense on them. 
In these cases, special purpose decoders are generally required to decode the instrument network, and more detailed analysis is required to fairly assess their performance.

Beyond purely topological codes, there is a lot of promise for fault tolerance in more general quantum LDPC codes~\cite{tillich2014quantum,gottesman2013fault,fawzi2018constant,fawzi2018efficient,breuckmann2021ldpc,hastings2021fiber,breuckmann2021balanced,panteleev2021asymptotically}. We have provided a proof-of-principle approach for fault-tolerant quantum computation based on concatenating surface codes with a general LDPC code. Such concatenated schemes may offer the advantages of the high thresholds of surface codes with the reduced overhead of constant-rate LDPC codes, but further analysis is required to determine the regimes in which they may outperform conventional topological code-based schemes. 

With recent advances in quantum hardware technology, and increased focus on large-scale, fault-tolerant computation, it is important that there is a unified language for fault tolerance that can span computational models and hardware platforms. We hope that the methods presented here provide a valuable step towards that goal.

\section{Ackowledgements}
We thank Daniel Dries, Terry Farrelly, and Daniel Litinski for detailed feedback and discussions on the draft, and 
Sara Bartolucci,
Patrick Birchall,
Hugo Cable,
Axel Dahlberg,
Andrew Doherty,
Megan Durney,
Mercedes Gimeno-Segovia,
Eric Johnston,
Konrad Kieling,
Isaac Kim,
Ye-Hua Liu,
Sam Morley-Short,
Andrea Olivo,
Sam Pallister,
Mihir Pant,
William Pol, 
Terry Rudolph,
Karthik Seetharam, 
Jake Smith,
Chris Sparrow,
Mark Steudtner,
Jordan Sullivan,
David Tuckett,
Andrzej P\'erez Veitia,
and all our colleagues at PsiQuantum for useful discussions.
R.V.M.'s current affiliation is Microsoft Station Q.

\section{Appendix}\label{secAppendix}

\subsection{Stabilizer states, operators, maps, and instruments}\label{secStabilizerOperatorsMapsAndInstruments}

The {\it stabilizer framework} \cite{gottesman1997stabilizer,gottesman2010introduction} has proven to be an extremely effective tool for describing a constrained form of fault-tolerant quantum computing.
In its most basic form, states are described by (Pauli) operators for which they are $+1$ eigenstates, rather than by their amplitudes in some computational reference basis.
Because it is possible to have highly entangled states be eigenstates of a set of commuting tensor product operators, this greatly enriches the set of states that can be described efficiently.
In general, in the Pauli stabilizer formalism, $n$-qubit pure states are described (up to a global phase) by $n$ commuting and independent Pauli product operators on $n$ qubits.
The number of classical bits required to provide such a representation is $\mathcal{O}(n^2)$ and allows representing $2^{\mathcal{O}(n^2)}$ distinct states.

The set of stabilizer states are preserved under the so-called {\it Clifford unitaries}.
Together, stabilizer states and Clifford unitaries admit a succinct algebraic description as a symplectic vector space over ${\mathbbm{Z}}_2$.
Within this constrained subset, this structure allows for efficient classical simulation as well as exhaustive theoretical analysis~\cite{gottesman1997stabilizer,gottesman2010introduction,aaronson2004improved}.
In this appendix, we describe extensions of this algebraic description and an extension to the set of operators, maps, and instruments that allow equally efficient simulation thanks to the underlying algebraic structure.
These extensions build on intuition gained from tensor network descriptions of stabilizer states (see, for example, Refs.~\cite{pastawski2015holographic, cao2021quantum,farrelly2021local}).
Although in general, stabilizer operators, need not be tensor product Pauli operators, this article makes the implicit assumption that they are, so the term {\it stabilizer} should be interpreted as {\it Pauli stabilizer}.

{\bf Notation.} Because the Pauli operators 
\begin{align*}
X={\begin{pmatrix}0&1\\1&0\end{pmatrix}},~
 Y={\begin{pmatrix}0&-i\\ i&0\end{pmatrix}},~   
 Z={\begin{pmatrix}1&0\\ 0&-1\end{pmatrix}},~
  I={\begin{pmatrix}1&0\\ 0&1\end{pmatrix}}
\end{align*}
play such a prominent role, we reserve these symbols for said operators. 
Furthermore, we liberally omit the tensor product operator symbol (``$\otimes$'') when specifying tensor products of such Pauli operators such that $X \otimes Y \otimes I \otimes Z$ may be simply denoted by $XYIZ$ (i.e., omission of an operator symbol among Pauli constants implies the tensor product rather than the usual matrix product).

\subsubsection{Stabilizer states}
\begin{defn}\label{defStabState}
A pure {\bf stabilizer state} $\ket{\psi}$ in an $n$-qubit Hilbert space is a state that can be specified (up to a scalar) as the common $+1$ eigenstate of a maximal Abelian subgroup ${\mathcal S}$ of $n$-qubit Pauli product operators (${\mathcal{P}}_n$) such that $-1 \not\in \mathcal{S}$.
\end{defn}
A maximal Pauli stabilizer group, $\mathcal{S}$ on $n$ qubits can be defined through $n$ independent and commuting Pauli operators $P_1, \ldots, P_n$, and is denoted as $ \mathcal{S}=\langle P_1, \ldots, P_n \rangle$. 

{\bf Example (Three-qubit GHZ state):}  The three-qubit entangled state $\ket{\psi} = (\ket{000}+\ket{111})/{\sqrt{2}}$ is a $+1$ eigenstate of the three independent Pauli product operators $Z  Z  I$, $I  Z  Z$ and $X  X  X$.
It is thus a stabilizer state stabilized by the group 
\begin{equation}
\mathcal{S}(\ket{\psi}) \equiv 
\langle Z  Z  I,~ I  Z  Z,~ X  X  X \rangle.
\end{equation}
    
\subsubsection{Stabilizer operators}
As mentioned earlier, the group of Clifford unitary operators (sometimes denoted as $\mathcal{C}_n$) is defined in such a way that it plays well with Pauli stabilizer states.
These are the $n$-qubit unitary operators that map Pauli product operators onto Pauli product operators:
\begin{align}\label{eqCliffordGroup}
	{\mathcal{C}}_n = \{ U \in \mathsf{SU}(2^n):  P \in {\mathcal{P}_n}  \implies U P U^\dagger \in  {\mathcal{P}_n} \}
\end{align}
The defining condition for elements of the Clifford group can be reexpressed as a set of $2n$ stabilizer conditions of the form 
\begin{align}\label{eqCliffordStabilizer}
	Q^\dagger U P = U \quad\text{ with }\quad Q:= UPU^\dagger
\end{align}
and $P$ ranging over all generators of the Pauli group $\mathcal{P}_n$.

In fact, this is a special case of what we call stabilizer operators (also called generalized Clifford operators).
\begin{defn}
An operator $O$ taking $k_{\text{in}}$ qubits as input, and $k_{\text{out}}$ qubits as output is a {\it stabilizer operator} if and only if the state $\ket{\psi(O)}  :=  (I^{\otimes k_{\text{in}}} \otimes O)\ket{\Omega_{k_{\text{in}}}}$ is a stabilizer state.
\end{defn}
Here, $\ket{ \Omega_{k_\text{in} } } = \sum_{j=0}^{2^{k_\text{in}}-1}  \ket{jj}$ is the (unnormalized) $2k_\text{in}$ qubit state used in the {\it operator state correspondence} (also known as the Choi-Jamio{\l}kowski isomorphism). 
In it, the first $n$ qubits and the remaining $k_\text{in}$ qubits are pairwise maximally entangled. 
It is equivalent (up to permutation of the tensor factors) to the tensor product of $k_\text{in}$ Bell states $\ket{\text{Bell}}^{\otimes k_\text{in}}$.

The stabilizer group $\mathcal{S}(O)$ is given by the stabilizer group of the corresponding state $\ket{\psi(O)}$ under the operator state correspondence.
This definition includes, for instance, stabilizer states, Clifford unitaries, and full and partial projectors onto stabilizer subspaces. In the following examples, we use the tensor product to partition the input and output spaces. 

{\bf Example (Identity):} For the single-qubit identity $I$, we have $\mathcal{S}(I)=\langle {X}\otimes {X}, {Z}\otimes {Z}\rangle$. 

{\bf Example (Phase gate):} The Phase gate
\begin{equation}
S\equiv  {\begin{pmatrix}1&0\\0&i\end{pmatrix}}
\end{equation} 
has a corresponding state $\ket{\psi(S)} = \ket{00} + i \ket{11}$.
Its corresponding stabilizer group is thus $\mathcal{S}(S)=\langle {X}\otimes {Y}, {Z}\otimes {Z}\rangle$. 

{\bf Example (Lowering operator):} The qubit lowering operator $\sigma^- := \ket{0}\bra{1}$ is also a stabilizer operator under this general definition. 
Its corresponding state is $\ket{\psi(\sigma^-)} = \ket{10}$, which is stabilized by $\mathcal{S}(\sigma^-) = \langle -Z\otimes I, I\otimes Z \rangle$.

{\bf Example (2-repetition encoding):} The encoding isometry $E := \ket{00}\bra{0} + \ket{11}\bra{1}$ that encodes a qubit onto the two-dimensional subspace stabilized by $ZZ$ of a two-qubit subspace in a way that the $X$ operator is mapped onto $XX$ and the $Z$ operator is mapped onto $IZ$ is a stabilizer operator.
Its corresponding state $\psi(I)=\ket{000}+\ket{111}$ is a three-qubit GHZ state up to normalization and its stabilizer group is $\mathcal{S}(E)= \langle I\otimes ZZ, X\otimes XX, Z\otimes IZ \rangle $.

{\bf Example (Partial projection):} The partial projection $\Pi$ onto a subspace stabilized by a Pauli stabilizer subgroup $\mathcal{G}_{\Pi}$ is a stabilizer operator. 
It is stabilized by $\mathcal{S}(\Pi) = \langle I \otimes G, G^T \otimes I, N^T \otimes N \rangle$, where $G$ is taken over a set of generators for   $\mathcal{G}_{\Pi}$ and $N$ is taken over the commutant of $\mathcal{G}_{\Pi}$ (i.e., $\{ N\in\mathcal{P}_n ~|~  NGN^\dagger = G \quad \forall G \in G_{\Pi}\}$).

\subsubsection{Stabilizer maps}
In principle, a general quantum channel $\Lambda$ can be expressed using the Kraus representation as a combination of quantum operators $K_j$ as
\begin{align}
	\Lambda(\rho) = \sum_j K_j \rho K^\dagger_j.
\end{align}	
The stabilizer formalism only specifies the stabilized operator up to a global scalar.
Requiring a channel to be trace preserving fixes one such magnitude up to an irrelevant global phase.
However, the relative magnitude of operators is important for specifying a channel with multiple Kraus operators.
For this reason, stabilizer channels are limited to a single Kraus operator.
This makes the trace preservation requirement particularly restrictive as it excludes all stabilizer operators other than the unitaries (corresponding to the Clifford group) and isometries from being lifted into a trace-preserving quantum channel interpretation.

In principle, it is possible to construct quantum channels from multiple stabilizer operators as Kraus operators; it becomes necessary to introduce scalars that significantly complicate the picture.
We find that, for the idealized fault-tolerant QIN, significant headway can be made by restricting the focus to stabilizer maps and instruments.
More general maps will become absolutely necessary when incorporating noise modeling.

\subsubsection{Stabilizer instruments}
Instruments allow modeling of operations that have both quantum and classical output. 
The possibility of extracting classical data is what enables entropy extraction in fault-tolerant protocols which is the main tool for noise mitigation.
While stabilizer operators are clearly more general than Clifford unitaries, the restriction on having maps be trace preserving and involve a single stabilizer operator leads to a very limited selection of valid stabilizer QINs.

Whereas, for general instruments, the structure of the classical outcomes is not prescribed, we define stabilizer instruments to posses a very specific $\mathbbm{Z}_2$-type linear structure on the classical outcomes.
In particular, a quantum instrument will be a quantum map for which a subset of the output qubits can be treated as a classical outcome register.
In a stabilizer quantum instrument, the map associated with each specific classical outcome will itself be a stabilizer operator.
\begin{defn} A {\bf stabilizer quantum instrument} from $k_\text{in}$ qubit inputs onto $k_\text{out}$ qubit outputs and $b$ classical outcome bits is specified by a stabilizer group $\mathcal{S} \subseteq \mathcal{P}_{k_\text{in} + k_\text{out} + b}$ such that
	\begin{itemize}
	\item outcome bits $b$ carry classical correlations ($\mathcal{S}|_b \subseteq \mathbf{Z}_b$),
	\item $\mathcal{S}$ can be completed into a maximal stabilizer group by including additional generators exclusively from $\mathbf{Z}_b$,
	\item the instrument is trace preserving ($\mathcal{S}\|_{\text{in}} = \langle I \rangle$). 
	\end{itemize}
\end{defn}

Here, $\mathbf{Z}_b$ denotes the group generated by $Z$-type operators in $b$ together with real phases and $\mathcal{S}\|_b$ is the subset of Pauli operators in $\mathcal{S}$ with support in subsystem $b$.
We obtain $\mathcal{S}|_b$ by restricting each Pauli operator in $\mathcal{S}$ to subsystem $b$. It is only defined up to phases.
Each distinct completion of $\mathcal{S}$ through elements of $\mathbf{Z}_b$ corresponds to a distinct stabilizer operator and corresponds to one of the terms composing the quantum instrument.
The generators of $\mathcal{S}\|_b \equiv \mathbf{Z}_b \cap \mathcal{S}$ are predetermined parity combinations of classical outcomes and correspond to checks in composite quantum instruments.
The number of distinct outcomes and stabilizer operators in the instrument is $2^b/|\mathcal{S}\|_b|$.
The outcome contains $b-\log |\mathcal{S}|_b|$ uniformly random bits of information that are uncorrelated with the input state or the transformation performed on it.

The trace-preserving condition is included to guarantee that the instrument does not postselect on a specific subspace.
It may be dropped if postselection is to be allowed.

{\bf Example (Single qubit destructive $M_Z$ measurement):} In a single qubit-measurement the $Z$ observable in a qubit is mapped onto a classical bit.
The (incomplete) stabilizer group defining the quantum instrument is $\mathcal{S}(M_Z) = \langle ZZ\rangle$, with the second tensor factor representing the classical outcome.
The stabilizer can be completed by adding either $IZ$ or $-IZ$.
The corresponding stabilizer operators are stabilized by $\langle ZI, IZ \rangle$ and $\langle -ZI, -IZ\rangle$.
The corresponding instrument $\{ \mathcal{E}_0, \mathcal{E}_1 \}$ has two terms corresponding to the two possible outcomes and computational basis projections
\begin{align*}
	\mathcal{E}_0(\rho) &= \langle 0 \vert \rho \vert 0 \rangle \\
	\mathcal{E}_1(\rho) &= \langle 1 \vert \rho \vert 1 \rangle.
\end{align*}

{\bf Example (Partial projective measurement):} Consider a projective measurement of $XX$ on two qubits, wherein the two qubits are retained. 
The ordering of qubits to represent the stabilizer will be input, output, classical outcome bits.
The (incomplete) stabilizer group defining the partial projective measurement is given by $\mathcal{S} = \langle XIXII, IXIXI, ZZZZI, XXIIZ \rangle$.
The last generator is indicating that the $XX$ observable is mapped onto the classical bit outcome whereas the first three are indicating that observables commuting with $XX$ are preserved.
The corresponding instrument $\{ \mathcal{E}_0, \mathcal{E}_1 \}$, has two terms corresponding to the two possible outcomes and computational basis projections
\begin{align*}
	\mathcal{E}_0(\rho) &= \Pi_{+XX} \rho \Pi_{+XX} \\
	\mathcal{E}_1(\rho) &= \Pi_{-XX} \rho \Pi_{-XX},
\end{align*}
with $\Pi_{\pm XX} := (II\pm XX)/{2}$ respectively being projectors onto the $\pm1$ eigenspaces of $XX$.

\subsection{Kitaev to Wen versions of the toric code}\label{appKitaevToWen}

\begin{figure}
	\centering
	\includegraphics[width=0.7\linewidth]{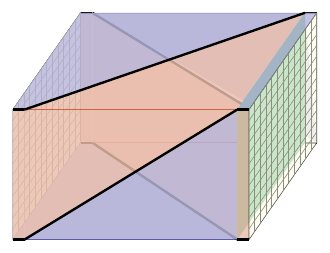} 
	\caption{Logical template for a Hadamard gate based on the (Kitaev-style) planar code. Note that while this geometry of planar code uses twice as many qubits than the Wen-style geometry to encode a qubit for a given distance, the volume of this gate is less than the version depicted in Fig.~\ref{figAllLogicalBlocks} for a fixed fault distance.}
	\label{figKitaevHadamard}
\end{figure}

The surface code can be defined on a variety of different lattice geometries. The original (CSS) toric code due to Kitaev~\cite{kitaev2003fault} associates qubits with the edges of an arbitrary 2D cell complex, with $X$-checks $X_v$ associated with the coboundary of a vertex $v$, and $Z$-checks $Z_p$ associated with the boundary of a 2-cell $p$, for every vertex $v$ and 2-cell $p$ of the cell complex. These checks are guaranteed to commute for an arbitrary 2D cell complex due to properties of the (co)boundary. 

If we consider a square lattice then the symmetry for the code is given by 
\begin{equation}\label{eqToricCodeLatticeSymKitaev}
S_{\text{Kitaev}}(g) = H^{\otimes n} T(\hat{u}),
\end{equation}
where $T(\hat{u})$ is the translation operator, translating a vertex to a plaquette in the Kitaev lattice, and $H$ is the Hadamard gate. In particular, this symmetry differs from that of the surface code due to Wen by a transversal Hadamard. To map between the Wen and Kitaev formulations of the surface code we can apply (to the stabilizers) a Hadamard transversally to half of the qubits in a bipartite way.

The Wen geometry requires half as many qubits than the Kitaev geometry to achieve the same distance~\cite{bombin2006topologicalencoding,bombin2007optimal,tomita2014low}. However, the logical blocks may require different volumes. For instance, for the planar code with boundaries introduced in Ref.~\cite{bravyi1998quantum}, one can perform a Hadamard using the protocol introduced in Ref.~\cite{dennis2002topological}, leading to the template displayed in Fig.~\ref{figKitaevHadamard}. Ref.~\cite{beverland2019role} illustrated that entropic effects can be significant when comparing the two codes for memory, as we have also seen in the context of gates. Therefore, one should carefully consider the noise model, decoder, and set of logic gates when determining the most efficient surface code geometry.

\subsection{Extending the feature labels to 1-cells}\label{secLogicalBlockTemplateExtension}
For a logical block template $(\mathcal{L}, F)$, we can extend the function $F: \mathcal{L}_2 \rightarrow L$ to the 1-cells $\mathcal{L}_1$. Namely, let the set of 1-cell labels be given by $\mathcal{F}_1= \{\text{Cornerline}, \text{TransparentCornerline}, \text{Twist} \}$.  For convenience, we denote PrimalBoundary by PB, DualBoundary by DB, DomainWall by DW, Cornerline by Cl, TransparentCornerline by TCl, and Twist by T.

Let $\delta c$ be the coboundary of $c$, consisting of the (at most four) faces that contain $c$. For a 1-cell $c\in \mathcal{L}_1$ we define $F(c)$ as 
\begin{widetext}
\begin{equation}
F(c) = \begin{cases}
\text{Cl} & \text{if } |F^{-1} (\text{PB})\cap \delta c| = |F^{-1} (\text{DB})\cap \delta c| = 1 \land |F^{-1} (\text{DW})\cap \delta c| = 0  \\ & \text{or } \{|F^{-1} (\text{PB})\cap \delta c|, ~|F^{-1} (\text{DB})\cap \delta c|\} = \{0, 2\} \land |F^{-1} (\text{DW})\cap \delta c| = 0, \\
\text{TCl} &\text{if } |F^{-1} (\text{PB})\cap \delta c| = |F^{-1} (\text{DB})\cap \delta c| = 1 \land |F^{-1} (\text{DW})\cap \delta c| = 1, \\
\text{T} & \text{if } |F^{-1} (\text{PB})\cap \delta c| = |F^{-1} (\text{DB})\cap \delta c| = 0 \land |F^{-1} (\text{DW})\cap \delta c| = 1.
\end{cases}
\end{equation}
\end{widetext}

\subsection{Converting a template to circuit-based measurement instructions}\label{secTemplateToCBQC}
In this section we explain how to convert a template $(\mathcal{L}, F)$ to a system $\Phi=(\mathcal{Q}, \mathcal{P}, \mathcal{O})$ of physical instructions on static qubits for circuit-based quantum computation. We walk through the phase gate example in the following section. Let the coordinates of the complex $\mathcal{L}$ be given by $(x,y,z)$.

\begin{figure*}
	\centering
	\includegraphics[width=0.9\linewidth]{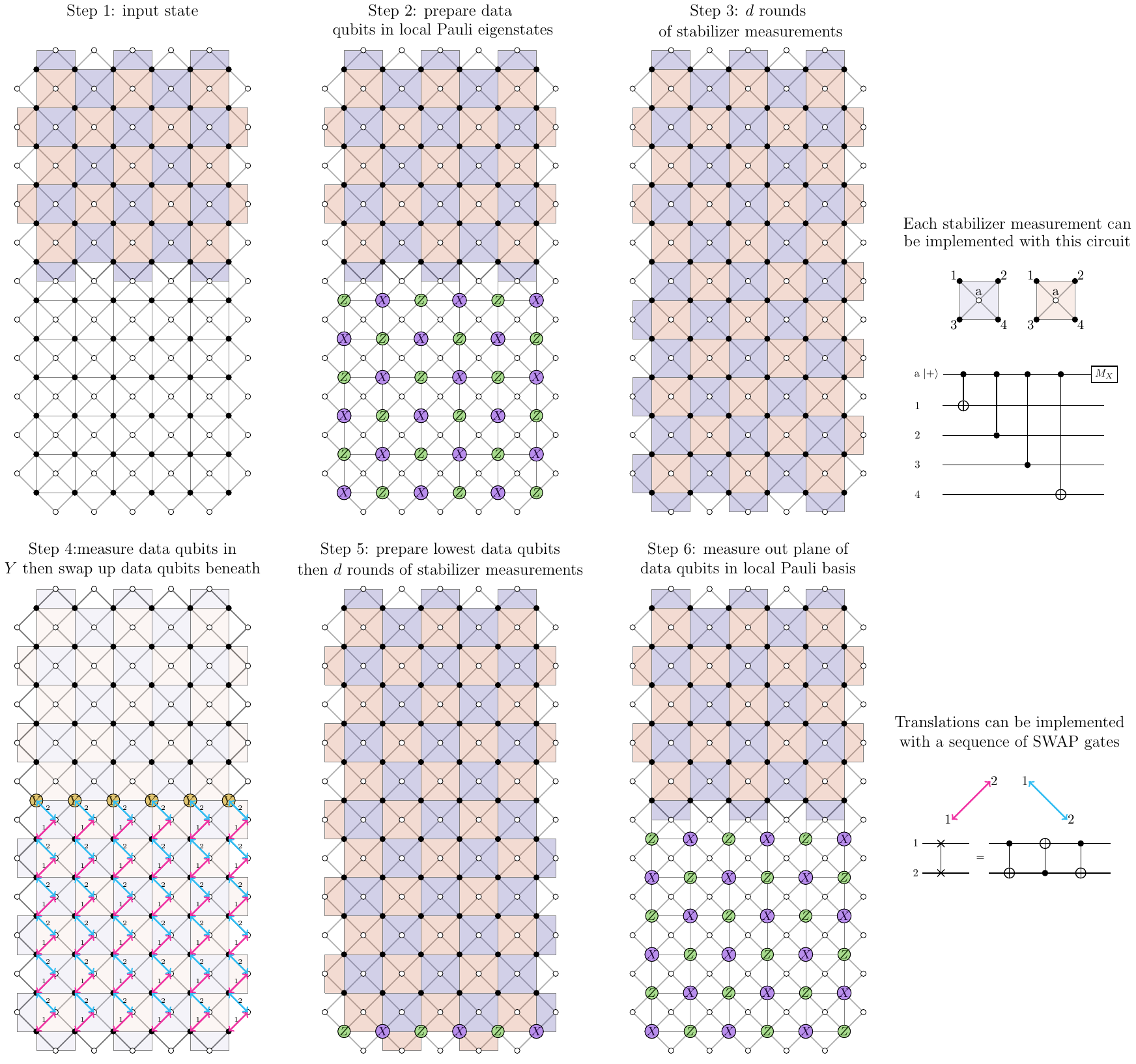} 
	\caption{Circuit-based instructions for the phase gate in Fig.~\ref{figAllLogicalBlocks}, assuming time slices are taken progressing from left to right. Data qubits are laid out in a square grid. We also depict the ancilla qubits required, which are placed on each face, and allow for the stabilizer measurements. Ancilla qubits are connected to neighbouring data qubits, between which two qubit gates (such as CNOTs) can be performed. At each step, the stabilizer measurements which are to be performed are highlighted---one may first measure stabilizers on plaquettes of one type before the other. In step 4, a row of data qubits is measured in the Pauli-$Y$ basis, after which all data qubits below are translated up. This translation can be achieved with a two-step process using SWAP gates as shown. One may progress through the block in Fig.~\ref{figAllLogicalBlocks} in other directions, e.g., from front to back or bottom to top, leading to different instruction sets. For example, if we progress from front to back, then 5-qubit twist operators need to be measured. }
	\label{figPhaseGateCBQC}
\end{figure*}

\textbf{The quantum system $\mathcal{Q}$.}
Logic block templates have no directional preference; the notions of space and time are on equal footing. To compile a template into physical instructions for CBQC we must break this symmetry (due to the static nature of qubits assumed in CBQC). In particular, to map a template to a sequence of stabilizer measurements on a 2D array of qubits, we first choose a coordinate direction, say $\hat{z}$, and define it as the physical time direction. For each time slice, we have a 2D subcomplex of the template, upon each vertex of which we place a qubit. Vertices with the same $x$ and $y$ coordinate but different $z$ coordinates correspond to the same qubit at different times. This defines the quantum system $\mathcal{Q}$. 

\textbf{The input and output ports $\mathcal{P}$.}
The ports are simply given by the set of qubits at the first and final time slices. In particular, if the template complex $\mathcal{L}$ has $z$ coordinates spanning $z\in[0,1,\ldots, T]$ then the set of qubits living on vertices at $z=0$ ($z=T$) define one or more surface codes forming the input (output) ports.

\begin{figure*}
	\centering
	\includegraphics[width=0.2\linewidth]{figAncillaMeasurement.pdf} \quad
	\includegraphics[width=0.2\linewidth]{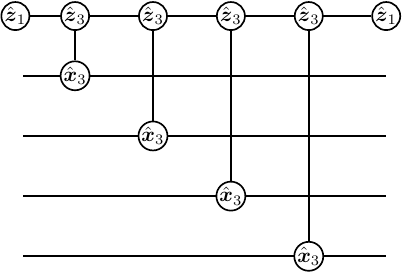} \quad
	\includegraphics[width=0.12\linewidth]{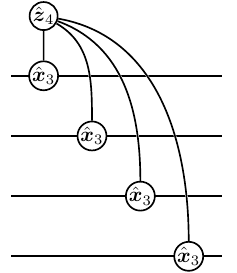} \quad
	\includegraphics[width=0.2\linewidth]{figSpiderNetworkCheck.pdf} \\ \vspace{0.1cm}
	\includegraphics[width=0.5\linewidth]{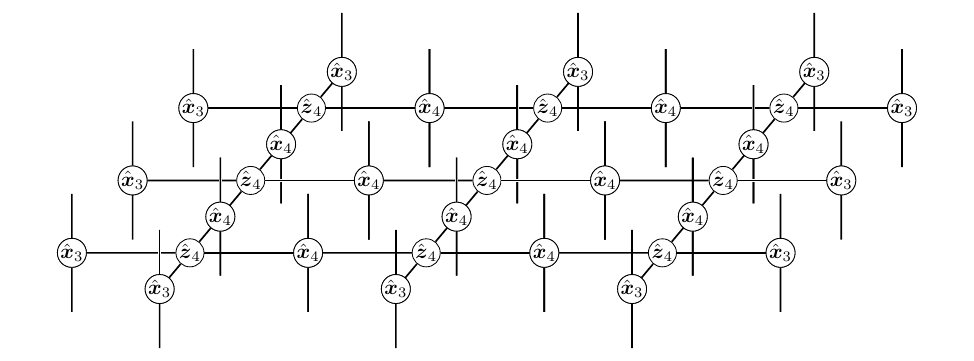} \quad
	\includegraphics[width=0.4\linewidth]{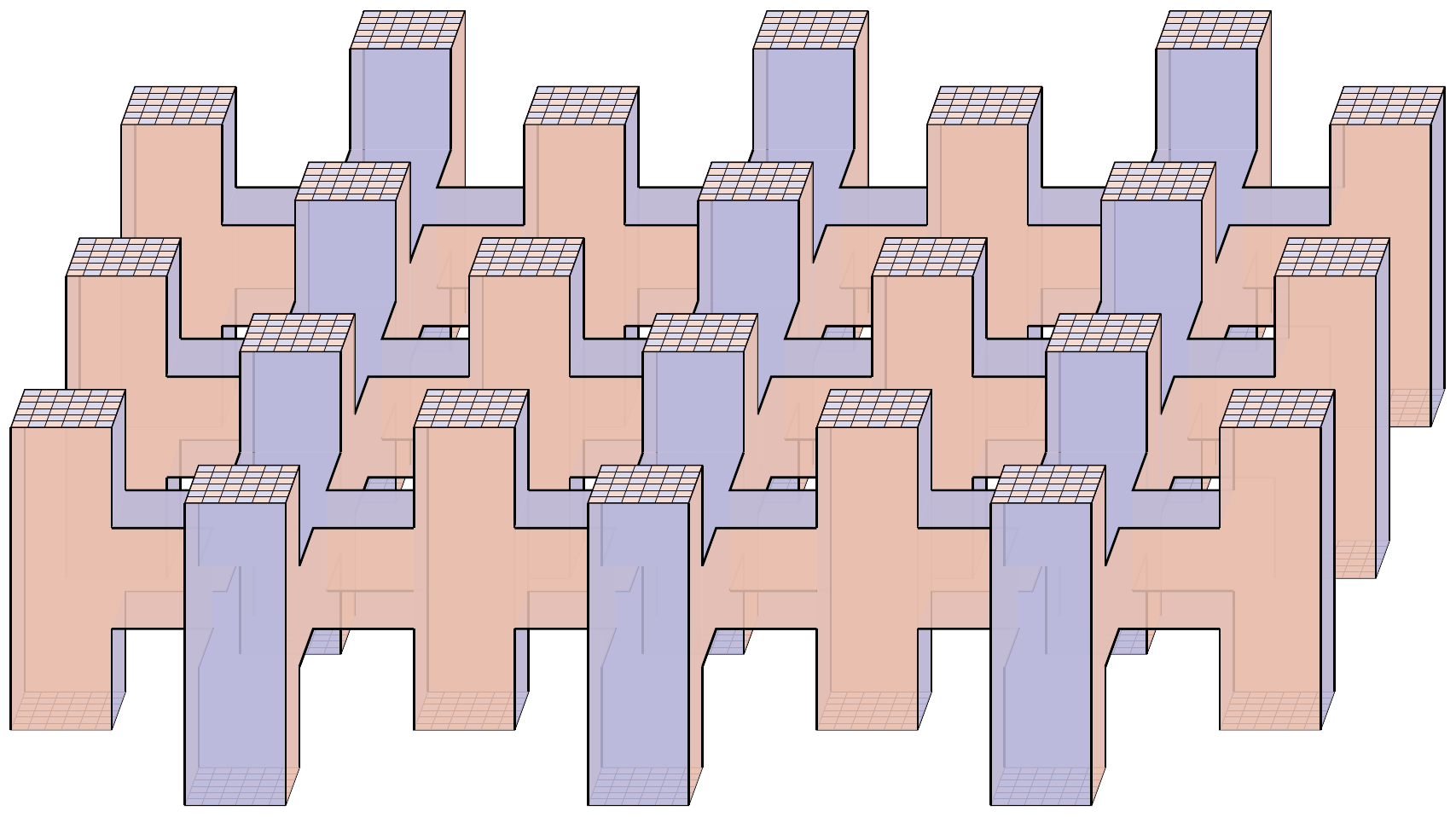}
	\caption{Converting a Clifford circuit to a spider network. (top) The CNOTs, $\ket{+}$ preparation and $M_X$ measurement operations can be represented by the spiders $\spiderx{k}$ and $\spiderz{k}$ as shown. One can simplify the spider operations by merging them as shown, giving the minimal network on the right. For the first three figures, time moves from left to right, whereas for the right-most figure, time moves from bottom to top. (bottom) Measuring stabilizers of a (Kitaev) surface code concatenated with itself. Time moves from bottom to top. On the left,  we depict a network of $\spiderx{4}$ and $\spiderz{k}$ spiders to implement a round of $X^{\otimes 4}$ surface-code stabilizers. On the right, we depict the corresponding network of logical blocks. To obtain the measurement pattern of the $Z^{\otimes 4}$ surface-code stabilizers, one can swap the role of the $\spiderx{4}$ and $\spiderz{k}$ spiders.}
	\label{figConvertingCircuitToSpider}
\end{figure*}

\textbf{The physical instructions $\mathcal{O}$.}
At a high level, each slice of the complex for a different $z$ coordinate corresponds to a different set of stabilizer measurements that are to be performed at a given timestep. For example, on a given time slice, each face in the absence of features corresponds to a bulk surface code stabilizer check measurement, as depicted in Fig.~\ref{figToricCode}. The features on the template determine stabilizer measurements to perform, and may have different meanings depending on whether they lie in the $x-y$ plane (a slice of constant time), or if they have a component in the time direction, as is due to the asymmetry between space and time.

\textit{Timelike features.}
First, we consider the operations corresponding to features propagating in the time direction. For a given time slice, the intersection pattern of domain walls, twists, and boundaries propagating in the time direction (i.e., a twist supported on a link in the $z$ direction, or a domain wall or boundary supported on a face normal to the $x$ or $y$ direction) defines a configuration of pointlike twists and 1D boundaries and domain walls on a 2D surface code, as depicted in Fig.~\ref{figToricCode}. For such features, one performs stabilizer measurements according to the corresponding 2D surface code features, as depicted by the examples in Fig.~\ref{figToricCode}. 

\textit{Spacelike features.}
Now we interpret twists, domain walls, and boundaries that are supported in a given time slice. Twists propagating in a spatial direction (i.e., along a $x$ or $y$ link) correspond to a sequence of single-qubit Pauli-$Y$ measurements that need be performed on all qubits supported on the twist line. Since spacelike twists are supported on the boundary of a timelike domain wall, such measurements allow one to transition to and from measuring the 2D stabilizer terms along a defect to the 2D bulk stabilizers (each of which is depicted in Fig.~\ref{figToricCode}) while maintaining a useful syndrome history. Domain walls in the spacelike direction (i.e., on faces normal to the $z$ direction) signal that one needs to apply the $\zz_2$ translation symmetry transformation---the natural direction to translate is toward the twist line. Since the purpose of this symmetry is to swap primal and dual plaquettes, rather than physically apply the translation symmetry to the qubits, one can simply keep track of the transformation and update future stabilizer measurements as appropriate. Finally, primal and dual measurement patterns correspond to checkerboard patterns of single-qubit measurements in the $X$ and $Z$ bases. Namely, for a primal (dual) boundary one needs to measure qubits in the single-qubit $X$ and $Z$ bases according to the restriction of primal (dual) checks to each individual qubit. In other words, the single-qubit measurement pattern should be such that primal (dual) checks can be recovered from the measurement outcomes---such measurement patterns can be thought of as transversal logical readouts.

\begin{figure*}
	\centering
	\includegraphics[width=0.66\linewidth]{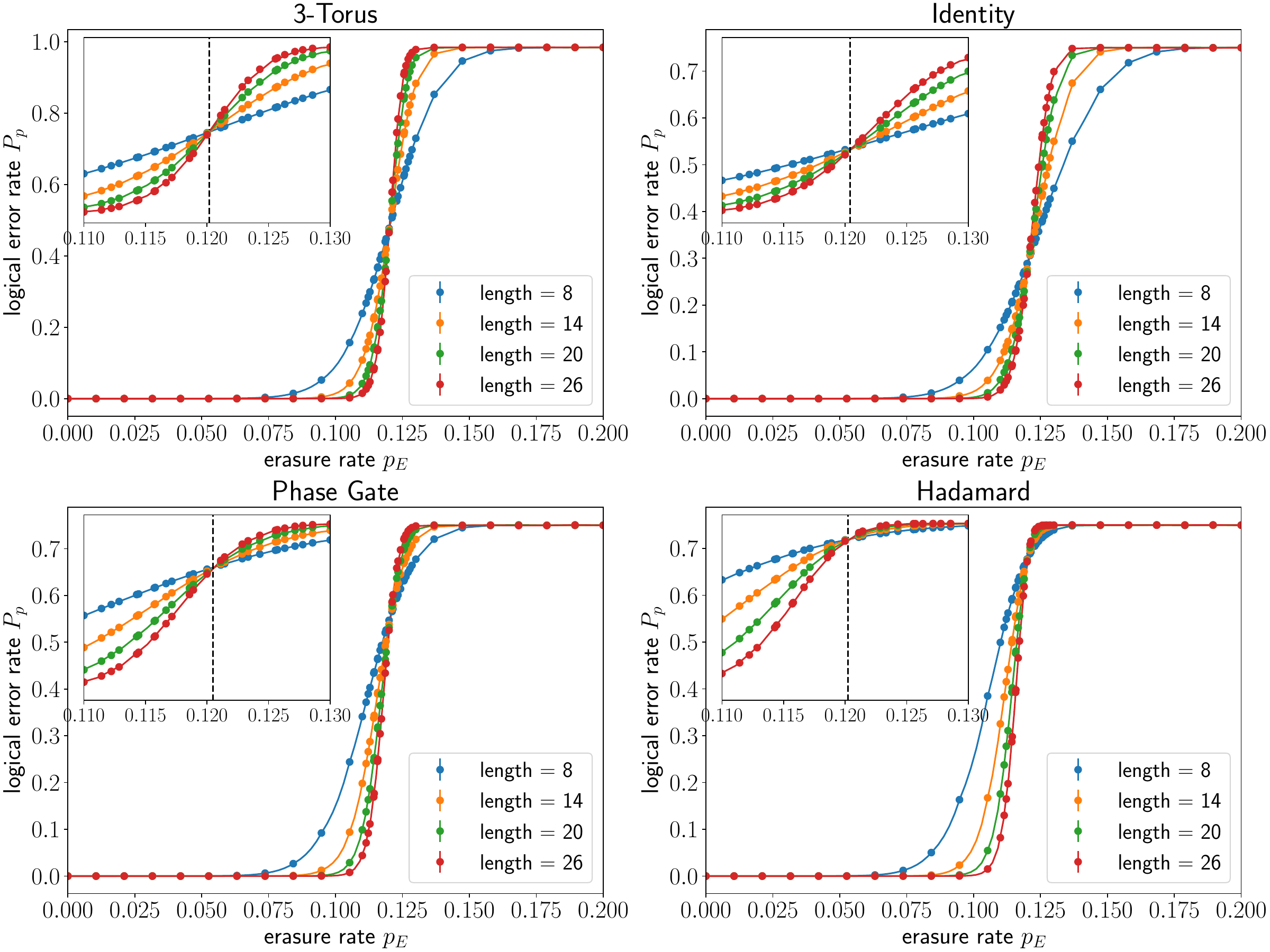}
	\includegraphics[width=0.99\linewidth]{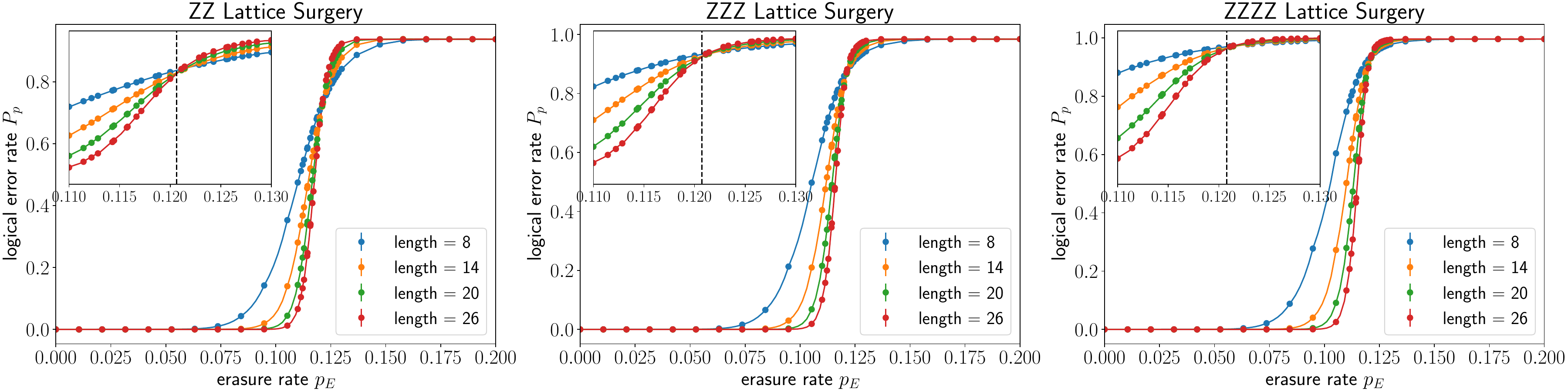}
	\caption{Threshold crossing plots for the different logical blocks against IID erasure noise. The logical error rate is defined with respect to any logical error (which is block dependent). In other words, a logical error occurs whenever \textit{any} of the logical membranes is incorrectly recovered. Each data point is the estimated logical error rate from $10^7$ to $10^8$ trials. The estimated thresholds are displayed by dashed lines.}
	\label{figThresholdsEr}
\end{figure*}

\begin{figure*}
	\centering
	\includegraphics[width=0.66\linewidth]{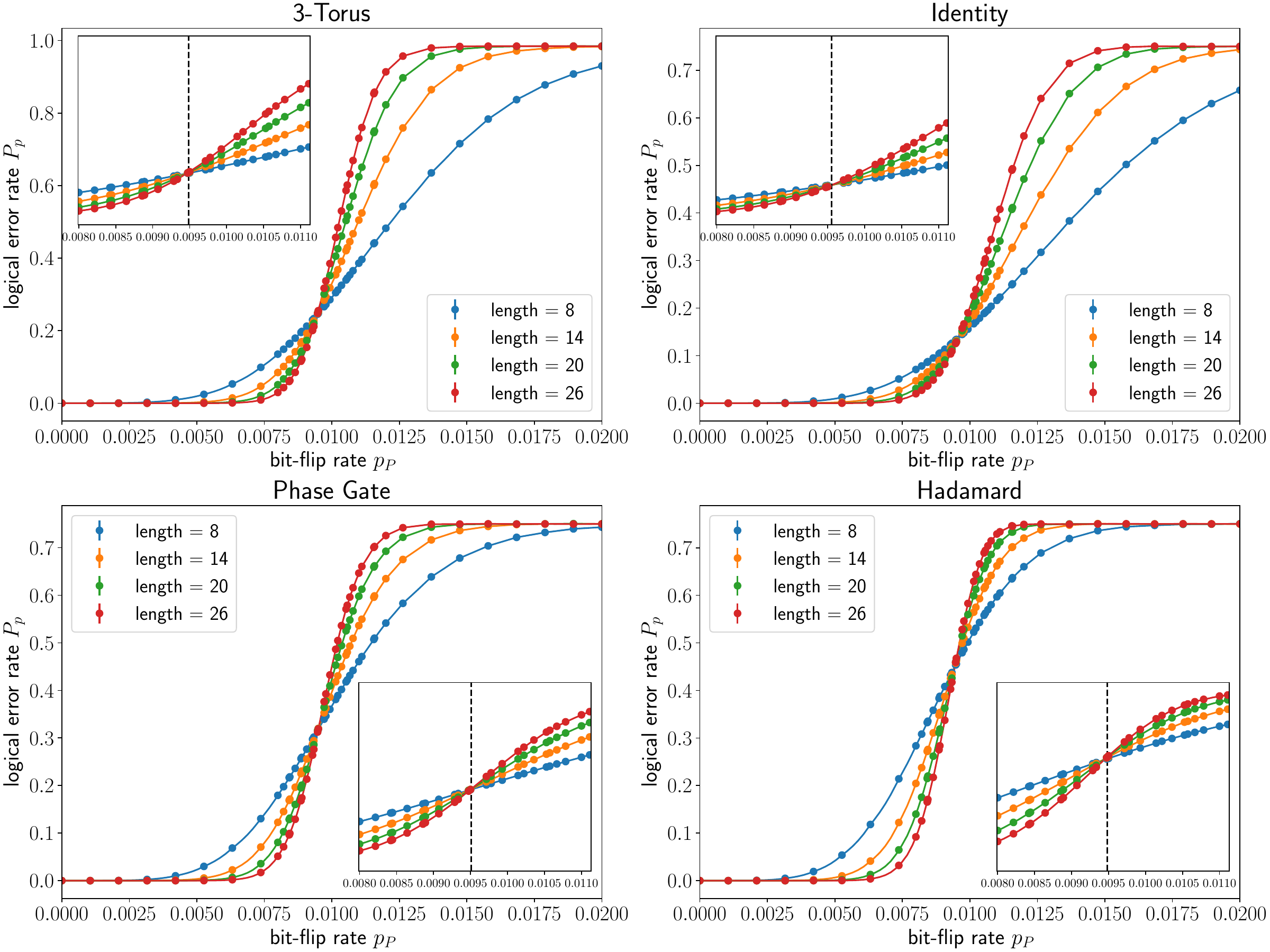}
	\includegraphics[width=0.99\linewidth]{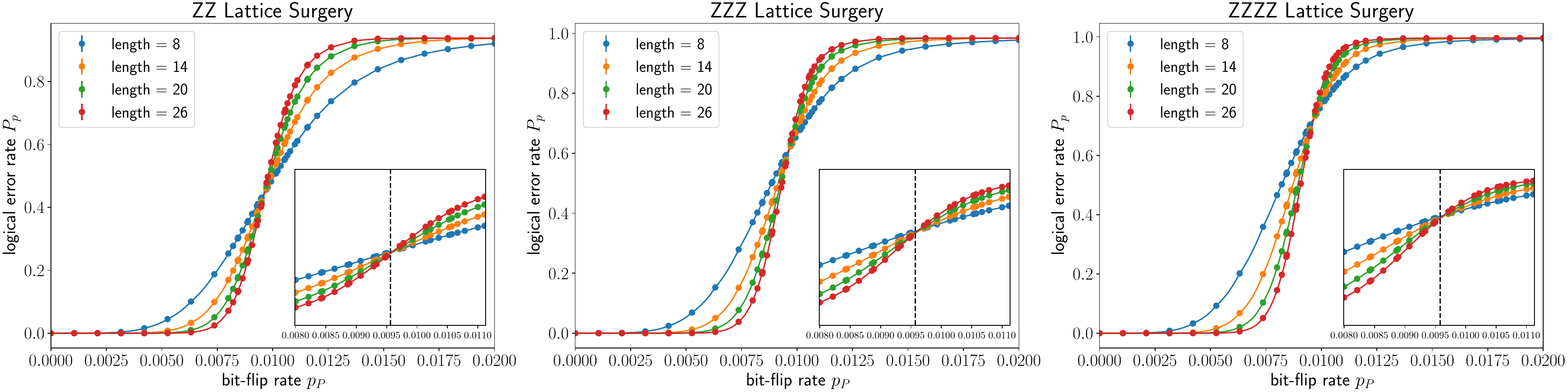}
	\caption{Threshold crossing plots as per Fig.~\ref{figThresholdsEr}, but under IID bit-flip noise.}
	\label{figThresholdsPa}
\end{figure*}

\textbf{Checks.}
Instrument checks may be identified at the level of the template and depend on the locations of features. 
Firstly, checks arise from repeated measurements (which should give the same outcome in the absence of errors). 
In the absence of features, we have a check for every repeated bulk stabilizer measurement and thus we have a check for every bulk 3-cell. 
Similarly, on a primal or dual timelike boundary we have a check for every repeated boundary stabilizer measurement, and these can be identified with boundary 2-cells (note that not every boundary 2-cell corresponds to a check). For spacelike primal (dual) boundaries, we have a check for each primal (dual) 2-cell; in the absence of errors, the stabilizer measurement must agree with the product of the four single-qubit measurements that comprise it. Finally, in the presence of domain walls and checks, we have a check for every pair of 3-cells sharing a defect or twist line: by design, the product of all measurement outcomes supported on the pair of 3-cells---including bulk stabilizers, domain wall stabilizers, twist stabilizers, and single-qubit $Y$ measurements---must deterministically multiply to $+1$. For example, in Fig.~\ref{figPhaseGateCBQC}, a pair of primal and dual stabilizers from step 3 (on either side of the location of the twist line in the following step) may be multiplied with two $Y$ measurements from step 4 on their common support, and one stabilizer from step 5, to produce a check.

\textbf{Logical membranes.}
Logical membranes correspond to representative logical operators being tracked through the surface codes at each timestep. To obtain the logical operator that the membrane corresponds to at time slice $\tau \in \{0,1,\ldots,T\}$, we take the restriction of a membrane to the subcomplex between times $0\leq z\leq \tau$ and take its boundary---this will give a set of Primal/Dual/PrimalDual labels on edges that corresponds to the representative logical Pauli operator at that time interval, as depicted in Fig.~\ref{figToricCode}. Components of a membrane in a plane of constant time correspond to stabilizer measurements that must be multiplied to give the equivalent representative on that slice (and thus their outcome is used to determine the Pauli frame). Recall that the membrane can locally be of primal type, dual type, or a composite primal-dual type, meaning when their projection on a single time slice corresponds to a primal ($X$-type), dual ($Z$-type), or composite ($Y$-type) logical string operator.

\vspace{10pt}

In FBQC the symmetry between space and time is restored, as there is no natural direction of time. This means that the fusions may be performed in any order; layer by layer (following the surface code analogy) or in a rastering pattern as in Ref.~\cite{bombin2021interleaving}. This decoupling of ``simulated time'' corresponding to the passage of time that logical qubits experience, and the ``physical time'' corresponding to the order in which physical operations are undertaken presents additional flexibility in FBQC. 

We remark that one can intuitively map between the different models of computation using ZX instruments of the diagrammatic ZX calculus~\cite{bombin2023unifying}.

\subsection{Converting a phase gate into circuit-based instructions}\label{AppPhaseGateCBQC}
We now present a concrete example of converting a template to circuit-based instructions. We give the phase gate of Fig.~\ref{figAllLogicalBlocks} as an example, as it illustrates many of the possible features in the construction. The instructions are depicted in Fig.~\ref{figPhaseGateCBQC}. Note that, when we propagate through the block in this way, no stabilizer measurements of weight higher than four are required. Importantly, if we progress through the block in a different way, we may require five-qubit measurements along the twist, as is the case when we propagate from front to back and the twist line is perpendicular to our time slices.

\subsection{Magic state preparation}\label{secMagicStatePreparation}
To complete a universal gate set, we need a non-Clifford operation. A common approach is to prepare noisy encoded magic states and then distill them~\cite{bravyi2005universal}. A standard magic state to prepare is the $T$-state $\ket{T} = (\ket{0} + e^{\frac{i \pi}{4}} \ket{1})/{\sqrt{2}}$. 
We show how to prepare the noisy encoded $\ket{T}$ states in Fig.~\ref{figFBQCMagicStatePreparation}, which is based on the preparation scheme in Ref.~\cite{lodyga2015simple}. 
We refer to the qubit in Fig.~\ref{figFBQCMagicStatePreparation} where the non-Clifford operation is performed as the \textit{injection site}.

The scheme is inherently noisy due to the existence of weight-1 and other low-weight logical errors on and around the injection site. Excluding errors on the injection site, all weight-1 Pauli errors are detectable. Therefore one can perform postselection (which depends on the erasure and syndrome pattern) to filter out noisy preparations. For example, with a simple error model of flipped measurement outcomes with rate $p$, the postselected logical error rate can be made to be of order $p + O(p^2)$. If the error rate $p$ is sufficiently small then the overhead for postselection can also be small~\cite{li2015magic, bombin2022fault}.

\begin{figure*}
	\centering
	\includegraphics[width=0.85\linewidth]{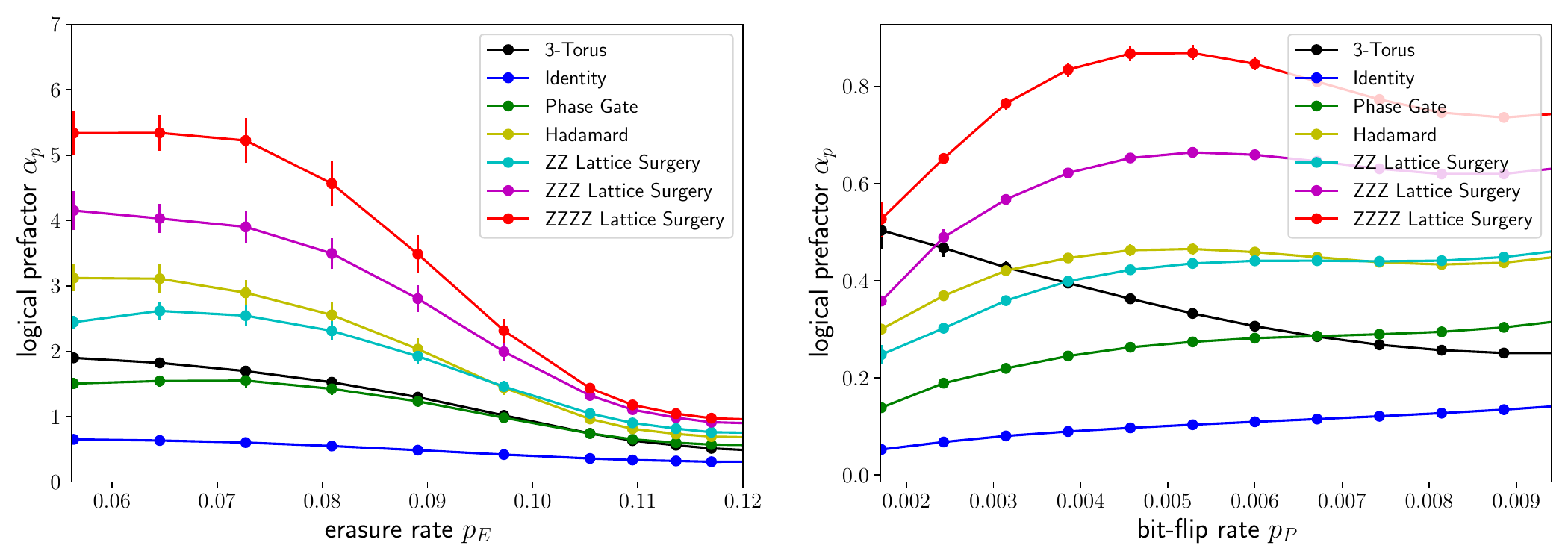}
	\caption{Estimated fits of the logical prefactor $\alpha_p$ to the logical block error rate $P_{p}(d) = \alpha_p e^{-\beta_p d}$ under IID erasure and IID bit-flip noise models for the identity, 3-torus, phase, Hadamard, and multi-qubit lattice surgery blocks.}
	\label{figLogicalPrefactor}
\end{figure*}

\begin{figure*}
	\centering
	\includegraphics[width=0.24\linewidth]{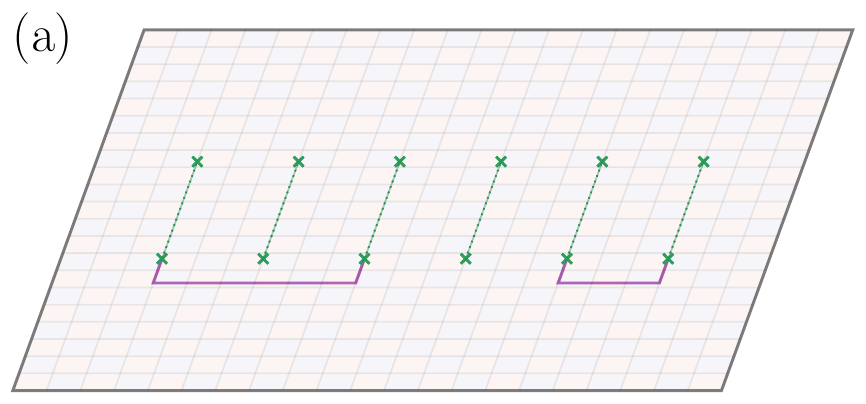} 
	\includegraphics[width=0.24\linewidth]{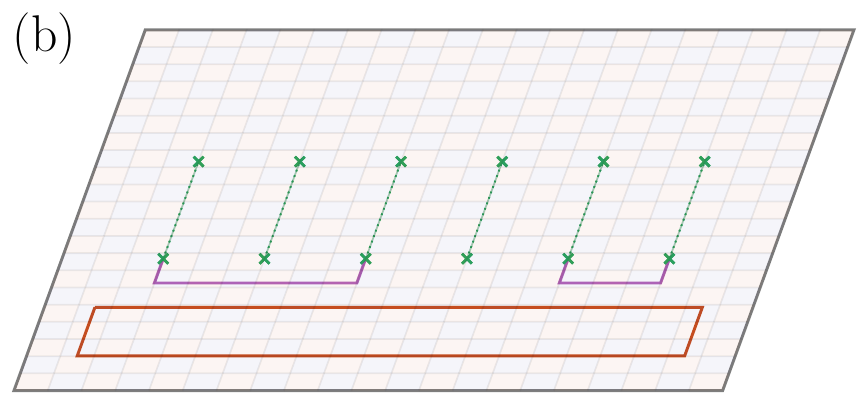} 
	\includegraphics[width=0.24\linewidth]{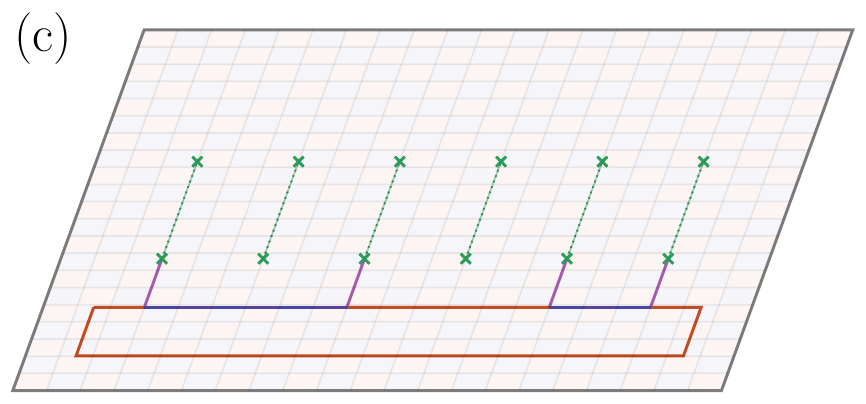}  
	\includegraphics[width=0.24\linewidth]{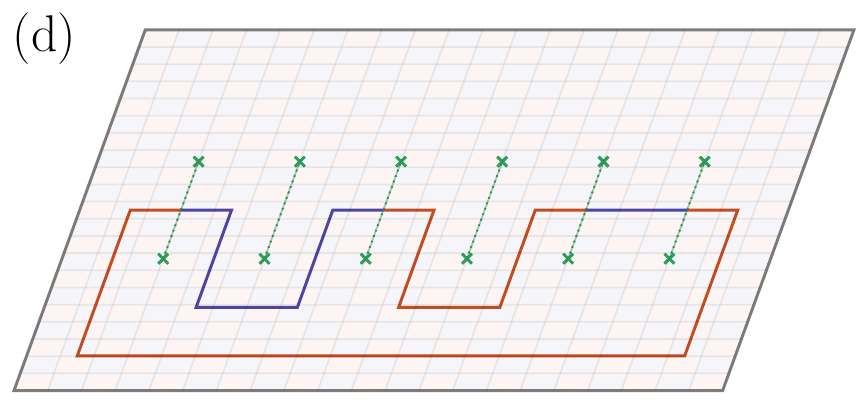}
	\caption{All Pauli logical operators for twist encoded qubits are traceable. (a) An arbitrary Pauli operator can be expressed as a product of composite primal-dual (Pauli-$Y$) strings connecting pairs of twists, as represented by the purple strings. (b) Multiply this logical by a trivial primal or dual loop (i.e., a stabilizer). (c) The results in a new string net configuration, involving primal, dual and composite primal-dual strings in general. (d) This string net can be further resolved to a new traceable string net consisting of only primal and dual strings enclosing some number of twists.}
	\label{figLogicalTraceability}
\end{figure*}

\begin{figure*}
	\centering
	\includegraphics[width=0.25\linewidth]{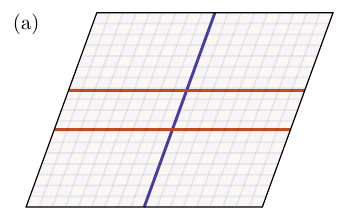} \quad
	\includegraphics[width=0.25\linewidth]{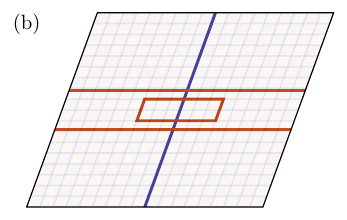} \quad
	\includegraphics[width=0.25\linewidth]{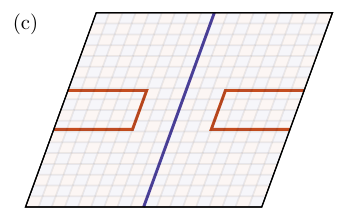}
	\caption{Commuting traceable operators can be resolved to only have intersections on strings of the same type. For simplicity we consider a setting absent of domain walls---the general case holds using the same argument up to local string relabellings. Label the two traceable operators $P$ and $Q$. (a) In a region, $P$ is depicted on the left by a single primal (blue) string, while $Q$ is a pair of dual (red) strings. For $P$ and $Q$ to commute, the primal strings of $P$ must overlap an even number of times with the dual strings of $Q$. (b) Consider pairing up the primal and dual intersections between $P$ and $Q$. For each intersection we can multiply by a trivial dual loop (i.e., a stabilizer) in the neighbourhood of the intersection. (c) This loop resolves the crossings, and so we are left only with intersections of the same type.}
	\label{figLogicalTraceableCommuting}
\end{figure*}

\begin{figure}
	\centering
	\includegraphics[width=0.95\linewidth]{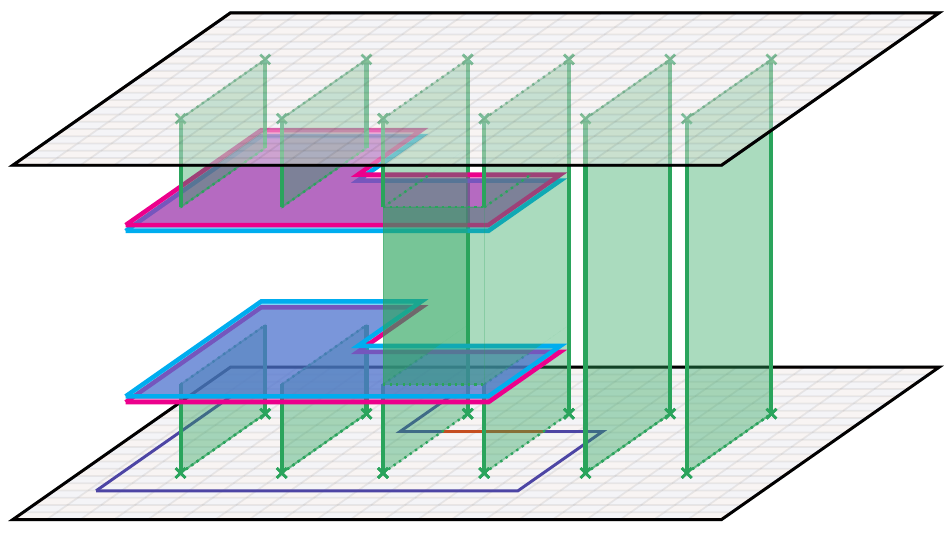} 
	\caption{Since a general Pauli measurement involves an even number of twists, we can always find a valid configuration of domain walls. As shown, we can pair twists and configure domain walls such that their boundary is the desired twists. This portal configuration is to measure the traceable operator depicted at the bottom.}
	\label{figTeleportedTwistResolvedDomainWall}
\end{figure}

\subsection{Toric code spider networks}\label{appSpiderNetworks}

In this section we present in more detail the concatenation schemes discussed in Sec.~\ref{secConcatenation}. In particular, in the top panels of Fig.~\ref{figConvertingCircuitToSpider} we demonstrate how to convert the Clifford circuit in Fig.~\ref{figToricCodeSpiderNetwork} to a network of toric code spiders. In the bottom panels of Fig.~\ref{figConvertingCircuitToSpider} we present the logical block for the spider network that measures stabilizers of the surface code concatenated with itself.

\subsection{FBQC instrument networks as gauge fixing on a subsystem code}\label{secFBQCsubsystem}
We remark that fusion-based quantum computation in this setting can be understood as gauge fixing on a subsystem code. This observation has previously been identified in the setting of MBQC~\cite{brown2020universal}, and we extend the discussion  to include FBQC in this section.

Subsystem (stabilizer) codes are a generalization of stabilizer codes, where some code  degrees  of freedom  are not used to encode information, and are referred to as gauge qubits. A  subsystem code~\cite{poulin2005stabilizer,bacon2006operator} is defined by a subgroup of the Pauli group $\mathcal{G}\subseteq \mathcal{P}_n$, which is not necessarily Abelian, known as the gauge group. For a given gauge group, stabilizers of the subsystem code are given by elements $\mathcal{Z}_{\mathcal{G}}(\mathcal{G})$, logical degrees of freedom are described by the group $\mathcal{Z}_{\mathcal{P}_n}(\mathcal{G})$, while $\mathcal{G}$ describes operations on the gauge qubits. In particular, we refer to an element of $\mathcal{Z}_{\mathcal{P}_n}(\mathcal{G})$  as a \textit{bare logical operator}---such operators act nontrivially only on logical degrees of freedom. One may multiply any bare logical operator by elements of $\mathcal{G}$ to obtain an operator known as a  \textit{dressed logical operator}, which has equivalent action on logical degrees of freedom.

We now express a FBQC instrument network in terms of a subsystem code. Namely, in FBQC, we may construct a gauge group $\mathcal{G}=\resstab\cup \measgp$. We can understand a fault-tolerant instrument in terms of \textit{gauge fixing}~\cite{paetznick2013universal}, whereby we start in one particular gauge of the subsystem code (i.e., in a joint eigenstate of some subset of $\mathcal{G}$) and project into a different gauge. In particular, the computation begins by starting in the gauge defined by $\resstab$, and performing measurements to fix onto the gauge defined by $\measgp$. The computation can be understood as proceeding by examining how bare logical operators are transformed. In terms of the subsystem code, the  bare logical operators are precisely  the logical membrane operators that we have previously studied.

\subsection{Decoding}\label{secDecoding}

In this section we briefly describe how to decode in the FBQC setting.

In order to extract useful logical data from the observed measurement data we must use a decoder. The decoder's job is to take as input the (potentially incomplete) measurement outcomes, and produce a recovery operation that is consistent with the observed syndrome. In particular, let $\trivundterror$ be the set of trivial undetectable errors (i.e., errors that commute with all check operators and membranes). For a Pauli error $E\in\mathcal{P}$, let $\sigma(E)$ be the syndrome (the outcomes of a generating set of checks). The decoder takes the syndrome and produces a recovery operator $R\in \mathcal{P}$ with the same syndrome. The decoding is successful whenever $ER \in \trivundterror$. In this case we necessarily have $[ER, M^{\alpha,\beta}] = 0$ for all logical membranes $M^{\alpha,\beta}$, implying that the logical Pauli frame observed is identical to the case without errors. 

The decoding problem is naturally expressed on the syndrome graph; we can immediately make use of many standard decoders. In particular, on the syndrome graph, the syndrome takes the form $\sigma(E) = \partial(\text{Supp}(E))$, where $\text{Supp}(E)$ is the set of edges on the syndrome graph that correspond to flipped (fusion) measurement outcomes and $\partial$ is the usual mod-2 boundary operator on the graph. The decoder must return a recovery operator $R\in \mathcal{P}$ such that $\sigma(E) = \sigma(R)$. MWPM~\cite{dennis2002topological,kolmogorov2009blossom} and UF~\cite{delfosse2017almost} are two popular decoding methods that produce a low-weight description of the observed syndrome. Such algorithms can be applied to the primal and dual syndrome graphs separately, where they match pairs of primal-type excitations and separately the dual-type excitations.

\subsection{Threshold analysis}\label{secThreshAnalysis}
We simulate two types of IID noise models; an erasure and a bit-flip model. In the context of FBQC, such errors arise from (i) photon loss and fusion failure, and (ii) Pauli errors on the qubits on resource states. To estimate the logical error rate on each 3D block at physical error rate $p$, we fix a block distance and perform many Monte Carlo trials. Each trial consists of generating a random sample of erasures or bit-flip errors on each edge of the corresponding 3D syndrome graph with some rate $p$, determining the syndrome, decoding based on the syndrome, and finally, checking if the decoding is successful. In particular, decoding is successful whenever the combined effect of error and recovery leads to the correct measurement outcome of the logical membrane. Note that different blocks have different numbers (and shapes) of logical membranes, and we declare success if and only if no logical errors occur (i.e., all membranes are correct). On the syndrome graph, an error sample results in a chain of flipped edges $E$. The recovery consists of a chain of flipped edges $R$ with the same boundary as $E$. The logical membrane is represented as a subset (a mask) of syndrome graph edges $M$, and to determine correct decoding, we only need to check if $|E\oplus R \cap M| \cong 0 \mod 2$. In other words, verify that the error and recovery intersects the membrane an even number of times. An example of a syndrome graph and logical mask is given in Fig.~\ref{figLatticeSurgerySyndromeGraph}.

To estimate the threshold, we find estimates of the logical error rate for different block distances $L\in\{8,14,20,26\}$ and for a range of physical error rates. For each block, we fit the logical error rate for each distance to the cumulative distribution function of the (rescaled and shifted) beta distribution. The threshold is estimated as the $p$ value for which the logical error rate curves of different distances intersect such that it is invariant under increasing distances. Figs.~\ref{figThresholdsEr} and \ref{figThresholdsPa} show the logical error rate fits for each logical block, and the crossing point at which the threshold is identified. Error bars for each data point are given by the standard error of the mean for the binomial distribution, from which we can identify error bars for the threshold crossing.

\subsection{Logical decay fits}\label{appLogicalPrefactor}
Here we outline how we obtain fitting parameters $\alpha_p$ and $\beta_p$ for the logical error rate $P_{p}(d)$ as a function of the block distance $d$, according to
\begin{equation}
P_{p}(d) = \alpha_p e^{-\beta_p d}.
\end{equation}
For a given physical error rate $p$, we estimate the logical error rate $P_{p}(d)$ for a variety of block distances, by running no less than $10^9$ decoder trials. Each estimate carries an error bar given by the standard error of the mean for the binomial distribution. We then perform an ordinary least-squares regression to the logarithm of the estimated logical error rate to directly infer $\alpha_p$ and $\beta_p$. The error bars for the estimates for $\alpha_p$ and $\beta_p$ are given by the heteroskedasticity robust standard errors.
Fig.~\ref{figNumericsDecay} contains numerical estimates of the logical decay $\beta_p$, while Fig.~\ref{figLogicalPrefactor} contains the numerical estimates of the logical prefactor $\alpha_p$.

\subsection{Twist encoding traceability}\label{appTwistTraceability}

In this section we show that all Pauli logical operators in the twist encoding are traceable. The argument is presented in Fig.~\ref{figLogicalTraceability}, where we demonstrate how to start with a general Pauli logical operator and transform it using stabilizer equivalences into traceable form. In Fig.~\ref{figLogicalTraceableCommuting} we show that commuting traceable operators can be made to have intersection only on strings of the same type. Finally, we remark that a general portal configuration to measure a Pauli operator always admits a compatible domain wall configuration. This is because the measurement of a Pauli logical operator always results in the measurement of an even number of twists. We present an illustrative example in Fig.~\ref{figTeleportedTwistResolvedDomainWall} of how such a domain wall can be found with the correct boundary conditions.

\subsection{Port  boundary conditions and block decoding simulations}\label{appPortBoundaryConditionsAndDecodingSims}

In order to simulate and calculate \emph{fault distance} of a modular component within a larger logical network a pragmatic boundary condition must be set, which allows evaluating the fault-tolerant performance of different logical blocks in isolation.
This choice is documented here as it plays a crucial role in specifying the numerical simulations that are being performed.

For the purpose of evaluating the fault-tolerant performance of a block and evaluating its \emph{fault distance}, each port is treated as though ideal stabilizer measurements had been performed on the stabilizer subgroup local to said port.
There is no additional noise attributed to the measurement outcomes, other than the noise already present in the qubits themselves.
The ideal nature of such stabilizer measurements makes it impractical in the context of a real-world quantum computer, since such operations would generally be noisy in practice. 
However, the qubits on which the stabilizer measurements are applied are themselves still subject to the underlying noise model.

\begin{figure}[h!]
	\centering
	\includegraphics[width=0.98\linewidth]{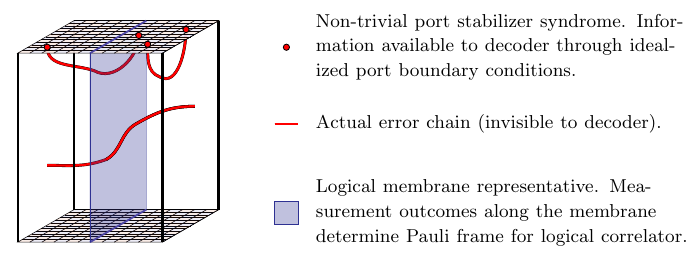} 
	\caption{Without port stabilizers, it would not be possible to associate a sensible fault distance to independent logical blocks.
	For any choice of the logical correlator, there would be low weight fault configurations which result in undetectable errors on the logical correlator.
	Incorporating stabilizer outcomes local to the ports is an idealization which allows us to evaluate sensible fault distance and fault-tolerance properties for the proposed logical blocks.
	}
	\label{figDecodingWithPortStabilizerBC}
\end{figure}

\begin{figure}
	\centering
	\includegraphics[width=0.9\linewidth]{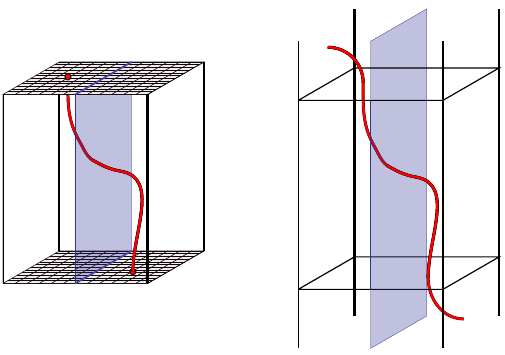} 
	\caption{The figure shows how a fault configuration which is undetectable in the context of a composite protocol (right), can become detectable in the context of the idealized boundary condition described by stabilizer measurements when the complete protocol is partitioned into logical blocks (left).
	}
	\label{figBCcapturedErrors}
\end{figure}

In the setting of topological fault tolerance, the stabilizer outcomes being extracted correspond to the geometrically local stabilizer generators of the surface code.
The situation studied corresponds to a well-defined quantum instrument inserted at said ports.
In terms of the blocks already considered, the chosen boundary condition corresponds to attaching a noiseless injection block (from Fig.~\ref{figFBQCMagicStatePreparation}) at every surface code port, with the noise rate being artificially set to zero for such blocks and the injection qubit left unmeasured (in other words, a noiseless encoding and unencoding isometry). These boundary conditions and their impact are displayed in Figs.~\ref{figDecodingWithPortStabilizerBC} and \ref{figBCcapturedErrors}.

The blocks we use in this article are chosen such that they individually have fault-tolerance properties with respect to the idealized port boundary condition, but also such that fault tolerance is preserved under arbitrary combinations of such blocks.
Whereas this statement is made as a claim in this article, a set of sufficient conditions satisfied by this set of blocks is presented in Ref.~\cite{modularDecoding} in the context of \textit{modular decoding}.

\begin{thebibliography}{133}%
\makeatletter
\providecommand \@ifxundefined [1]{%
 \@ifx{#1\undefined}
}%
\providecommand \@ifnum [1]{%
 \ifnum #1\expandafter \@firstoftwo
 \else \expandafter \@secondoftwo
 \fi
}%
\providecommand \@ifx [1]{%
 \ifx #1\expandafter \@firstoftwo
 \else \expandafter \@secondoftwo
 \fi
}%
\providecommand \natexlab [1]{#1}%
\providecommand \enquote  [1]{``#1''}%
\providecommand \bibnamefont  [1]{#1}%
\providecommand \bibfnamefont [1]{#1}%
\providecommand \citenamefont [1]{#1}%
\providecommand \href@noop [0]{\@secondoftwo}%
\providecommand \href [0]{\begingroup \@sanitize@url \@href}%
\providecommand \@href[1]{\@@startlink{#1}\@@href}%
\providecommand \@@href[1]{\endgroup#1\@@endlink}%
\providecommand \@sanitize@url [0]{\catcode `\\12\catcode `\$12\catcode
  `\&12\catcode `\#12\catcode `\^12\catcode `\_12\catcode `\%12\relax}%
\providecommand \@@startlink[1]{}%
\providecommand \@@endlink[0]{}%
\providecommand \url  [0]{\begingroup\@sanitize@url \@url }%
\providecommand \@url [1]{\endgroup\@href {#1}{\urlprefix }}%
\providecommand \urlprefix  [0]{URL }%
\providecommand \Eprint [0]{\href }%
\providecommand \doibase [0]{http://dx.doi.org/}%
\providecommand \selectlanguage [0]{\@gobble}%
\providecommand \bibinfo  [0]{\@secondoftwo}%
\providecommand \bibfield  [0]{\@secondoftwo}%
\providecommand \translation [1]{[#1]}%
\providecommand \BibitemOpen [0]{}%
\providecommand \bibitemStop [0]{}%
\providecommand \bibitemNoStop [0]{.\EOS\space}%
\providecommand \EOS [0]{\spacefactor3000\relax}%
\providecommand \BibitemShut  [1]{\csname bibitem#1\endcsname}%
\let\auto@bib@innerbib\@empty
\bibitem [{\citenamefont {Kitaev}(2003)}]{kitaev2003fault}%
  \BibitemOpen
  \bibfield  {author} {\bibinfo {author} {\bibfnamefont {A.~Y.}\ \bibnamefont
  {Kitaev}},\ }\bibfield  {title} {\emph {\bibinfo {title} {Fault-tolerant
  quantum computation by anyons},\ }}\href@noop {} {\bibfield  {journal}
  {\bibinfo  {journal} {Annals of Physics}\ }\textbf {\bibinfo {volume}
  {303}},\ \bibinfo {pages} {2} (\bibinfo {year} {2003})}\BibitemShut {NoStop}%
\bibitem [{\citenamefont {Kitaev}(1997)}]{kitaev1997quantum}%
  \BibitemOpen
  \bibfield  {author} {\bibinfo {author} {\bibfnamefont {A.~Y.}\ \bibnamefont
  {Kitaev}},\ }in\ \href@noop {} {\emph {\bibinfo {booktitle} {Quantum
  Communication, Computing, and Measurement}}}\ (\bibinfo  {publisher}
  {Springer},\ \bibinfo {year} {1997})\ pp.\ \bibinfo {pages}
  {181--188}\BibitemShut {NoStop}%
\bibitem [{\citenamefont {Dennis}\ \emph {et~al.}(2002)\citenamefont {Dennis},
  \citenamefont {Kitaev}, \citenamefont {Landahl},\ and\ \citenamefont
  {Preskill}}]{dennis2002topological}%
  \BibitemOpen
  \bibfield  {author} {\bibinfo {author} {\bibfnamefont {E.}~\bibnamefont
  {Dennis}}, \bibinfo {author} {\bibfnamefont {A.}~\bibnamefont {Kitaev}},
  \bibinfo {author} {\bibfnamefont {A.}~\bibnamefont {Landahl}}, \ and\
  \bibinfo {author} {\bibfnamefont {J.}~\bibnamefont {Preskill}},\ }\bibfield
  {title} {\emph {\bibinfo {title} {Topological quantum memory},\ }}\href@noop
  {} {\bibfield  {journal} {\bibinfo  {journal} {Journal of Mathematical
  Physics}\ }\textbf {\bibinfo {volume} {43}},\ \bibinfo {pages} {4452}
  (\bibinfo {year} {2002})}\BibitemShut {NoStop}%
\bibitem [{\citenamefont {Raussendorf}\ and\ \citenamefont
  {Harrington}(2007)}]{raussendorf2007fault}%
  \BibitemOpen
  \bibfield  {author} {\bibinfo {author} {\bibfnamefont {R.}~\bibnamefont
  {Raussendorf}}\ and\ \bibinfo {author} {\bibfnamefont {J.}~\bibnamefont
  {Harrington}},\ }\bibfield  {title} {\emph {\bibinfo {title} {Fault-tolerant
  quantum computation with high threshold in two dimensions},\ }}\href@noop {}
  {\bibfield  {journal} {\bibinfo  {journal} {Physical review letters}\
  }\textbf {\bibinfo {volume} {98}},\ \bibinfo {pages} {190504} (\bibinfo
  {year} {2007})}\BibitemShut {NoStop}%
\bibitem [{\citenamefont {Raussendorf}\ \emph {et~al.}(2007)\citenamefont
  {Raussendorf}, \citenamefont {Harrington},\ and\ \citenamefont
  {Goyal}}]{raussendorf2007topological}%
  \BibitemOpen
  \bibfield  {author} {\bibinfo {author} {\bibfnamefont {R.}~\bibnamefont
  {Raussendorf}}, \bibinfo {author} {\bibfnamefont {J.}~\bibnamefont
  {Harrington}}, \ and\ \bibinfo {author} {\bibfnamefont {K.}~\bibnamefont
  {Goyal}},\ }\bibfield  {title} {\emph {\bibinfo {title} {Topological
  fault-tolerance in cluster state quantum computation},\ }}\href@noop {}
  {\bibfield  {journal} {\bibinfo  {journal} {New Journal of Physics}\ }\textbf
  {\bibinfo {volume} {9}},\ \bibinfo {pages} {199} (\bibinfo {year}
  {2007})}\BibitemShut {NoStop}%
\bibitem [{\citenamefont {Bolt}\ \emph {et~al.}(2016)\citenamefont {Bolt},
  \citenamefont {Duclos-Cianci}, \citenamefont {Poulin},\ and\ \citenamefont
  {Stace}}]{bolt2016foliated}%
  \BibitemOpen
  \bibfield  {author} {\bibinfo {author} {\bibfnamefont {A.}~\bibnamefont
  {Bolt}}, \bibinfo {author} {\bibfnamefont {G.}~\bibnamefont {Duclos-Cianci}},
  \bibinfo {author} {\bibfnamefont {D.}~\bibnamefont {Poulin}}, \ and\ \bibinfo
  {author} {\bibfnamefont {T.}~\bibnamefont {Stace}},\ }\bibfield  {title}
  {\emph {\bibinfo {title} {Foliated quantum error-correcting codes},\
  }}\href@noop {} {\bibfield  {journal} {\bibinfo  {journal} {Physical review
  letters}\ }\textbf {\bibinfo {volume} {117}},\ \bibinfo {pages} {070501}
  (\bibinfo {year} {2016})}\BibitemShut {NoStop}%
\bibitem [{\citenamefont {Nickerson}\ and\ \citenamefont
  {Bomb{\'\i}n}(2018)}]{nickerson2018measurement}%
  \BibitemOpen
  \bibfield  {author} {\bibinfo {author} {\bibfnamefont {N.}~\bibnamefont
  {Nickerson}}\ and\ \bibinfo {author} {\bibfnamefont {H.}~\bibnamefont
  {Bomb{\'\i}n}},\ }\bibfield  {title} {\emph {\bibinfo {title} {Measurement
  based fault tolerance beyond foliation},\ }}\href@noop {} {\bibfield
  {journal} {\bibinfo  {journal} {arXiv preprint arXiv:1810.09621}\ } (\bibinfo
  {year} {2018})}\BibitemShut {NoStop}%
\bibitem [{\citenamefont {Brown}\ and\ \citenamefont
  {Roberts}(2020)}]{brown2020universal}%
  \BibitemOpen
  \bibfield  {author} {\bibinfo {author} {\bibfnamefont {B.~J.}\ \bibnamefont
  {Brown}}\ and\ \bibinfo {author} {\bibfnamefont {S.}~\bibnamefont
  {Roberts}},\ }\bibfield  {title} {\emph {\bibinfo {title} {Universal
  fault-tolerant measurement-based quantum computation},\ }}\href@noop {}
  {\bibfield  {journal} {\bibinfo  {journal} {Physical Review Research}\
  }\textbf {\bibinfo {volume} {2}},\ \bibinfo {pages} {033305} (\bibinfo {year}
  {2020})}\BibitemShut {NoStop}%
\bibitem [{\citenamefont {Wang}\ \emph {et~al.}(2003)\citenamefont {Wang},
  \citenamefont {Harrington},\ and\ \citenamefont
  {Preskill}}]{wang2003confinement}%
  \BibitemOpen
  \bibfield  {author} {\bibinfo {author} {\bibfnamefont {C.}~\bibnamefont
  {Wang}}, \bibinfo {author} {\bibfnamefont {J.}~\bibnamefont {Harrington}}, \
  and\ \bibinfo {author} {\bibfnamefont {J.}~\bibnamefont {Preskill}},\
  }\bibfield  {title} {\emph {\bibinfo {title} {Confinement-higgs transition in
  a disordered gauge theory and the accuracy threshold for quantum memory},\
  }}\href@noop {} {\bibfield  {journal} {\bibinfo  {journal} {Annals of
  Physics}\ }\textbf {\bibinfo {volume} {303}},\ \bibinfo {pages} {31}
  (\bibinfo {year} {2003})}\BibitemShut {NoStop}%
\bibitem [{\citenamefont {Stace}\ \emph {et~al.}(2009)\citenamefont {Stace},
  \citenamefont {Barrett},\ and\ \citenamefont
  {Doherty}}]{stace2009thresholds}%
  \BibitemOpen
  \bibfield  {author} {\bibinfo {author} {\bibfnamefont {T.~M.}\ \bibnamefont
  {Stace}}, \bibinfo {author} {\bibfnamefont {S.~D.}\ \bibnamefont {Barrett}},
  \ and\ \bibinfo {author} {\bibfnamefont {A.~C.}\ \bibnamefont {Doherty}},\
  }\bibfield  {title} {\emph {\bibinfo {title} {Thresholds for topological
  codes in the presence of loss},\ }}\href@noop {} {\bibfield  {journal}
  {\bibinfo  {journal} {Physical review letters}\ }\textbf {\bibinfo {volume}
  {102}},\ \bibinfo {pages} {200501} (\bibinfo {year} {2009})}\BibitemShut
  {NoStop}%
\bibitem [{\citenamefont {Duclos-Cianci}\ and\ \citenamefont
  {Poulin}(2010)}]{duclos2010fast}%
  \BibitemOpen
  \bibfield  {author} {\bibinfo {author} {\bibfnamefont {G.}~\bibnamefont
  {Duclos-Cianci}}\ and\ \bibinfo {author} {\bibfnamefont {D.}~\bibnamefont
  {Poulin}},\ }\bibfield  {title} {\emph {\bibinfo {title} {Fast decoders for
  topological quantum codes},\ }}\href@noop {} {\bibfield  {journal} {\bibinfo
  {journal} {Physical review letters}\ }\textbf {\bibinfo {volume} {104}},\
  \bibinfo {pages} {050504} (\bibinfo {year} {2010})}\BibitemShut {NoStop}%
\bibitem [{\citenamefont {Bombin}\ \emph
  {et~al.}(2012{\natexlab{a}})\citenamefont {Bombin}, \citenamefont {Andrist},
  \citenamefont {Ohzeki}, \citenamefont {Katzgraber},\ and\ \citenamefont
  {Mart{\'\i}n-Delgado}}]{bombin2012strong}%
  \BibitemOpen
  \bibfield  {author} {\bibinfo {author} {\bibfnamefont {H.}~\bibnamefont
  {Bombin}}, \bibinfo {author} {\bibfnamefont {R.~S.}\ \bibnamefont {Andrist}},
  \bibinfo {author} {\bibfnamefont {M.}~\bibnamefont {Ohzeki}}, \bibinfo
  {author} {\bibfnamefont {H.~G.}\ \bibnamefont {Katzgraber}}, \ and\ \bibinfo
  {author} {\bibfnamefont {M.~A.}\ \bibnamefont {Mart{\'\i}n-Delgado}},\
  }\bibfield  {title} {\emph {\bibinfo {title} {Strong resilience of
  topological codes to depolarization},\ }}\href@noop {} {\bibfield  {journal}
  {\bibinfo  {journal} {Physical Review X}\ }\textbf {\bibinfo {volume} {2}},\
  \bibinfo {pages} {021004} (\bibinfo {year} {2012}{\natexlab{a}})}\BibitemShut
  {NoStop}%
\bibitem [{\citenamefont {Fowler}\ \emph {et~al.}(2012)\citenamefont {Fowler},
  \citenamefont {Mariantoni}, \citenamefont {Martinis},\ and\ \citenamefont
  {Cleland}}]{fowler2012surface}%
  \BibitemOpen
  \bibfield  {author} {\bibinfo {author} {\bibfnamefont {A.~G.}\ \bibnamefont
  {Fowler}}, \bibinfo {author} {\bibfnamefont {M.}~\bibnamefont {Mariantoni}},
  \bibinfo {author} {\bibfnamefont {J.~M.}\ \bibnamefont {Martinis}}, \ and\
  \bibinfo {author} {\bibfnamefont {A.~N.}\ \bibnamefont {Cleland}},\
  }\bibfield  {title} {\emph {\bibinfo {title} {Surface codes: Towards
  practical large-scale quantum computation},\ }}\href@noop {} {\bibfield
  {journal} {\bibinfo  {journal} {Physical Review A}\ }\textbf {\bibinfo
  {volume} {86}},\ \bibinfo {pages} {032324} (\bibinfo {year}
  {2012})}\BibitemShut {NoStop}%
\bibitem [{\citenamefont {Watson}\ and\ \citenamefont
  {Barrett}(2014)}]{watson2014logical}%
  \BibitemOpen
  \bibfield  {author} {\bibinfo {author} {\bibfnamefont {F.~H.}\ \bibnamefont
  {Watson}}\ and\ \bibinfo {author} {\bibfnamefont {S.~D.}\ \bibnamefont
  {Barrett}},\ }\bibfield  {title} {\emph {\bibinfo {title} {Logical error rate
  scaling of the toric code},\ }}\href@noop {} {\bibfield  {journal} {\bibinfo
  {journal} {New Journal of Physics}\ }\textbf {\bibinfo {volume} {16}},\
  \bibinfo {pages} {093045} (\bibinfo {year} {2014})}\BibitemShut {NoStop}%
\bibitem [{\citenamefont {Bravyi}\ \emph {et~al.}(2014)\citenamefont {Bravyi},
  \citenamefont {Suchara},\ and\ \citenamefont {Vargo}}]{bravyi2014efficient}%
  \BibitemOpen
  \bibfield  {author} {\bibinfo {author} {\bibfnamefont {S.}~\bibnamefont
  {Bravyi}}, \bibinfo {author} {\bibfnamefont {M.}~\bibnamefont {Suchara}}, \
  and\ \bibinfo {author} {\bibfnamefont {A.}~\bibnamefont {Vargo}},\ }\bibfield
   {title} {\emph {\bibinfo {title} {Efficient algorithms for maximum
  likelihood decoding in the surface code},\ }}\href@noop {} {\bibfield
  {journal} {\bibinfo  {journal} {Physical Review A}\ }\textbf {\bibinfo
  {volume} {90}},\ \bibinfo {pages} {032326} (\bibinfo {year}
  {2014})}\BibitemShut {NoStop}%
\bibitem [{\citenamefont {Darmawan}\ and\ \citenamefont
  {Poulin}(2017)}]{darmawan2017tensor}%
  \BibitemOpen
  \bibfield  {author} {\bibinfo {author} {\bibfnamefont {A.~S.}\ \bibnamefont
  {Darmawan}}\ and\ \bibinfo {author} {\bibfnamefont {D.}~\bibnamefont
  {Poulin}},\ }\bibfield  {title} {\emph {\bibinfo {title} {Tensor-network
  simulations of the surface code under realistic noise},\ }}\href@noop {}
  {\bibfield  {journal} {\bibinfo  {journal} {Physical review letters}\
  }\textbf {\bibinfo {volume} {119}},\ \bibinfo {pages} {040502} (\bibinfo
  {year} {2017})}\BibitemShut {NoStop}%
\bibitem [{\citenamefont {Bomb{\'\i}n}\ and\ \citenamefont
  {Martin-Delgado}(2009)}]{bombin2009quantum}%
  \BibitemOpen
  \bibfield  {author} {\bibinfo {author} {\bibfnamefont {H.}~\bibnamefont
  {Bomb{\'\i}n}}\ and\ \bibinfo {author} {\bibfnamefont {M.~A.}\ \bibnamefont
  {Martin-Delgado}},\ }\bibfield  {title} {\emph {\bibinfo {title} {Quantum
  measurements and gates by code deformation},\ }}\href@noop {} {\bibfield
  {journal} {\bibinfo  {journal} {Journal of Physics A: Mathematical and
  Theoretical}\ }\textbf {\bibinfo {volume} {42}},\ \bibinfo {pages} {095302}
  (\bibinfo {year} {2009})}\BibitemShut {NoStop}%
\bibitem [{\citenamefont {Bomb{\'\i}n}(2010)}]{bombin2010topological}%
  \BibitemOpen
  \bibfield  {author} {\bibinfo {author} {\bibfnamefont {H.}~\bibnamefont
  {Bomb{\'\i}n}},\ }\bibfield  {title} {\emph {\bibinfo {title} {Topological
  order with a twist: Ising anyons from an abelian model},\ }}\href@noop {}
  {\bibfield  {journal} {\bibinfo  {journal} {Physical review letters}\
  }\textbf {\bibinfo {volume} {105}},\ \bibinfo {pages} {030403} (\bibinfo
  {year} {2010})}\BibitemShut {NoStop}%
\bibitem [{\citenamefont {Landahl}\ \emph {et~al.}(2011)\citenamefont
  {Landahl}, \citenamefont {Anderson},\ and\ \citenamefont
  {Rice}}]{landahl2011fault}%
  \BibitemOpen
  \bibfield  {author} {\bibinfo {author} {\bibfnamefont {A.~J.}\ \bibnamefont
  {Landahl}}, \bibinfo {author} {\bibfnamefont {J.~T.}\ \bibnamefont
  {Anderson}}, \ and\ \bibinfo {author} {\bibfnamefont {P.~R.}\ \bibnamefont
  {Rice}},\ }\bibfield  {title} {\emph {\bibinfo {title} {Fault-tolerant
  quantum computing with color codes},\ }}\href
  {https://arxiv.org/abs/1108.5738} {\bibfield  {journal} {\bibinfo  {journal}
  {arXiv preprint arXiv:1108.5738}\ } (\bibinfo {year} {2011})}\BibitemShut
  {NoStop}%
\bibitem [{\citenamefont {Horsman}\ \emph {et~al.}(2012)\citenamefont
  {Horsman}, \citenamefont {Fowler}, \citenamefont {Devitt},\ and\
  \citenamefont {Van~Meter}}]{horsman2012surface}%
  \BibitemOpen
  \bibfield  {author} {\bibinfo {author} {\bibfnamefont {C.}~\bibnamefont
  {Horsman}}, \bibinfo {author} {\bibfnamefont {A.~G.}\ \bibnamefont {Fowler}},
  \bibinfo {author} {\bibfnamefont {S.}~\bibnamefont {Devitt}}, \ and\ \bibinfo
  {author} {\bibfnamefont {R.}~\bibnamefont {Van~Meter}},\ }\bibfield  {title}
  {\emph {\bibinfo {title} {Surface code quantum computing by lattice
  surgery},\ }}\href@noop {} {\bibfield  {journal} {\bibinfo  {journal} {New
  Journal of Physics}\ }\textbf {\bibinfo {volume} {14}},\ \bibinfo {pages}
  {123011} (\bibinfo {year} {2012})}\BibitemShut {NoStop}%
\bibitem [{\citenamefont {Fowler}(2012)}]{fowler2012time}%
  \BibitemOpen
  \bibfield  {author} {\bibinfo {author} {\bibfnamefont {A.~G.}\ \bibnamefont
  {Fowler}},\ }\bibfield  {title} {\emph {\bibinfo {title} {Time-optimal
  quantum computation},\ }}\href@noop {} {\bibfield  {journal} {\bibinfo
  {journal} {arXiv preprint arXiv:1210.4626}\ } (\bibinfo {year}
  {2012})}\BibitemShut {NoStop}%
\bibitem [{\citenamefont {Barkeshli}\ \emph
  {et~al.}(2013{\natexlab{a}})\citenamefont {Barkeshli}, \citenamefont {Jian},\
  and\ \citenamefont {Qi}}]{barkeshli2013classification}%
  \BibitemOpen
  \bibfield  {author} {\bibinfo {author} {\bibfnamefont {M.}~\bibnamefont
  {Barkeshli}}, \bibinfo {author} {\bibfnamefont {C.-M.}\ \bibnamefont {Jian}},
  \ and\ \bibinfo {author} {\bibfnamefont {X.-L.}\ \bibnamefont {Qi}},\
  }\bibfield  {title} {\emph {\bibinfo {title} {Classification of topological
  defects in abelian topological states},\ }}\href@noop {} {\bibfield
  {journal} {\bibinfo  {journal} {Physical Review B}\ }\textbf {\bibinfo
  {volume} {88}},\ \bibinfo {pages} {241103} (\bibinfo {year}
  {2013}{\natexlab{a}})}\BibitemShut {NoStop}%
\bibitem [{\citenamefont {Barkeshli}\ \emph
  {et~al.}(2013{\natexlab{b}})\citenamefont {Barkeshli}, \citenamefont {Jian},\
  and\ \citenamefont {Qi}}]{barkeshli2013twist}%
  \BibitemOpen
  \bibfield  {author} {\bibinfo {author} {\bibfnamefont {M.}~\bibnamefont
  {Barkeshli}}, \bibinfo {author} {\bibfnamefont {C.-M.}\ \bibnamefont {Jian}},
  \ and\ \bibinfo {author} {\bibfnamefont {X.-L.}\ \bibnamefont {Qi}},\
  }\bibfield  {title} {\emph {\bibinfo {title} {Twist defects and projective
  non-abelian braiding statistics},\ }}\href@noop {} {\bibfield  {journal}
  {\bibinfo  {journal} {Physical Review B}\ }\textbf {\bibinfo {volume} {87}},\
  \bibinfo {pages} {045130} (\bibinfo {year} {2013}{\natexlab{b}})}\BibitemShut
  {NoStop}%
\bibitem [{\citenamefont {Hastings}\ and\ \citenamefont
  {Geller}(2014)}]{hastings2014reduced}%
  \BibitemOpen
  \bibfield  {author} {\bibinfo {author} {\bibfnamefont {M.~B.}\ \bibnamefont
  {Hastings}}\ and\ \bibinfo {author} {\bibfnamefont {A.}~\bibnamefont
  {Geller}},\ }\bibfield  {title} {\emph {\bibinfo {title} {Reduced space-time
  and time costs using dislocation codes and arbitrary ancillas},\ }}\href@noop
  {} {\bibfield  {journal} {\bibinfo  {journal} {arXiv preprint
  arXiv:1408.3379}\ } (\bibinfo {year} {2014})}\BibitemShut {NoStop}%
\bibitem [{\citenamefont {Yoshida}(2015)}]{yoshida2015topological}%
  \BibitemOpen
  \bibfield  {author} {\bibinfo {author} {\bibfnamefont {B.}~\bibnamefont
  {Yoshida}},\ }\bibfield  {title} {\emph {\bibinfo {title} {Topological color
  code and symmetry-protected topological phases},\ }}\href@noop {} {\bibfield
  {journal} {\bibinfo  {journal} {Physical Review B}\ }\textbf {\bibinfo
  {volume} {91}},\ \bibinfo {pages} {245131} (\bibinfo {year}
  {2015})}\BibitemShut {NoStop}%
\bibitem [{\citenamefont {Terhal}(2015)}]{terhal2015quantum}%
  \BibitemOpen
  \bibfield  {author} {\bibinfo {author} {\bibfnamefont {B.~M.}\ \bibnamefont
  {Terhal}},\ }\bibfield  {title} {\emph {\bibinfo {title} {Quantum error
  correction for quantum memories},\ }}\href@noop {} {\bibfield  {journal}
  {\bibinfo  {journal} {Reviews of Modern Physics}\ }\textbf {\bibinfo {volume}
  {87}},\ \bibinfo {pages} {307} (\bibinfo {year} {2015})}\BibitemShut
  {NoStop}%
\bibitem [{\citenamefont {Yoder}\ and\ \citenamefont
  {Kim}(2017)}]{yoder2017surface}%
  \BibitemOpen
  \bibfield  {author} {\bibinfo {author} {\bibfnamefont {T.~J.}\ \bibnamefont
  {Yoder}}\ and\ \bibinfo {author} {\bibfnamefont {I.~H.}\ \bibnamefont
  {Kim}},\ }\bibfield  {title} {\emph {\bibinfo {title} {The surface code with
  a twist},\ }}\href@noop {} {\bibfield  {journal} {\bibinfo  {journal}
  {Quantum}\ }\textbf {\bibinfo {volume} {1}},\ \bibinfo {pages} {2} (\bibinfo
  {year} {2017})}\BibitemShut {NoStop}%
\bibitem [{\citenamefont {Brown}\ \emph {et~al.}(2017)\citenamefont {Brown},
  \citenamefont {Laubscher}, \citenamefont {Kesselring},\ and\ \citenamefont
  {Wootton}}]{brown2017poking}%
  \BibitemOpen
  \bibfield  {author} {\bibinfo {author} {\bibfnamefont {B.~J.}\ \bibnamefont
  {Brown}}, \bibinfo {author} {\bibfnamefont {K.}~\bibnamefont {Laubscher}},
  \bibinfo {author} {\bibfnamefont {M.~S.}\ \bibnamefont {Kesselring}}, \ and\
  \bibinfo {author} {\bibfnamefont {J.~R.}\ \bibnamefont {Wootton}},\
  }\bibfield  {title} {\emph {\bibinfo {title} {Poking holes and cutting
  corners to achieve clifford gates with the surface code},\ }}\href@noop {}
  {\bibfield  {journal} {\bibinfo  {journal} {Physical Review X}\ }\textbf
  {\bibinfo {volume} {7}},\ \bibinfo {pages} {021029} (\bibinfo {year}
  {2017})}\BibitemShut {NoStop}%
\bibitem [{\citenamefont {Yoshida}(2017)}]{yoshida2017gapped}%
  \BibitemOpen
  \bibfield  {author} {\bibinfo {author} {\bibfnamefont {B.}~\bibnamefont
  {Yoshida}},\ }\bibfield  {title} {\emph {\bibinfo {title} {Gapped boundaries,
  group cohomology and fault-tolerant logical gates},\ }}\href@noop {}
  {\bibfield  {journal} {\bibinfo  {journal} {Annals of Physics}\ }\textbf
  {\bibinfo {volume} {377}},\ \bibinfo {pages} {387} (\bibinfo {year}
  {2017})}\BibitemShut {NoStop}%
\bibitem [{\citenamefont {Roberts}\ \emph {et~al.}(2017)\citenamefont
  {Roberts}, \citenamefont {Yoshida}, \citenamefont {Kubica},\ and\
  \citenamefont {Bartlett}}]{roberts2017symmetry}%
  \BibitemOpen
  \bibfield  {author} {\bibinfo {author} {\bibfnamefont {S.}~\bibnamefont
  {Roberts}}, \bibinfo {author} {\bibfnamefont {B.}~\bibnamefont {Yoshida}},
  \bibinfo {author} {\bibfnamefont {A.}~\bibnamefont {Kubica}}, \ and\ \bibinfo
  {author} {\bibfnamefont {S.~D.}\ \bibnamefont {Bartlett}},\ }\bibfield
  {title} {\emph {\bibinfo {title} {Symmetry-protected topological order at
  nonzero temperature},\ }}\href@noop {} {\bibfield  {journal} {\bibinfo
  {journal} {Physical Review A}\ }\textbf {\bibinfo {volume} {96}},\ \bibinfo
  {pages} {022306} (\bibinfo {year} {2017})}\BibitemShut {NoStop}%
\bibitem [{\citenamefont {Bombin}(2018{\natexlab{a}})}]{bombin2018transversal}%
  \BibitemOpen
  \bibfield  {author} {\bibinfo {author} {\bibfnamefont {H.}~\bibnamefont
  {Bombin}},\ }\bibfield  {title} {\emph {\bibinfo {title} {Transversal gates
  and error propagation in 3d topological codes},\ }}\href
  {https://arxiv.org/abs/1810.09575} {\bibfield  {journal} {\bibinfo  {journal}
  {arXiv preprint arXiv:1810.09575}\ } (\bibinfo {year}
  {2018}{\natexlab{a}})}\BibitemShut {NoStop}%
\bibitem [{\citenamefont {Bombin}(2018{\natexlab{b}})}]{bombin20182d}%
  \BibitemOpen
  \bibfield  {author} {\bibinfo {author} {\bibfnamefont {H.}~\bibnamefont
  {Bombin}},\ }\bibfield  {title} {\emph {\bibinfo {title} {2d quantum
  computation with 3d topological codes},\ }}\href
  {https://arxiv.org/abs/1810.09571} {\bibfield  {journal} {\bibinfo  {journal}
  {arXiv preprint arXiv:1810.09571}\ } (\bibinfo {year}
  {2018}{\natexlab{b}})}\BibitemShut {NoStop}%
\bibitem [{\citenamefont {Lavasani}\ and\ \citenamefont
  {Barkeshli}(2018)}]{lavasani2018low}%
  \BibitemOpen
  \bibfield  {author} {\bibinfo {author} {\bibfnamefont {A.}~\bibnamefont
  {Lavasani}}\ and\ \bibinfo {author} {\bibfnamefont {M.}~\bibnamefont
  {Barkeshli}},\ }\bibfield  {title} {\emph {\bibinfo {title} {Low overhead
  clifford gates from joint measurements in surface, color, and hyperbolic
  codes},\ }}\href@noop {} {\bibfield  {journal} {\bibinfo  {journal} {Physical
  Review A}\ }\textbf {\bibinfo {volume} {98}},\ \bibinfo {pages} {052319}
  (\bibinfo {year} {2018})}\BibitemShut {NoStop}%
\bibitem [{\citenamefont {Lavasani}\ \emph {et~al.}(2019)\citenamefont
  {Lavasani}, \citenamefont {Zhu},\ and\ \citenamefont
  {Barkeshli}}]{lavasani2019universal}%
  \BibitemOpen
  \bibfield  {author} {\bibinfo {author} {\bibfnamefont {A.}~\bibnamefont
  {Lavasani}}, \bibinfo {author} {\bibfnamefont {G.}~\bibnamefont {Zhu}}, \
  and\ \bibinfo {author} {\bibfnamefont {M.}~\bibnamefont {Barkeshli}},\
  }\bibfield  {title} {\emph {\bibinfo {title} {Universal logical gates with
  constant overhead: instantaneous dehn twists for hyperbolic quantum codes},\
  }}\href@noop {} {\bibfield  {journal} {\bibinfo  {journal} {arXiv preprint
  arXiv:1901.11029}\ } (\bibinfo {year} {2019})}\BibitemShut {NoStop}%
\bibitem [{\citenamefont {Webster}\ and\ \citenamefont
  {Bartlett}(2020)}]{webster2020fault}%
  \BibitemOpen
  \bibfield  {author} {\bibinfo {author} {\bibfnamefont {P.}~\bibnamefont
  {Webster}}\ and\ \bibinfo {author} {\bibfnamefont {S.~D.}\ \bibnamefont
  {Bartlett}},\ }\bibfield  {title} {\emph {\bibinfo {title} {Fault-tolerant
  quantum gates with defects in topological stabilizer codes},\ }}\href@noop {}
  {\bibfield  {journal} {\bibinfo  {journal} {Physical Review A}\ }\textbf
  {\bibinfo {volume} {102}},\ \bibinfo {pages} {022403} (\bibinfo {year}
  {2020})}\BibitemShut {NoStop}%
\bibitem [{\citenamefont {Hanks}\ \emph {et~al.}(2020)\citenamefont {Hanks},
  \citenamefont {Estarellas}, \citenamefont {Munro},\ and\ \citenamefont
  {Nemoto}}]{hanks2020effective}%
  \BibitemOpen
  \bibfield  {author} {\bibinfo {author} {\bibfnamefont {M.}~\bibnamefont
  {Hanks}}, \bibinfo {author} {\bibfnamefont {M.~P.}\ \bibnamefont
  {Estarellas}}, \bibinfo {author} {\bibfnamefont {W.~J.}\ \bibnamefont
  {Munro}}, \ and\ \bibinfo {author} {\bibfnamefont {K.}~\bibnamefont
  {Nemoto}},\ }\bibfield  {title} {\emph {\bibinfo {title} {Effective
  compression of quantum braided circuits aided by zx-calculus},\ }}\href@noop
  {} {\bibfield  {journal} {\bibinfo  {journal} {Physical Review X}\ }\textbf
  {\bibinfo {volume} {10}},\ \bibinfo {pages} {041030} (\bibinfo {year}
  {2020})}\BibitemShut {NoStop}%
\bibitem [{\citenamefont {Roberts}\ and\ \citenamefont
  {Williamson}(2020)}]{roberts20203}%
  \BibitemOpen
  \bibfield  {author} {\bibinfo {author} {\bibfnamefont {S.}~\bibnamefont
  {Roberts}}\ and\ \bibinfo {author} {\bibfnamefont {D.~J.}\ \bibnamefont
  {Williamson}},\ }\bibfield  {title} {\emph {\bibinfo {title} {3-fermion
  topological quantum computation},\ }}\href {https://arxiv.org/abs/2011.04693}
  {\bibfield  {journal} {\bibinfo  {journal} {arXiv preprint arXiv:2011.04693}\
  } (\bibinfo {year} {2020})}\BibitemShut {NoStop}%
\bibitem [{\citenamefont {Zhu}\ \emph {et~al.}(2021)\citenamefont {Zhu},
  \citenamefont {Jochym-O'Connor},\ and\ \citenamefont
  {Dua}}]{zhu2021topological}%
  \BibitemOpen
  \bibfield  {author} {\bibinfo {author} {\bibfnamefont {G.}~\bibnamefont
  {Zhu}}, \bibinfo {author} {\bibfnamefont {T.}~\bibnamefont
  {Jochym-O'Connor}}, \ and\ \bibinfo {author} {\bibfnamefont {A.}~\bibnamefont
  {Dua}},\ }\bibfield  {title} {\emph {\bibinfo {title} {Topological order,
  quantum codes and quantum computation on fractal geometries},\ }}\href@noop
  {} {\bibfield  {journal} {\bibinfo  {journal} {arXiv preprint
  arXiv:2108.00018}\ } (\bibinfo {year} {2021})}\BibitemShut {NoStop}%
\bibitem [{\citenamefont {Chamberland}\ and\ \citenamefont
  {Campbell}(2021)}]{chamberland2021universal}%
  \BibitemOpen
  \bibfield  {author} {\bibinfo {author} {\bibfnamefont {C.}~\bibnamefont
  {Chamberland}}\ and\ \bibinfo {author} {\bibfnamefont {E.~T.}\ \bibnamefont
  {Campbell}},\ }\bibfield  {title} {\emph {\bibinfo {title} {Universal quantum
  computing with twist-free and temporally encoded lattice surgery},\
  }}\href@noop {} {\bibfield  {journal} {\bibinfo  {journal} {arXiv preprint
  arXiv:2109.02746}\ } (\bibinfo {year} {2021})}\BibitemShut {NoStop}%
\bibitem [{\citenamefont {Landahl}\ and\ \citenamefont
  {Morrison}(2021)}]{landahl2021logical}%
  \BibitemOpen
  \bibfield  {author} {\bibinfo {author} {\bibfnamefont {A.~J.}\ \bibnamefont
  {Landahl}}\ and\ \bibinfo {author} {\bibfnamefont {B.~C.}\ \bibnamefont
  {Morrison}},\ }\bibfield  {title} {\emph {\bibinfo {title} {Logical majorana
  fermions for fault-tolerant quantum simulation},\ }}\href@noop {} {\bibfield
  {journal} {\bibinfo  {journal} {arXiv preprint arXiv:2110.10280}\ } (\bibinfo
  {year} {2021})}\BibitemShut {NoStop}%
\bibitem [{\citenamefont {Beverland}\ \emph {et~al.}(2021)\citenamefont
  {Beverland}, \citenamefont {Kubica},\ and\ \citenamefont
  {Svore}}]{beverland2021cost}%
  \BibitemOpen
  \bibfield  {author} {\bibinfo {author} {\bibfnamefont {M.~E.}\ \bibnamefont
  {Beverland}}, \bibinfo {author} {\bibfnamefont {A.}~\bibnamefont {Kubica}}, \
  and\ \bibinfo {author} {\bibfnamefont {K.~M.}\ \bibnamefont {Svore}},\
  }\bibfield  {title} {\emph {\bibinfo {title} {Cost of universality: A
  comparative study of the overhead of state distillation and code switching
  with color codes},\ }}\href@noop {} {\bibfield  {journal} {\bibinfo
  {journal} {PRX Quantum}\ }\textbf {\bibinfo {volume} {2}},\ \bibinfo {pages}
  {020341} (\bibinfo {year} {2021})}\BibitemShut {NoStop}%
\bibitem [{\citenamefont {Fowler}(2013)}]{fowler2013accurate}%
  \BibitemOpen
  \bibfield  {author} {\bibinfo {author} {\bibfnamefont {A.~G.}\ \bibnamefont
  {Fowler}},\ }\bibfield  {title} {\emph {\bibinfo {title} {Accurate
  simulations of planar topological codes cannot use cyclic boundaries},\
  }}\href@noop {} {\bibfield  {journal} {\bibinfo  {journal} {Physical Review
  A}\ }\textbf {\bibinfo {volume} {87}},\ \bibinfo {pages} {062320} (\bibinfo
  {year} {2013})}\BibitemShut {NoStop}%
\bibitem [{\citenamefont {Beverland}\ \emph {et~al.}(2019)\citenamefont
  {Beverland}, \citenamefont {Brown}, \citenamefont {Kastoryano},\ and\
  \citenamefont {Marolleau}}]{beverland2019role}%
  \BibitemOpen
  \bibfield  {author} {\bibinfo {author} {\bibfnamefont {M.~E.}\ \bibnamefont
  {Beverland}}, \bibinfo {author} {\bibfnamefont {B.~J.}\ \bibnamefont
  {Brown}}, \bibinfo {author} {\bibfnamefont {M.~J.}\ \bibnamefont
  {Kastoryano}}, \ and\ \bibinfo {author} {\bibfnamefont {Q.}~\bibnamefont
  {Marolleau}},\ }\bibfield  {title} {\emph {\bibinfo {title} {The role of
  entropy in topological quantum error correction},\ }}\href@noop {} {\bibfield
   {journal} {\bibinfo  {journal} {Journal of Statistical Mechanics: Theory and
  Experiment}\ }\textbf {\bibinfo {volume} {2019}},\ \bibinfo {pages} {073404}
  (\bibinfo {year} {2019})}\BibitemShut {NoStop}%
\bibitem [{\citenamefont {Satzinger}\ \emph {et~al.}(2021)\citenamefont
  {Satzinger}, \citenamefont {Liu}, \citenamefont {Smith}, \citenamefont
  {Knapp}, \citenamefont {Newman}, \citenamefont {Jones}, \citenamefont {Chen},
  \citenamefont {Quintana}, \citenamefont {Mi}, \citenamefont {Dunsworth},
  \citenamefont {Gidney}, \citenamefont {Aleiner}, \citenamefont {Arute},
  \citenamefont {Arya}, \citenamefont {Atalaya}, \citenamefont {Babbush},
  \citenamefont {Bardin}, \citenamefont {Barends}, \citenamefont {Basso},
  \citenamefont {Bengtsson}, \citenamefont {Bilmes}, \citenamefont {Broughton},
  \citenamefont {Buckley}, \citenamefont {Buell}, \citenamefont {Burkett},
  \citenamefont {Bushnell}, \citenamefont {Chiaro}, \citenamefont {Collins},
  \citenamefont {Courtney}, \citenamefont {Demura}, \citenamefont {Derk},
  \citenamefont {Eppens}, \citenamefont {Erickson}, \citenamefont {Faoro},
  \citenamefont {Farhi}, \citenamefont {Fowler}, \citenamefont {Foxen},
  \citenamefont {Giustina}, \citenamefont {Greene}, \citenamefont {Gross},
  \citenamefont {Harrigan}, \citenamefont {Harrington}, \citenamefont {Hilton},
  \citenamefont {Hong}, \citenamefont {Huang}, \citenamefont {Huggins},
  \citenamefont {Ioffe}, \citenamefont {Isakov}, \citenamefont {Jeffrey},
  \citenamefont {Jiang}, \citenamefont {Kafri}, \citenamefont {Kechedzhi},
  \citenamefont {Khattar}, \citenamefont {Kim}, \citenamefont {Klimov},
  \citenamefont {Korotkov}, \citenamefont {Kostritsa}, \citenamefont
  {Landhuis}, \citenamefont {Laptev}, \citenamefont {Locharla}, \citenamefont
  {Lucero}, \citenamefont {Martin}, \citenamefont {McClean}, \citenamefont
  {McEwen}, \citenamefont {Miao}, \citenamefont {Mohseni}, \citenamefont
  {Montazeri}, \citenamefont {Mruczkiewicz}, \citenamefont {Mutus},
  \citenamefont {Naaman}, \citenamefont {Neeley}, \citenamefont {Neill},
  \citenamefont {Niu}, \citenamefont {O’Brien}, \citenamefont {Opremcak},
  \citenamefont {Pató}, \citenamefont {Petukhov}, \citenamefont {Rubin},
  \citenamefont {Sank}, \citenamefont {Shvarts}, \citenamefont {Strain},
  \citenamefont {Szalay}, \citenamefont {Villalonga}, \citenamefont {White},
  \citenamefont {Yao}, \citenamefont {Yeh}, \citenamefont {Yoo}, \citenamefont
  {Zalcman}, \citenamefont {Neven}, \citenamefont {Boixo}, \citenamefont
  {Megrant}, \citenamefont {Chen}, \citenamefont {Kelly}, \citenamefont
  {Smelyanskiy}, \citenamefont {Kitaev}, \citenamefont {Knap}, \citenamefont
  {Pollmann},\ and\ \citenamefont {Roushan}}]{doi:10.1126/science.abi8378}%
  \BibitemOpen
  \bibfield  {author} {\bibinfo {author} {\bibfnamefont {K.~J.}\ \bibnamefont
  {Satzinger}}, \bibinfo {author} {\bibfnamefont {Y.-J.}\ \bibnamefont {Liu}},
  \bibinfo {author} {\bibfnamefont {A.}~\bibnamefont {Smith}}, \bibinfo
  {author} {\bibfnamefont {C.}~\bibnamefont {Knapp}}, \bibinfo {author}
  {\bibfnamefont {M.}~\bibnamefont {Newman}}, \bibinfo {author} {\bibfnamefont
  {C.}~\bibnamefont {Jones}}, \bibinfo {author} {\bibfnamefont
  {Z.}~\bibnamefont {Chen}}, \bibinfo {author} {\bibfnamefont {C.}~\bibnamefont
  {Quintana}}, \bibinfo {author} {\bibfnamefont {X.}~\bibnamefont {Mi}},
  \bibinfo {author} {\bibfnamefont {A.}~\bibnamefont {Dunsworth}}, \bibinfo
  {author} {\bibfnamefont {C.}~\bibnamefont {Gidney}}, \bibinfo {author}
  {\bibfnamefont {I.}~\bibnamefont {Aleiner}}, \bibinfo {author} {\bibfnamefont
  {F.}~\bibnamefont {Arute}}, \bibinfo {author} {\bibfnamefont
  {K.}~\bibnamefont {Arya}}, \bibinfo {author} {\bibfnamefont {J.}~\bibnamefont
  {Atalaya}}, \bibinfo {author} {\bibfnamefont {R.}~\bibnamefont {Babbush}},
  \bibinfo {author} {\bibfnamefont {J.~C.}\ \bibnamefont {Bardin}}, \bibinfo
  {author} {\bibfnamefont {R.}~\bibnamefont {Barends}}, \bibinfo {author}
  {\bibfnamefont {J.}~\bibnamefont {Basso}}, \bibinfo {author} {\bibfnamefont
  {A.}~\bibnamefont {Bengtsson}}, \bibinfo {author} {\bibfnamefont
  {A.}~\bibnamefont {Bilmes}}, \bibinfo {author} {\bibfnamefont
  {M.}~\bibnamefont {Broughton}}, \bibinfo {author} {\bibfnamefont {B.~B.}\
  \bibnamefont {Buckley}}, \bibinfo {author} {\bibfnamefont {D.~A.}\
  \bibnamefont {Buell}}, \bibinfo {author} {\bibfnamefont {B.}~\bibnamefont
  {Burkett}}, \bibinfo {author} {\bibfnamefont {N.}~\bibnamefont {Bushnell}},
  \bibinfo {author} {\bibfnamefont {B.}~\bibnamefont {Chiaro}}, \bibinfo
  {author} {\bibfnamefont {R.}~\bibnamefont {Collins}}, \bibinfo {author}
  {\bibfnamefont {W.}~\bibnamefont {Courtney}}, \bibinfo {author}
  {\bibfnamefont {S.}~\bibnamefont {Demura}}, \bibinfo {author} {\bibfnamefont
  {A.~R.}\ \bibnamefont {Derk}}, \bibinfo {author} {\bibfnamefont
  {D.}~\bibnamefont {Eppens}}, \bibinfo {author} {\bibfnamefont
  {C.}~\bibnamefont {Erickson}}, \bibinfo {author} {\bibfnamefont
  {L.}~\bibnamefont {Faoro}}, \bibinfo {author} {\bibfnamefont
  {E.}~\bibnamefont {Farhi}}, \bibinfo {author} {\bibfnamefont {A.~G.}\
  \bibnamefont {Fowler}}, \bibinfo {author} {\bibfnamefont {B.}~\bibnamefont
  {Foxen}}, \bibinfo {author} {\bibfnamefont {M.}~\bibnamefont {Giustina}},
  \bibinfo {author} {\bibfnamefont {A.}~\bibnamefont {Greene}}, \bibinfo
  {author} {\bibfnamefont {J.~A.}\ \bibnamefont {Gross}}, \bibinfo {author}
  {\bibfnamefont {M.~P.}\ \bibnamefont {Harrigan}}, \bibinfo {author}
  {\bibfnamefont {S.~D.}\ \bibnamefont {Harrington}}, \bibinfo {author}
  {\bibfnamefont {J.}~\bibnamefont {Hilton}}, \bibinfo {author} {\bibfnamefont
  {S.}~\bibnamefont {Hong}}, \bibinfo {author} {\bibfnamefont {T.}~\bibnamefont
  {Huang}}, \bibinfo {author} {\bibfnamefont {W.~J.}\ \bibnamefont {Huggins}},
  \bibinfo {author} {\bibfnamefont {L.~B.}\ \bibnamefont {Ioffe}}, \bibinfo
  {author} {\bibfnamefont {S.~V.}\ \bibnamefont {Isakov}}, \bibinfo {author}
  {\bibfnamefont {E.}~\bibnamefont {Jeffrey}}, \bibinfo {author} {\bibfnamefont
  {Z.}~\bibnamefont {Jiang}}, \bibinfo {author} {\bibfnamefont
  {D.}~\bibnamefont {Kafri}}, \bibinfo {author} {\bibfnamefont
  {K.}~\bibnamefont {Kechedzhi}}, \bibinfo {author} {\bibfnamefont
  {T.}~\bibnamefont {Khattar}}, \bibinfo {author} {\bibfnamefont
  {S.}~\bibnamefont {Kim}}, \bibinfo {author} {\bibfnamefont {P.~V.}\
  \bibnamefont {Klimov}}, \bibinfo {author} {\bibfnamefont {A.~N.}\
  \bibnamefont {Korotkov}}, \bibinfo {author} {\bibfnamefont {F.}~\bibnamefont
  {Kostritsa}}, \bibinfo {author} {\bibfnamefont {D.}~\bibnamefont {Landhuis}},
  \bibinfo {author} {\bibfnamefont {P.}~\bibnamefont {Laptev}}, \bibinfo
  {author} {\bibfnamefont {A.}~\bibnamefont {Locharla}}, \bibinfo {author}
  {\bibfnamefont {E.}~\bibnamefont {Lucero}}, \bibinfo {author} {\bibfnamefont
  {O.}~\bibnamefont {Martin}}, \bibinfo {author} {\bibfnamefont {J.~R.}\
  \bibnamefont {McClean}}, \bibinfo {author} {\bibfnamefont {M.}~\bibnamefont
  {McEwen}}, \bibinfo {author} {\bibfnamefont {K.~C.}\ \bibnamefont {Miao}},
  \bibinfo {author} {\bibfnamefont {M.}~\bibnamefont {Mohseni}}, \bibinfo
  {author} {\bibfnamefont {S.}~\bibnamefont {Montazeri}}, \bibinfo {author}
  {\bibfnamefont {W.}~\bibnamefont {Mruczkiewicz}}, \bibinfo {author}
  {\bibfnamefont {J.}~\bibnamefont {Mutus}}, \bibinfo {author} {\bibfnamefont
  {O.}~\bibnamefont {Naaman}}, \bibinfo {author} {\bibfnamefont
  {M.}~\bibnamefont {Neeley}}, \bibinfo {author} {\bibfnamefont
  {C.}~\bibnamefont {Neill}}, \bibinfo {author} {\bibfnamefont {M.~Y.}\
  \bibnamefont {Niu}}, \bibinfo {author} {\bibfnamefont {T.~E.}\ \bibnamefont
  {O’Brien}}, \bibinfo {author} {\bibfnamefont {A.}~\bibnamefont {Opremcak}},
  \bibinfo {author} {\bibfnamefont {B.}~\bibnamefont {Pató}}, \bibinfo
  {author} {\bibfnamefont {A.}~\bibnamefont {Petukhov}}, \bibinfo {author}
  {\bibfnamefont {N.~C.}\ \bibnamefont {Rubin}}, \bibinfo {author}
  {\bibfnamefont {D.}~\bibnamefont {Sank}}, \bibinfo {author} {\bibfnamefont
  {V.}~\bibnamefont {Shvarts}}, \bibinfo {author} {\bibfnamefont
  {D.}~\bibnamefont {Strain}}, \bibinfo {author} {\bibfnamefont
  {M.}~\bibnamefont {Szalay}}, \bibinfo {author} {\bibfnamefont
  {B.}~\bibnamefont {Villalonga}}, \bibinfo {author} {\bibfnamefont {T.~C.}\
  \bibnamefont {White}}, \bibinfo {author} {\bibfnamefont {Z.}~\bibnamefont
  {Yao}}, \bibinfo {author} {\bibfnamefont {P.}~\bibnamefont {Yeh}}, \bibinfo
  {author} {\bibfnamefont {J.}~\bibnamefont {Yoo}}, \bibinfo {author}
  {\bibfnamefont {A.}~\bibnamefont {Zalcman}}, \bibinfo {author} {\bibfnamefont
  {H.}~\bibnamefont {Neven}}, \bibinfo {author} {\bibfnamefont
  {S.}~\bibnamefont {Boixo}}, \bibinfo {author} {\bibfnamefont
  {A.}~\bibnamefont {Megrant}}, \bibinfo {author} {\bibfnamefont
  {Y.}~\bibnamefont {Chen}}, \bibinfo {author} {\bibfnamefont {J.}~\bibnamefont
  {Kelly}}, \bibinfo {author} {\bibfnamefont {V.}~\bibnamefont {Smelyanskiy}},
  \bibinfo {author} {\bibfnamefont {A.}~\bibnamefont {Kitaev}}, \bibinfo
  {author} {\bibfnamefont {M.}~\bibnamefont {Knap}}, \bibinfo {author}
  {\bibfnamefont {F.}~\bibnamefont {Pollmann}}, \ and\ \bibinfo {author}
  {\bibfnamefont {P.}~\bibnamefont {Roushan}},\ }\bibfield  {title} {\emph
  {\bibinfo {title} {Realizing topologically ordered states on a quantum
  processor},\ }}\href {\doibase 10.1126/science.abi8378} {\bibfield  {journal}
  {\bibinfo  {journal} {Science}\ }\textbf {\bibinfo {volume} {374}},\ \bibinfo
  {pages} {1237} (\bibinfo {year} {2021})},\ \Eprint
  {http://arxiv.org/abs/https://www.science.org/doi/pdf/10.1126/science.abi8378}
  {https://www.science.org/doi/pdf/10.1126/science.abi8378} \BibitemShut
  {NoStop}%
\bibitem [{\citenamefont {Egan}\ \emph {et~al.}(2021)\citenamefont {Egan},
  \citenamefont {Debroy}, \citenamefont {Noel}, \citenamefont {Risinger},
  \citenamefont {Zhu}, \citenamefont {Biswas}, \citenamefont {Newman},
  \citenamefont {Li}, \citenamefont {Brown}, \citenamefont {Cetina} \emph
  {et~al.}}]{egan2021fault}%
  \BibitemOpen
  \bibfield  {author} {\bibinfo {author} {\bibfnamefont {L.}~\bibnamefont
  {Egan}}, \bibinfo {author} {\bibfnamefont {D.~M.}\ \bibnamefont {Debroy}},
  \bibinfo {author} {\bibfnamefont {C.}~\bibnamefont {Noel}}, \bibinfo {author}
  {\bibfnamefont {A.}~\bibnamefont {Risinger}}, \bibinfo {author}
  {\bibfnamefont {D.}~\bibnamefont {Zhu}}, \bibinfo {author} {\bibfnamefont
  {D.}~\bibnamefont {Biswas}}, \bibinfo {author} {\bibfnamefont
  {M.}~\bibnamefont {Newman}}, \bibinfo {author} {\bibfnamefont
  {M.}~\bibnamefont {Li}}, \bibinfo {author} {\bibfnamefont {K.~R.}\
  \bibnamefont {Brown}}, \bibinfo {author} {\bibfnamefont {M.}~\bibnamefont
  {Cetina}},  \emph {et~al.},\ }\bibfield  {title} {\emph {\bibinfo {title}
  {Fault-tolerant control of an error-corrected qubit},\ }}\href@noop {}
  {\bibfield  {journal} {\bibinfo  {journal} {Nature}\ }\textbf {\bibinfo
  {volume} {598}},\ \bibinfo {pages} {281} (\bibinfo {year}
  {2021})}\BibitemShut {NoStop}%
\bibitem [{\citenamefont {Ryan-Anderson}\ \emph {et~al.}(2021)\citenamefont
  {Ryan-Anderson}, \citenamefont {Bohnet}, \citenamefont {Lee}, \citenamefont
  {Gresh}, \citenamefont {Hankin}, \citenamefont {Gaebler}, \citenamefont
  {Francois}, \citenamefont {Chernoguzov}, \citenamefont {Lucchetti},
  \citenamefont {Brown} \emph {et~al.}}]{ryan2021realization}%
  \BibitemOpen
  \bibfield  {author} {\bibinfo {author} {\bibfnamefont {C.}~\bibnamefont
  {Ryan-Anderson}}, \bibinfo {author} {\bibfnamefont {J.}~\bibnamefont
  {Bohnet}}, \bibinfo {author} {\bibfnamefont {K.}~\bibnamefont {Lee}},
  \bibinfo {author} {\bibfnamefont {D.}~\bibnamefont {Gresh}}, \bibinfo
  {author} {\bibfnamefont {A.}~\bibnamefont {Hankin}}, \bibinfo {author}
  {\bibfnamefont {J.}~\bibnamefont {Gaebler}}, \bibinfo {author} {\bibfnamefont
  {D.}~\bibnamefont {Francois}}, \bibinfo {author} {\bibfnamefont
  {A.}~\bibnamefont {Chernoguzov}}, \bibinfo {author} {\bibfnamefont
  {D.}~\bibnamefont {Lucchetti}}, \bibinfo {author} {\bibfnamefont
  {N.}~\bibnamefont {Brown}},  \emph {et~al.},\ }\bibfield  {title} {\emph
  {\bibinfo {title} {Realization of real-time fault-tolerant quantum error
  correction},\ }}\href@noop {} {\bibfield  {journal} {\bibinfo  {journal}
  {arXiv preprint arXiv:2107.07505}\ } (\bibinfo {year} {2021})}\BibitemShut
  {NoStop}%
\bibitem [{\citenamefont {Postler}\ \emph {et~al.}(2021)\citenamefont
  {Postler}, \citenamefont {Heu{\ss}en}, \citenamefont {Pogorelov},
  \citenamefont {Rispler}, \citenamefont {Feldker}, \citenamefont {Meth},
  \citenamefont {Marciniak}, \citenamefont {Stricker}, \citenamefont
  {Ringbauer}, \citenamefont {Blatt} \emph
  {et~al.}}]{postler2021demonstration}%
  \BibitemOpen
  \bibfield  {author} {\bibinfo {author} {\bibfnamefont {L.}~\bibnamefont
  {Postler}}, \bibinfo {author} {\bibfnamefont {S.}~\bibnamefont {Heu{\ss}en}},
  \bibinfo {author} {\bibfnamefont {I.}~\bibnamefont {Pogorelov}}, \bibinfo
  {author} {\bibfnamefont {M.}~\bibnamefont {Rispler}}, \bibinfo {author}
  {\bibfnamefont {T.}~\bibnamefont {Feldker}}, \bibinfo {author} {\bibfnamefont
  {M.}~\bibnamefont {Meth}}, \bibinfo {author} {\bibfnamefont {C.~D.}\
  \bibnamefont {Marciniak}}, \bibinfo {author} {\bibfnamefont {R.}~\bibnamefont
  {Stricker}}, \bibinfo {author} {\bibfnamefont {M.}~\bibnamefont {Ringbauer}},
  \bibinfo {author} {\bibfnamefont {R.}~\bibnamefont {Blatt}},  \emph
  {et~al.},\ }\bibfield  {title} {\emph {\bibinfo {title} {Demonstration of
  fault-tolerant universal quantum gate operations},\ }}\href@noop {}
  {\bibfield  {journal} {\bibinfo  {journal} {arXiv preprint arXiv:2111.12654}\
  } (\bibinfo {year} {2021})}\BibitemShut {NoStop}%
\bibitem [{\citenamefont {Bartolucci}\ \emph {et~al.}(2023)\citenamefont
  {Bartolucci}, \citenamefont {Birchall}, \citenamefont {Bombin}, \citenamefont
  {Cable}, \citenamefont {Dawson}, \citenamefont {Gimeno-Segovia},
  \citenamefont {Johnston}, \citenamefont {Kieling}, \citenamefont {Nickerson},
  \citenamefont {Pant} \emph {et~al.}}]{bartolucci2021fusion}%
  \BibitemOpen
  \bibfield  {author} {\bibinfo {author} {\bibfnamefont {S.}~\bibnamefont
  {Bartolucci}}, \bibinfo {author} {\bibfnamefont {P.}~\bibnamefont
  {Birchall}}, \bibinfo {author} {\bibfnamefont {H.}~\bibnamefont {Bombin}},
  \bibinfo {author} {\bibfnamefont {H.}~\bibnamefont {Cable}}, \bibinfo
  {author} {\bibfnamefont {C.}~\bibnamefont {Dawson}}, \bibinfo {author}
  {\bibfnamefont {M.}~\bibnamefont {Gimeno-Segovia}}, \bibinfo {author}
  {\bibfnamefont {E.}~\bibnamefont {Johnston}}, \bibinfo {author}
  {\bibfnamefont {K.}~\bibnamefont {Kieling}}, \bibinfo {author} {\bibfnamefont
  {N.}~\bibnamefont {Nickerson}}, \bibinfo {author} {\bibfnamefont
  {M.}~\bibnamefont {Pant}},  \emph {et~al.},\ }\bibfield  {title} {\emph
  {\bibinfo {title} {Fusion-based quantum computation},\ }}\href@noop {}
  {\bibfield  {journal} {\bibinfo  {journal} {Nature Communications}\ }\textbf
  {\bibinfo {volume} {14}},\ \bibinfo {pages} {912} (\bibinfo {year}
  {2023})}\BibitemShut {NoStop}%
\bibitem [{\citenamefont {Fujii}(2015)}]{fujii2015quantum}%
  \BibitemOpen
  \bibfield  {author} {\bibinfo {author} {\bibfnamefont {K.}~\bibnamefont
  {Fujii}},\ }\href {\doibase https://doi.org/10.1007/978-981-287-996-7} {\emph
  {\bibinfo {title} {Quantum Computation with Topological Codes: from qubit to
  topological fault-tolerance}}},\ \bibinfo {series} {SpringerBriefs in
  Mathematical Physics}, Vol.~\bibinfo {volume} {8}\ (\bibinfo  {publisher}
  {Springer},\ \bibinfo {address} {Singapore},\ \bibinfo {year}
  {2015})\BibitemShut {NoStop}%
\bibitem [{\citenamefont {Bombin}\ \emph {et~al.}(2021)\citenamefont {Bombin},
  \citenamefont {Kim}, \citenamefont {Litinski}, \citenamefont {Nickerson},
  \citenamefont {Pant}, \citenamefont {Pastawski}, \citenamefont {Roberts},\
  and\ \citenamefont {Rudolph}}]{bombin2021interleaving}%
  \BibitemOpen
  \bibfield  {author} {\bibinfo {author} {\bibfnamefont {H.}~\bibnamefont
  {Bombin}}, \bibinfo {author} {\bibfnamefont {I.~H.}\ \bibnamefont {Kim}},
  \bibinfo {author} {\bibfnamefont {D.}~\bibnamefont {Litinski}}, \bibinfo
  {author} {\bibfnamefont {N.}~\bibnamefont {Nickerson}}, \bibinfo {author}
  {\bibfnamefont {M.}~\bibnamefont {Pant}}, \bibinfo {author} {\bibfnamefont
  {F.}~\bibnamefont {Pastawski}}, \bibinfo {author} {\bibfnamefont
  {S.}~\bibnamefont {Roberts}}, \ and\ \bibinfo {author} {\bibfnamefont
  {T.}~\bibnamefont {Rudolph}},\ }\bibfield  {title} {\emph {\bibinfo {title}
  {Interleaving: Modular architectures for fault-tolerant photonic quantum
  computing},\ }}\href {https://arxiv.org/abs/2103.08612} {\bibfield  {journal}
  {\bibinfo  {journal} {arXiv preprint arXiv:2103.08612}\ } (\bibinfo {year}
  {2021})}\BibitemShut {NoStop}%
\bibitem [{\citenamefont {Bravyi}\ and\ \citenamefont
  {Kitaev}(1998)}]{bravyi1998quantum}%
  \BibitemOpen
  \bibfield  {author} {\bibinfo {author} {\bibfnamefont {S.~B.}\ \bibnamefont
  {Bravyi}}\ and\ \bibinfo {author} {\bibfnamefont {A.~Y.}\ \bibnamefont
  {Kitaev}},\ }\bibfield  {title} {\emph {\bibinfo {title} {Quantum codes on a
  lattice with boundary},\ }}\href@noop {} {\bibfield  {journal} {\bibinfo
  {journal} {arXiv preprint quant-ph/9811052}\ } (\bibinfo {year}
  {1998})}\BibitemShut {NoStop}%
\bibitem [{\citenamefont {Nussinov}\ and\ \citenamefont
  {Ortiz}(2009)}]{nussinov2009symmetry}%
  \BibitemOpen
  \bibfield  {author} {\bibinfo {author} {\bibfnamefont {Z.}~\bibnamefont
  {Nussinov}}\ and\ \bibinfo {author} {\bibfnamefont {G.}~\bibnamefont
  {Ortiz}},\ }\bibfield  {title} {\emph {\bibinfo {title} {A symmetry principle
  for topological quantum order},\ }}\href@noop {} {\bibfield  {journal}
  {\bibinfo  {journal} {Annals of Physics}\ }\textbf {\bibinfo {volume}
  {324}},\ \bibinfo {pages} {977} (\bibinfo {year} {2009})}\BibitemShut
  {NoStop}%
\bibitem [{\citenamefont {Bomb{\'\i}n}\ and\ \citenamefont
  {Martin-Delgado}(2007)}]{bombin2007optimal}%
  \BibitemOpen
  \bibfield  {author} {\bibinfo {author} {\bibfnamefont {H.}~\bibnamefont
  {Bomb{\'\i}n}}\ and\ \bibinfo {author} {\bibfnamefont {M.~A.}\ \bibnamefont
  {Martin-Delgado}},\ }\bibfield  {title} {\emph {\bibinfo {title} {Optimal
  resources for topological two-dimensional stabilizer codes: Comparative
  study},\ }}\href@noop {} {\bibfield  {journal} {\bibinfo  {journal} {Physical
  Review A}\ }\textbf {\bibinfo {volume} {76}},\ \bibinfo {pages} {012305}
  (\bibinfo {year} {2007})}\BibitemShut {NoStop}%
\bibitem [{\citenamefont {Litinski}(2019{\natexlab{a}})}]{litinski2019game}%
  \BibitemOpen
  \bibfield  {author} {\bibinfo {author} {\bibfnamefont {D.}~\bibnamefont
  {Litinski}},\ }\bibfield  {title} {\emph {\bibinfo {title} {A game of surface
  codes: Large-scale quantum computing with lattice surgery},\ }}\href@noop {}
  {\bibfield  {journal} {\bibinfo  {journal} {Quantum}\ }\textbf {\bibinfo
  {volume} {3}},\ \bibinfo {pages} {128} (\bibinfo {year}
  {2019}{\natexlab{a}})}\BibitemShut {NoStop}%
\bibitem [{\citenamefont {Tillich}\ and\ \citenamefont
  {Z{\'e}mor}(2014)}]{tillich2014quantum}%
  \BibitemOpen
  \bibfield  {author} {\bibinfo {author} {\bibfnamefont {J.-P.}\ \bibnamefont
  {Tillich}}\ and\ \bibinfo {author} {\bibfnamefont {G.}~\bibnamefont
  {Z{\'e}mor}},\ }\bibfield  {title} {\emph {\bibinfo {title} {Quantum ldpc
  codes with positive rate and minimum distance proportional to the square root
  of the blocklength},\ }}\href@noop {} {\bibfield  {journal} {\bibinfo
  {journal} {IEEE Transactions on Information Theory}\ }\textbf {\bibinfo
  {volume} {60}},\ \bibinfo {pages} {1193} (\bibinfo {year}
  {2014})}\BibitemShut {NoStop}%
\bibitem [{\citenamefont {Gottesman}(2013)}]{gottesman2013fault}%
  \BibitemOpen
  \bibfield  {author} {\bibinfo {author} {\bibfnamefont {D.}~\bibnamefont
  {Gottesman}},\ }\bibfield  {title} {\emph {\bibinfo {title} {Fault-tolerant
  quantum computation with constant overhead},\ }}\href
  {https://arxiv.org/abs/1310.2984} {\bibfield  {journal} {\bibinfo  {journal}
  {arXiv preprint arXiv:1310.2984}\ } (\bibinfo {year} {2013})}\BibitemShut
  {NoStop}%
\bibitem [{\citenamefont {Fawzi}\ \emph
  {et~al.}(2018{\natexlab{a}})\citenamefont {Fawzi}, \citenamefont
  {Grospellier},\ and\ \citenamefont {Leverrier}}]{fawzi2018constant}%
  \BibitemOpen
  \bibfield  {author} {\bibinfo {author} {\bibfnamefont {O.}~\bibnamefont
  {Fawzi}}, \bibinfo {author} {\bibfnamefont {A.}~\bibnamefont {Grospellier}},
  \ and\ \bibinfo {author} {\bibfnamefont {A.}~\bibnamefont {Leverrier}},\ }in\
  \href@noop {} {\emph {\bibinfo {booktitle} {2018 IEEE 59th Annual Symposium
  on Foundations of Computer Science (FOCS)}}}\ (\bibinfo {organization}
  {IEEE},\ \bibinfo {year} {2018})\ pp.\ \bibinfo {pages}
  {743--754}\BibitemShut {NoStop}%
\bibitem [{\citenamefont {Fawzi}\ \emph
  {et~al.}(2018{\natexlab{b}})\citenamefont {Fawzi}, \citenamefont
  {Grospellier},\ and\ \citenamefont {Leverrier}}]{fawzi2018efficient}%
  \BibitemOpen
  \bibfield  {author} {\bibinfo {author} {\bibfnamefont {O.}~\bibnamefont
  {Fawzi}}, \bibinfo {author} {\bibfnamefont {A.}~\bibnamefont {Grospellier}},
  \ and\ \bibinfo {author} {\bibfnamefont {A.}~\bibnamefont {Leverrier}},\ }in\
  \href@noop {} {\emph {\bibinfo {booktitle} {Proceedings of the 50th Annual
  ACM SIGACT Symposium on Theory of Computing}}}\ (\bibinfo {year} {2018})\
  pp.\ \bibinfo {pages} {521--534}\BibitemShut {NoStop}%
\bibitem [{\citenamefont {Breuckmann}\ and\ \citenamefont
  {Eberhardt}(2021{\natexlab{a}})}]{breuckmann2021ldpc}%
  \BibitemOpen
  \bibfield  {author} {\bibinfo {author} {\bibfnamefont {N.~P.}\ \bibnamefont
  {Breuckmann}}\ and\ \bibinfo {author} {\bibfnamefont {J.~N.}\ \bibnamefont
  {Eberhardt}},\ }\bibfield  {title} {\emph {\bibinfo {title} {Quantum
  low-density parity-check codes},\ }}\href {\doibase
  10.1103/PRXQuantum.2.040101} {\bibfield  {journal} {\bibinfo  {journal} {PRX
  Quantum}\ }\textbf {\bibinfo {volume} {2}},\ \bibinfo {pages} {040101}
  (\bibinfo {year} {2021}{\natexlab{a}})}\BibitemShut {NoStop}%
\bibitem [{\citenamefont {Hastings}\ \emph {et~al.}(2021)\citenamefont
  {Hastings}, \citenamefont {Haah},\ and\ \citenamefont
  {O'Donnell}}]{hastings2021fiber}%
  \BibitemOpen
  \bibfield  {author} {\bibinfo {author} {\bibfnamefont {M.~B.}\ \bibnamefont
  {Hastings}}, \bibinfo {author} {\bibfnamefont {J.}~\bibnamefont {Haah}}, \
  and\ \bibinfo {author} {\bibfnamefont {R.}~\bibnamefont {O'Donnell}},\ }in\
  \href@noop {} {\emph {\bibinfo {booktitle} {Proceedings of the 53rd Annual
  ACM SIGACT Symposium on Theory of Computing}}}\ (\bibinfo {year} {2021})\
  pp.\ \bibinfo {pages} {1276--1288}\BibitemShut {NoStop}%
\bibitem [{\citenamefont {Breuckmann}\ and\ \citenamefont
  {Eberhardt}(2021{\natexlab{b}})}]{breuckmann2021balanced}%
  \BibitemOpen
  \bibfield  {author} {\bibinfo {author} {\bibfnamefont {N.~P.}\ \bibnamefont
  {Breuckmann}}\ and\ \bibinfo {author} {\bibfnamefont {J.~N.}\ \bibnamefont
  {Eberhardt}},\ }\bibfield  {title} {\emph {\bibinfo {title} {Balanced product
  quantum codes},\ }}\href@noop {} {\bibfield  {journal} {\bibinfo  {journal}
  {IEEE Transactions on Information Theory}\ }\textbf {\bibinfo {volume}
  {67}},\ \bibinfo {pages} {6653} (\bibinfo {year}
  {2021}{\natexlab{b}})}\BibitemShut {NoStop}%
\bibitem [{\citenamefont {Panteleev}\ and\ \citenamefont
  {Kalachev}(2021)}]{panteleev2021asymptotically}%
  \BibitemOpen
  \bibfield  {author} {\bibinfo {author} {\bibfnamefont {P.}~\bibnamefont
  {Panteleev}}\ and\ \bibinfo {author} {\bibfnamefont {G.}~\bibnamefont
  {Kalachev}},\ }\bibfield  {title} {\emph {\bibinfo {title} {Asymptotically
  good quantum and locally testable classical ldpc codes},\ }}\href@noop {}
  {\bibfield  {journal} {\bibinfo  {journal} {arXiv preprint arXiv:2111.03654}\
  } (\bibinfo {year} {2021})}\BibitemShut {NoStop}%
\bibitem [{\citenamefont {Gottesman}(1997)}]{gottesman1997stabilizer}%
  \BibitemOpen
  \bibfield  {author} {\bibinfo {author} {\bibfnamefont {D.}~\bibnamefont
  {Gottesman}},\ }\emph {\bibinfo {title} {Stabilizer codes and quantum error
  correction}},\ \href {\doibase
  https://doi.org/10.48550/arXiv.quant-ph/9705052} {Ph.D. thesis},\ \bibinfo
  {school} {California Institute of Technology}, \bibinfo {address} {Pasadena,
  California} (\bibinfo {year} {1997})\BibitemShut {NoStop}%
\bibitem [{\citenamefont {Bombin}\ and\ \citenamefont
  {Martin-Delgado}(2006{\natexlab{a}})}]{bombin2006topological}%
  \BibitemOpen
  \bibfield  {author} {\bibinfo {author} {\bibfnamefont {H.}~\bibnamefont
  {Bombin}}\ and\ \bibinfo {author} {\bibfnamefont {M.~A.}\ \bibnamefont
  {Martin-Delgado}},\ }\bibfield  {title} {\emph {\bibinfo {title} {Topological
  quantum distillation},\ }}\href@noop {} {\bibfield  {journal} {\bibinfo
  {journal} {Physical review letters}\ }\textbf {\bibinfo {volume} {97}},\
  \bibinfo {pages} {180501} (\bibinfo {year} {2006}{\natexlab{a}})}\BibitemShut
  {NoStop}%
\bibitem [{\citenamefont {Paetznick}\ and\ \citenamefont
  {Reichardt}(2013)}]{paetznick2013universal}%
  \BibitemOpen
  \bibfield  {author} {\bibinfo {author} {\bibfnamefont {A.}~\bibnamefont
  {Paetznick}}\ and\ \bibinfo {author} {\bibfnamefont {B.~W.}\ \bibnamefont
  {Reichardt}},\ }\bibfield  {title} {\emph {\bibinfo {title} {Universal
  fault-tolerant quantum computation with only transversal gates and error
  correction},\ }}\href@noop {} {\bibfield  {journal} {\bibinfo  {journal}
  {Physical review letters}\ }\textbf {\bibinfo {volume} {111}},\ \bibinfo
  {pages} {090505} (\bibinfo {year} {2013})}\BibitemShut {NoStop}%
\bibitem [{\citenamefont {Newman}\ \emph {et~al.}(2020)\citenamefont {Newman},
  \citenamefont {de~Castro},\ and\ \citenamefont
  {Brown}}]{newman2020generating}%
  \BibitemOpen
  \bibfield  {author} {\bibinfo {author} {\bibfnamefont {M.}~\bibnamefont
  {Newman}}, \bibinfo {author} {\bibfnamefont {L.~A.}\ \bibnamefont
  {de~Castro}}, \ and\ \bibinfo {author} {\bibfnamefont {K.~R.}\ \bibnamefont
  {Brown}},\ }\bibfield  {title} {\emph {\bibinfo {title} {Generating
  fault-tolerant cluster states from crystal structures},\ }}\href@noop {}
  {\bibfield  {journal} {\bibinfo  {journal} {Quantum}\ }\textbf {\bibinfo
  {volume} {4}},\ \bibinfo {pages} {295} (\bibinfo {year} {2020})}\BibitemShut
  {NoStop}%
\bibitem [{\citenamefont {Hastings}\ and\ \citenamefont
  {Haah}(2021)}]{hastings2021dynamically}%
  \BibitemOpen
  \bibfield  {author} {\bibinfo {author} {\bibfnamefont {M.~B.}\ \bibnamefont
  {Hastings}}\ and\ \bibinfo {author} {\bibfnamefont {J.}~\bibnamefont
  {Haah}},\ }\bibfield  {title} {\emph {\bibinfo {title} {Dynamically generated
  logical qubits},\ }}\href@noop {} {\bibfield  {journal} {\bibinfo  {journal}
  {Quantum}\ }\textbf {\bibinfo {volume} {5}},\ \bibinfo {pages} {564}
  (\bibinfo {year} {2021})}\BibitemShut {NoStop}%
\bibitem [{\citenamefont {Kitaev}(2006)}]{kitaev2006anyons}%
  \BibitemOpen
  \bibfield  {author} {\bibinfo {author} {\bibfnamefont {A.}~\bibnamefont
  {Kitaev}},\ }\bibfield  {title} {\emph {\bibinfo {title} {Anyons in an
  exactly solved model and beyond},\ }}\href@noop {} {\bibfield  {journal}
  {\bibinfo  {journal} {Annals of Physics}\ }\textbf {\bibinfo {volume}
  {321}},\ \bibinfo {pages} {2} (\bibinfo {year} {2006})}\BibitemShut {NoStop}%
\bibitem [{\citenamefont {Preskill}(1999)}]{preskill1999lecture}%
  \BibitemOpen
  \bibfield  {author} {\bibinfo {author} {\bibfnamefont {J.}~\bibnamefont
  {Preskill}},\ }\bibfield  {title} {\emph {\bibinfo {title} {Lecture notes for
  physics 219: Quantum computation},\ }}\href@noop {} {\bibfield  {journal}
  {\bibinfo  {journal} {Caltech Lecture Notes}\ } (\bibinfo {year}
  {1999})}\BibitemShut {NoStop}%
\bibitem [{\citenamefont {Levin}(2013)}]{levin2013protected}%
  \BibitemOpen
  \bibfield  {author} {\bibinfo {author} {\bibfnamefont {M.}~\bibnamefont
  {Levin}},\ }\bibfield  {title} {\emph {\bibinfo {title} {Protected edge modes
  without symmetry},\ }}\href@noop {} {\bibfield  {journal} {\bibinfo
  {journal} {Physical Review X}\ }\textbf {\bibinfo {volume} {3}},\ \bibinfo
  {pages} {021009} (\bibinfo {year} {2013})}\BibitemShut {NoStop}%
\bibitem [{\citenamefont {Liu}\ \emph {et~al.}(2017)\citenamefont {Liu},
  \citenamefont {Wozniakowski},\ and\ \citenamefont {Jaffe}}]{liu2017quon}%
  \BibitemOpen
  \bibfield  {author} {\bibinfo {author} {\bibfnamefont {Z.}~\bibnamefont
  {Liu}}, \bibinfo {author} {\bibfnamefont {A.}~\bibnamefont {Wozniakowski}}, \
  and\ \bibinfo {author} {\bibfnamefont {A.~M.}\ \bibnamefont {Jaffe}},\
  }\bibfield  {title} {\emph {\bibinfo {title} {Quon 3d language for quantum
  information},\ }}\href@noop {} {\bibfield  {journal} {\bibinfo  {journal}
  {Proceedings of the National Academy of Sciences}\ }\textbf {\bibinfo
  {volume} {114}},\ \bibinfo {pages} {2497} (\bibinfo {year}
  {2017})}\BibitemShut {NoStop}%
\bibitem [{\citenamefont {Barkeshli}\ \emph {et~al.}(2019)\citenamefont
  {Barkeshli}, \citenamefont {Bonderson}, \citenamefont {Cheng},\ and\
  \citenamefont {Wang}}]{barkeshli2019symmetry}%
  \BibitemOpen
  \bibfield  {author} {\bibinfo {author} {\bibfnamefont {M.}~\bibnamefont
  {Barkeshli}}, \bibinfo {author} {\bibfnamefont {P.}~\bibnamefont
  {Bonderson}}, \bibinfo {author} {\bibfnamefont {M.}~\bibnamefont {Cheng}}, \
  and\ \bibinfo {author} {\bibfnamefont {Z.}~\bibnamefont {Wang}},\ }\bibfield
  {title} {\emph {\bibinfo {title} {Symmetry fractionalization, defects, and
  gauging of topological phases},\ }}\href@noop {} {\bibfield  {journal}
  {\bibinfo  {journal} {Physical Review B}\ }\textbf {\bibinfo {volume}
  {100}},\ \bibinfo {pages} {115147} (\bibinfo {year} {2019})}\BibitemShut
  {NoStop}%
\bibitem [{\citenamefont {Wen}(2003)}]{wen2003quantum}%
  \BibitemOpen
  \bibfield  {author} {\bibinfo {author} {\bibfnamefont {X.-G.}\ \bibnamefont
  {Wen}},\ }\bibfield  {title} {\emph {\bibinfo {title} {Quantum orders in an
  exact soluble model},\ }}\href@noop {} {\bibfield  {journal} {\bibinfo
  {journal} {Physical review letters}\ }\textbf {\bibinfo {volume} {90}},\
  \bibinfo {pages} {016803} (\bibinfo {year} {2003})}\BibitemShut {NoStop}%
\bibitem [{\citenamefont {Bombin}\ and\ \citenamefont
  {Martin-Delgado}(2006{\natexlab{b}})}]{bombin2006topologicalencoding}%
  \BibitemOpen
  \bibfield  {author} {\bibinfo {author} {\bibfnamefont {H.}~\bibnamefont
  {Bombin}}\ and\ \bibinfo {author} {\bibfnamefont {M.}~\bibnamefont
  {Martin-Delgado}},\ }\bibfield  {title} {\emph {\bibinfo {title} {Topological
  quantum error correction with optimal encoding rate},\ }}\href@noop {}
  {\bibfield  {journal} {\bibinfo  {journal} {Physical Review A}\ }\textbf
  {\bibinfo {volume} {73}},\ \bibinfo {pages} {062303} (\bibinfo {year}
  {2006}{\natexlab{b}})}\BibitemShut {NoStop}%
\bibitem [{\citenamefont {Tomita}\ and\ \citenamefont
  {Svore}(2014)}]{tomita2014low}%
  \BibitemOpen
  \bibfield  {author} {\bibinfo {author} {\bibfnamefont {Y.}~\bibnamefont
  {Tomita}}\ and\ \bibinfo {author} {\bibfnamefont {K.~M.}\ \bibnamefont
  {Svore}},\ }\bibfield  {title} {\emph {\bibinfo {title} {Low-distance surface
  codes under realistic quantum noise},\ }}\href@noop {} {\bibfield  {journal}
  {\bibinfo  {journal} {Physical Review A}\ }\textbf {\bibinfo {volume} {90}},\
  \bibinfo {pages} {062320} (\bibinfo {year} {2014})}\BibitemShut {NoStop}%
\bibitem [{Note1()}]{Note1}%
  \BibitemOpen
  \bibinfo {note} {The two types of anyons of the surface code are also often
  also referred to as $e$-type and $m$-type, or alternatively $X$-type and
  $Z$-type.}\BibitemShut {Stop}%
\bibitem [{\citenamefont {Beverland}\ \emph {et~al.}()\citenamefont
  {Beverland}, \citenamefont {Buerschaper}, \citenamefont {Koenig},
  \citenamefont {Pastawski}, \citenamefont {Preskill},\ and\ \citenamefont
  {Sijher}}]{beverland2016protected}%
  \BibitemOpen
  \bibfield  {author} {\bibinfo {author} {\bibfnamefont {M.~E.}\ \bibnamefont
  {Beverland}}, \bibinfo {author} {\bibfnamefont {O.}~\bibnamefont
  {Buerschaper}}, \bibinfo {author} {\bibfnamefont {R.}~\bibnamefont {Koenig}},
  \bibinfo {author} {\bibfnamefont {F.}~\bibnamefont {Pastawski}}, \bibinfo
  {author} {\bibfnamefont {J.}~\bibnamefont {Preskill}}, \ and\ \bibinfo
  {author} {\bibfnamefont {S.}~\bibnamefont {Sijher}},\ }\bibfield  {title}
  {\emph {\bibinfo {title} {Protected gates for topological quantum field
  theories},\ }}\href@noop {} {\bibinfo  {journal} {Journal of Mathematical
  Physics}\ }\BibitemShut {NoStop}%
\bibitem [{\citenamefont {Kitaev}\ and\ \citenamefont
  {Kong}(2012)}]{kitaev2012models}%
  \BibitemOpen
\bibfield  {journal} {  }\bibfield  {author} {\bibinfo {author} {\bibfnamefont
  {A.}~\bibnamefont {Kitaev}}\ and\ \bibinfo {author} {\bibfnamefont
  {L.}~\bibnamefont {Kong}},\ }\bibfield  {title} {\emph {\bibinfo {title}
  {Models for gapped boundaries and domain walls},\ }}\href {\doibase
  10.1007/s00220-012-1500-5} {\bibfield  {journal} {\bibinfo  {journal}
  {Communications in Mathematical Physics}\ }\textbf {\bibinfo {volume}
  {313}},\ \bibinfo {pages} {351} (\bibinfo {year} {2012})}\BibitemShut
  {NoStop}%
\bibitem [{\citenamefont {Levin}\ and\ \citenamefont
  {Gu}(2012)}]{levin2012braiding}%
  \BibitemOpen
  \bibfield  {author} {\bibinfo {author} {\bibfnamefont {M.}~\bibnamefont
  {Levin}}\ and\ \bibinfo {author} {\bibfnamefont {Z.-C.}\ \bibnamefont {Gu}},\
  }\bibfield  {title} {\emph {\bibinfo {title} {Braiding statistics approach to
  symmetry-protected topological phases},\ }}\href@noop {} {\bibfield
  {journal} {\bibinfo  {journal} {Physical Review B}\ }\textbf {\bibinfo
  {volume} {86}},\ \bibinfo {pages} {115109} (\bibinfo {year}
  {2012})}\BibitemShut {NoStop}%
\bibitem [{\citenamefont {Lan}\ \emph {et~al.}(2015)\citenamefont {Lan},
  \citenamefont {Wang},\ and\ \citenamefont {Wen}}]{lan2015gapped}%
  \BibitemOpen
  \bibfield  {author} {\bibinfo {author} {\bibfnamefont {T.}~\bibnamefont
  {Lan}}, \bibinfo {author} {\bibfnamefont {J.~C.}\ \bibnamefont {Wang}}, \
  and\ \bibinfo {author} {\bibfnamefont {X.-G.}\ \bibnamefont {Wen}},\
  }\bibfield  {title} {\emph {\bibinfo {title} {Gapped domain walls, gapped
  boundaries, and topological degeneracy},\ }}\href {\doibase
  10.1103/PhysRevLett.114.076402} {\bibfield  {journal} {\bibinfo  {journal}
  {Physical Review Letters}\ }\textbf {\bibinfo {volume} {114}},\ \bibinfo
  {pages} {076402} (\bibinfo {year} {2015})}\BibitemShut {NoStop}%
\bibitem [{Note2()}]{Note2}%
  \BibitemOpen
  \bibinfo {note} {Note that generally, the term ``domain wall'' refers to any
  boundary between two topological phases. Here, we specifically use it to
  refer to the primal-dual swapping boundary.}\BibitemShut {Stop}%
\bibitem [{\citenamefont {Freedman}\ and\ \citenamefont
  {Meyer}(2001)}]{freedman2001projective}%
  \BibitemOpen
  \bibfield  {author} {\bibinfo {author} {\bibfnamefont {M.~H.}\ \bibnamefont
  {Freedman}}\ and\ \bibinfo {author} {\bibfnamefont {D.~A.}\ \bibnamefont
  {Meyer}},\ }\bibfield  {title} {\emph {\bibinfo {title} {Projective plane and
  planar quantum codes},\ }}\href@noop {} {\bibfield  {journal} {\bibinfo
  {journal} {Foundations of Computational Mathematics}\ }\textbf {\bibinfo
  {volume} {1}},\ \bibinfo {pages} {325} (\bibinfo {year} {2001})}\BibitemShut
  {NoStop}%
\bibitem [{Note3()}]{Note3}%
  \BibitemOpen
  \bibinfo {note} {For example, the 2D color code (which is locally equivalent
  to two copies of the surface code code~\cite {bombin2012universal}) has a
  symmetry group containing 72 elements~\cite
  {yoshida2015topological,scruby2020hierarchy}, compared to the $\protect
  \mathbbm {Z}_2$ symmetry of a single surface code.}\BibitemShut {Stop}%
\bibitem [{Note4()}]{Note4}%
  \BibitemOpen
  \bibinfo {note} {This cell complex is distinct from the cell complex commonly
  used in the context of fault-tolerant topological MBQC~\cite
  {raussendorf2007fault,raussendorf2007topological, nickerson2018measurement}
  in which the checks correspond to 3-cells and 0-cells of the
  complex.}\BibitemShut {Stop}%
\bibitem [{Note5()}]{Note5}%
  \BibitemOpen
  \bibinfo {note} {In condensed matter language, this check operator group can
  be understood as a $\protect \mathbbm {Z}_2\times \protect \mathbbm {Z}_2$
  1-form symmetry~\cite
  {gaiotto2015generalized,kapustin2017higher,roberts2020symmetry}.}\BibitemShut
  {Stop}%
\bibitem [{\citenamefont {Bravyi}\ and\ \citenamefont
  {Kitaev}(2005)}]{bravyi2005universal}%
  \BibitemOpen
  \bibfield  {author} {\bibinfo {author} {\bibfnamefont {S.}~\bibnamefont
  {Bravyi}}\ and\ \bibinfo {author} {\bibfnamefont {A.}~\bibnamefont
  {Kitaev}},\ }\bibfield  {title} {\emph {\bibinfo {title} {Universal quantum
  computation with ideal clifford gates and noisy ancillas},\ }}\href@noop {}
  {\bibfield  {journal} {\bibinfo  {journal} {Physical Review A}\ }\textbf
  {\bibinfo {volume} {71}},\ \bibinfo {pages} {022316} (\bibinfo {year}
  {2005})}\BibitemShut {NoStop}%
\bibitem [{\citenamefont {Bravyi}\ and\ \citenamefont
  {Haah}(2012)}]{bravyi2012magic}%
  \BibitemOpen
  \bibfield  {author} {\bibinfo {author} {\bibfnamefont {S.}~\bibnamefont
  {Bravyi}}\ and\ \bibinfo {author} {\bibfnamefont {J.}~\bibnamefont {Haah}},\
  }\bibfield  {title} {\emph {\bibinfo {title} {Magic-state distillation with
  low overhead},\ }}\href@noop {} {\bibfield  {journal} {\bibinfo  {journal}
  {Physical Review A}\ }\textbf {\bibinfo {volume} {86}},\ \bibinfo {pages}
  {052329} (\bibinfo {year} {2012})}\BibitemShut {NoStop}%
\bibitem [{\citenamefont {Litinski}(2019{\natexlab{b}})}]{litinski2019magic}%
  \BibitemOpen
  \bibfield  {author} {\bibinfo {author} {\bibfnamefont {D.}~\bibnamefont
  {Litinski}},\ }\bibfield  {title} {\emph {\bibinfo {title} {Magic state
  distillation: Not as costly as you think},\ }}\href@noop {} {\bibfield
  {journal} {\bibinfo  {journal} {Quantum}\ }\textbf {\bibinfo {volume} {3}},\
  \bibinfo {pages} {205} (\bibinfo {year} {2019}{\natexlab{b}})}\BibitemShut
  {NoStop}%
\bibitem [{\citenamefont {Kim}\ \emph {et~al.}(2022)\citenamefont {Kim},
  \citenamefont {Liu}, \citenamefont {Pallister}, \citenamefont {Pol},
  \citenamefont {Roberts},\ and\ \citenamefont {Lee}}]{kim2021faulttolerant}%
  \BibitemOpen
  \bibfield  {author} {\bibinfo {author} {\bibfnamefont {I.~H.}\ \bibnamefont
  {Kim}}, \bibinfo {author} {\bibfnamefont {Y.-H.}\ \bibnamefont {Liu}},
  \bibinfo {author} {\bibfnamefont {S.}~\bibnamefont {Pallister}}, \bibinfo
  {author} {\bibfnamefont {W.}~\bibnamefont {Pol}}, \bibinfo {author}
  {\bibfnamefont {S.}~\bibnamefont {Roberts}}, \ and\ \bibinfo {author}
  {\bibfnamefont {E.}~\bibnamefont {Lee}},\ }\bibfield  {title} {\emph
  {\bibinfo {title} {Fault-tolerant resource estimate for quantum chemical
  simulations: Case study on li-ion battery electrolyte molecules},\ }}\href
  {\doibase 10.1103/PhysRevResearch.4.023019} {\bibfield  {journal} {\bibinfo
  {journal} {Phys. Rev. Research}\ }\textbf {\bibinfo {volume} {4}},\ \bibinfo
  {pages} {023019} (\bibinfo {year} {2022})}\BibitemShut {NoStop}%
\bibitem [{\citenamefont {Kubica}\ \emph {et~al.}(2015)\citenamefont {Kubica},
  \citenamefont {Yoshida},\ and\ \citenamefont
  {Pastawski}}]{kubica2015unfolding}%
  \BibitemOpen
  \bibfield  {author} {\bibinfo {author} {\bibfnamefont {A.}~\bibnamefont
  {Kubica}}, \bibinfo {author} {\bibfnamefont {B.}~\bibnamefont {Yoshida}}, \
  and\ \bibinfo {author} {\bibfnamefont {F.}~\bibnamefont {Pastawski}},\
  }\bibfield  {title} {\emph {\bibinfo {title} {Unfolding the color code},\
  }}\href@noop {} {\bibfield  {journal} {\bibinfo  {journal} {New Journal of
  Physics}\ }\textbf {\bibinfo {volume} {17}},\ \bibinfo {pages} {083026}
  (\bibinfo {year} {2015})}\BibitemShut {NoStop}%
\bibitem [{\citenamefont {Moussa}(2016)}]{moussa2016transversal}%
  \BibitemOpen
  \bibfield  {author} {\bibinfo {author} {\bibfnamefont {J.~E.}\ \bibnamefont
  {Moussa}},\ }\bibfield  {title} {\emph {\bibinfo {title} {Transversal
  clifford gates on folded surface codes},\ }}\href@noop {} {\bibfield
  {journal} {\bibinfo  {journal} {Physical Review A}\ }\textbf {\bibinfo
  {volume} {94}},\ \bibinfo {pages} {042316} (\bibinfo {year}
  {2016})}\BibitemShut {NoStop}%
\bibitem [{\citenamefont {Elliott}\ \emph {et~al.}(2009)\citenamefont
  {Elliott}, \citenamefont {Eastin},\ and\ \citenamefont
  {Caves}}]{elliott2009graphical}%
  \BibitemOpen
  \bibfield  {author} {\bibinfo {author} {\bibfnamefont {M.~B.}\ \bibnamefont
  {Elliott}}, \bibinfo {author} {\bibfnamefont {B.}~\bibnamefont {Eastin}}, \
  and\ \bibinfo {author} {\bibfnamefont {C.~M.}\ \bibnamefont {Caves}},\
  }\bibfield  {title} {\emph {\bibinfo {title} {Graphical description of pauli
  measurements on stabilizer states},\ }}\href@noop {} {\bibfield  {journal}
  {\bibinfo  {journal} {Journal of Physics A: Mathematical and Theoretical}\
  }\textbf {\bibinfo {volume} {43}},\ \bibinfo {pages} {025301} (\bibinfo
  {year} {2009})}\BibitemShut {NoStop}%
\bibitem [{\citenamefont {Coecke}\ and\ \citenamefont
  {Duncan}(2008)}]{coecke2008interacting}%
  \BibitemOpen
  \bibfield  {author} {\bibinfo {author} {\bibfnamefont {B.}~\bibnamefont
  {Coecke}}\ and\ \bibinfo {author} {\bibfnamefont {R.}~\bibnamefont
  {Duncan}},\ }in\ \href@noop {} {\emph {\bibinfo {booktitle} {International
  Colloquium on Automata, Languages, and Programming}}}\ (\bibinfo
  {organization} {Springer},\ \bibinfo {year} {2008})\ pp.\ \bibinfo {pages}
  {298--310}\BibitemShut {NoStop}%
\bibitem [{\citenamefont {van~de Wetering}(2020)}]{van2020zx}%
  \BibitemOpen
  \bibfield  {author} {\bibinfo {author} {\bibfnamefont {J.}~\bibnamefont
  {van~de Wetering}},\ }\bibfield  {title} {\emph {\bibinfo {title}
  {Zx-calculus for the working quantum computer scientist},\ }}\href@noop {}
  {\bibfield  {journal} {\bibinfo  {journal} {arXiv preprint arXiv:2012.13966}\
  } (\bibinfo {year} {2020})}\BibitemShut {NoStop}%
\bibitem [{\citenamefont {de~Beaudrap}\ and\ \citenamefont
  {Horsman}(2020)}]{de2020zx}%
  \BibitemOpen
  \bibfield  {author} {\bibinfo {author} {\bibfnamefont {N.}~\bibnamefont
  {de~Beaudrap}}\ and\ \bibinfo {author} {\bibfnamefont {D.}~\bibnamefont
  {Horsman}},\ }\bibfield  {title} {\emph {\bibinfo {title} {The zx calculus is
  a language for surface code lattice surgery},\ }}\href@noop {} {\bibfield
  {journal} {\bibinfo  {journal} {Quantum}\ }\textbf {\bibinfo {volume} {4}},\
  \bibinfo {pages} {218} (\bibinfo {year} {2020})}\BibitemShut {NoStop}%
\bibitem [{\citenamefont {Calderbank}\ and\ \citenamefont
  {Shor}(1996)}]{calderbank1996good}%
  \BibitemOpen
  \bibfield  {author} {\bibinfo {author} {\bibfnamefont {A.~R.}\ \bibnamefont
  {Calderbank}}\ and\ \bibinfo {author} {\bibfnamefont {P.~W.}\ \bibnamefont
  {Shor}},\ }\bibfield  {title} {\emph {\bibinfo {title} {Good quantum
  error-correcting codes exist},\ }}\href@noop {} {\bibfield  {journal}
  {\bibinfo  {journal} {Physical Review A}\ }\textbf {\bibinfo {volume} {54}},\
  \bibinfo {pages} {1098} (\bibinfo {year} {1996})}\BibitemShut {NoStop}%
\bibitem [{\citenamefont {Steane}(1996)}]{steane1996multiple}%
  \BibitemOpen
  \bibfield  {author} {\bibinfo {author} {\bibfnamefont {A.}~\bibnamefont
  {Steane}},\ }\bibfield  {title} {\emph {\bibinfo {title} {Multiple-particle
  interference and quantum error correction},\ }}\href@noop {} {\bibfield
  {journal} {\bibinfo  {journal} {Proceedings of the Royal Society of London.
  Series A: Mathematical, Physical and Engineering Sciences}\ }\textbf
  {\bibinfo {volume} {452}},\ \bibinfo {pages} {2551} (\bibinfo {year}
  {1996})}\BibitemShut {NoStop}%
\bibitem [{\citenamefont {Bravyi}\ \emph {et~al.}(2010)\citenamefont {Bravyi},
  \citenamefont {Poulin},\ and\ \citenamefont {Terhal}}]{bravyi2010tradeoffs}%
  \BibitemOpen
  \bibfield  {author} {\bibinfo {author} {\bibfnamefont {S.}~\bibnamefont
  {Bravyi}}, \bibinfo {author} {\bibfnamefont {D.}~\bibnamefont {Poulin}}, \
  and\ \bibinfo {author} {\bibfnamefont {B.}~\bibnamefont {Terhal}},\
  }\bibfield  {title} {\emph {\bibinfo {title} {Tradeoffs for reliable quantum
  information storage in 2d systems},\ }}\href@noop {} {\bibfield  {journal}
  {\bibinfo  {journal} {Physical review letters}\ }\textbf {\bibinfo {volume}
  {104}},\ \bibinfo {pages} {050503} (\bibinfo {year} {2010})}\BibitemShut
  {NoStop}%
\bibitem [{\citenamefont {Krishna}\ and\ \citenamefont
  {Poulin}(2021)}]{krishna2021fault}%
  \BibitemOpen
  \bibfield  {author} {\bibinfo {author} {\bibfnamefont {A.}~\bibnamefont
  {Krishna}}\ and\ \bibinfo {author} {\bibfnamefont {D.}~\bibnamefont
  {Poulin}},\ }\bibfield  {title} {\emph {\bibinfo {title} {Fault-tolerant
  gates on hypergraph product codes},\ }}\href@noop {} {\bibfield  {journal}
  {\bibinfo  {journal} {Physical Review X}\ }\textbf {\bibinfo {volume} {11}},\
  \bibinfo {pages} {011023} (\bibinfo {year} {2021})}\BibitemShut {NoStop}%
\bibitem [{\citenamefont {Cohen}\ \emph {et~al.}(2021)\citenamefont {Cohen},
  \citenamefont {Kim}, \citenamefont {Bartlett},\ and\ \citenamefont
  {Brown}}]{cohen2021low}%
  \BibitemOpen
  \bibfield  {author} {\bibinfo {author} {\bibfnamefont {L.~Z.}\ \bibnamefont
  {Cohen}}, \bibinfo {author} {\bibfnamefont {I.~H.}\ \bibnamefont {Kim}},
  \bibinfo {author} {\bibfnamefont {S.~D.}\ \bibnamefont {Bartlett}}, \ and\
  \bibinfo {author} {\bibfnamefont {B.~J.}\ \bibnamefont {Brown}},\ }\bibfield
  {title} {\emph {\bibinfo {title} {Low-overhead fault-tolerant quantum
  computing using long-range connectivity},\ }}\href@noop {} {\bibfield
  {journal} {\bibinfo  {journal} {arXiv preprint arXiv:2110.10794}\ } (\bibinfo
  {year} {2021})}\BibitemShut {NoStop}%
\bibitem [{Note6()}]{Note6}%
  \BibitemOpen
  \bibinfo {note} {More generally, the outer code encoding circuit for an
  arbitrary state is Clifford for a general CSS code, but it may not be
  constant depth.}\BibitemShut {Stop}%
\bibitem [{\citenamefont {Browne}\ and\ \citenamefont
  {Rudolph}(2005)}]{browne2005resource}%
  \BibitemOpen
  \bibfield  {author} {\bibinfo {author} {\bibfnamefont {D.~E.}\ \bibnamefont
  {Browne}}\ and\ \bibinfo {author} {\bibfnamefont {T.}~\bibnamefont
  {Rudolph}},\ }\bibfield  {title} {\emph {\bibinfo {title} {Resource-efficient
  linear optical quantum computation},\ }}\href@noop {} {\bibfield  {journal}
  {\bibinfo  {journal} {Physical Review Letters}\ }\textbf {\bibinfo {volume}
  {95}},\ \bibinfo {pages} {010501} (\bibinfo {year} {2005})}\BibitemShut
  {NoStop}%
\bibitem [{\citenamefont {Gimeno-Segovia}\ \emph {et~al.}(2015)\citenamefont
  {Gimeno-Segovia}, \citenamefont {Shadbolt}, \citenamefont {Browne},\ and\
  \citenamefont {Rudolph}}]{gimeno2015three}%
  \BibitemOpen
  \bibfield  {author} {\bibinfo {author} {\bibfnamefont {M.}~\bibnamefont
  {Gimeno-Segovia}}, \bibinfo {author} {\bibfnamefont {P.}~\bibnamefont
  {Shadbolt}}, \bibinfo {author} {\bibfnamefont {D.~E.}\ \bibnamefont
  {Browne}}, \ and\ \bibinfo {author} {\bibfnamefont {T.}~\bibnamefont
  {Rudolph}},\ }\bibfield  {title} {\emph {\bibinfo {title} {From three-photon
  greenberger-horne-zeilinger states to ballistic universal quantum
  computation},\ }}\href@noop {} {\bibfield  {journal} {\bibinfo  {journal}
  {Physical review letters}\ }\textbf {\bibinfo {volume} {115}},\ \bibinfo
  {pages} {020502} (\bibinfo {year} {2015})}\BibitemShut {NoStop}%
\bibitem [{\citenamefont {Hein}\ \emph {et~al.}(2004)\citenamefont {Hein},
  \citenamefont {Eisert},\ and\ \citenamefont {Briegel}}]{hein2004graph}%
  \BibitemOpen
  \bibfield  {author} {\bibinfo {author} {\bibfnamefont {M.}~\bibnamefont
  {Hein}}, \bibinfo {author} {\bibfnamefont {J.}~\bibnamefont {Eisert}}, \ and\
  \bibinfo {author} {\bibfnamefont {H.}~\bibnamefont {Briegel}},\ }\bibfield
  {title} {\emph {\bibinfo {title} {Multiparty entanglement in graph states},\
  }}\href@noop {} {\bibfield  {journal} {\bibinfo  {journal} {Physical Review
  A}\ }\textbf {\bibinfo {volume} {69}},\ \bibinfo {pages} {062311} (\bibinfo
  {year} {2004})}\BibitemShut {NoStop}%
\bibitem [{Note7()}]{Note7}%
  \BibitemOpen
  \bibinfo {note} {In Ref.~\cite {bartolucci2021fusion} this group is termed
  the \protect \emph {fusion group} and denoted $F$. It is renamed the
  measurement group $\protect \mathcal {M}$ here to note the inclusion of
  single-qubit measurements when required.}\BibitemShut {Stop}%
\bibitem [{\citenamefont {{\L}odyga}\ \emph {et~al.}(2015)\citenamefont
  {{\L}odyga}, \citenamefont {Mazurek}, \citenamefont {Grudka},\ and\
  \citenamefont {Horodecki}}]{lodyga2015simple}%
  \BibitemOpen
  \bibfield  {author} {\bibinfo {author} {\bibfnamefont {J.}~\bibnamefont
  {{\L}odyga}}, \bibinfo {author} {\bibfnamefont {P.}~\bibnamefont {Mazurek}},
  \bibinfo {author} {\bibfnamefont {A.}~\bibnamefont {Grudka}}, \ and\ \bibinfo
  {author} {\bibfnamefont {M.}~\bibnamefont {Horodecki}},\ }\bibfield  {title}
  {\emph {\bibinfo {title} {Simple scheme for encoding and decoding a qubit in
  unknown state for various topological codes},\ }}\href@noop {} {\bibfield
  {journal} {\bibinfo  {journal} {Scientific reports}\ }\textbf {\bibinfo
  {volume} {5}},\ \bibinfo {pages} {8975} (\bibinfo {year} {2015})}\BibitemShut
  {NoStop}%
\bibitem [{Note8()}]{Note8}%
  \BibitemOpen
  \bibinfo {note} {Note that this is distinct from phenomenological noise model
  that is commonly used to model measurement and Pauli errors in a code based
  fault tolerance scheme. Our model is closer to a gate error model in that it
  accounts for large scale entanglement generation from finite sized
  resources.}\BibitemShut {Stop}%
\bibitem [{Note9()}]{Note9}%
  \BibitemOpen
  \bibinfo {note} {A potentially related observation is found in Ref.~\cite
  {farrelly2020parallel} for a different family of codes, whereby different
  logical qubits can be decoded independently while remaining nearly globally
  optimal.}\BibitemShut {Stop}%
\bibitem [{\citenamefont {Kolmogorov}(2009)}]{kolmogorov2009blossom}%
  \BibitemOpen
  \bibfield  {author} {\bibinfo {author} {\bibfnamefont {V.}~\bibnamefont
  {Kolmogorov}},\ }\bibfield  {title} {\emph {\bibinfo {title} {Blossom v: a
  new implementation of a minimum cost perfect matching algorithm},\
  }}\href@noop {} {\bibfield  {journal} {\bibinfo  {journal} {Mathematical
  Programming Computation}\ }\textbf {\bibinfo {volume} {1}},\ \bibinfo {pages}
  {43} (\bibinfo {year} {2009})}\BibitemShut {NoStop}%
\bibitem [{\citenamefont {Delfosse}\ and\ \citenamefont
  {Nickerson}(2017)}]{delfosse2017almost}%
  \BibitemOpen
  \bibfield  {author} {\bibinfo {author} {\bibfnamefont {N.}~\bibnamefont
  {Delfosse}}\ and\ \bibinfo {author} {\bibfnamefont {N.~H.}\ \bibnamefont
  {Nickerson}},\ }\bibfield  {title} {\emph {\bibinfo {title} {Almost-linear
  time decoding algorithm for topological codes},\ }}\href@noop {} {\bibfield
  {journal} {\bibinfo  {journal} {arXiv preprint arXiv:1709.06218}\ } (\bibinfo
  {year} {2017})}\BibitemShut {NoStop}%
\bibitem [{\citenamefont {Henkel}\ and\ \citenamefont
  {Sch{\"u}tz}(1994)}]{henkel1994boundary}%
  \BibitemOpen
  \bibfield  {author} {\bibinfo {author} {\bibfnamefont {M.}~\bibnamefont
  {Henkel}}\ and\ \bibinfo {author} {\bibfnamefont {G.}~\bibnamefont
  {Sch{\"u}tz}},\ }\bibfield  {title} {\emph {\bibinfo {title}
  {Boundary-induced phase transitions in equilibrium and non-equilibrium
  systems},\ }}\href@noop {} {\bibfield  {journal} {\bibinfo  {journal}
  {Physica A: Statistical Mechanics and its Applications}\ }\textbf {\bibinfo
  {volume} {206}},\ \bibinfo {pages} {187} (\bibinfo {year}
  {1994})}\BibitemShut {NoStop}%
\bibitem [{\citenamefont {Bravyi}\ and\ \citenamefont
  {Vargo}(2013)}]{bravyi2013simulation}%
  \BibitemOpen
  \bibfield  {author} {\bibinfo {author} {\bibfnamefont {S.}~\bibnamefont
  {Bravyi}}\ and\ \bibinfo {author} {\bibfnamefont {A.}~\bibnamefont {Vargo}},\
  }\bibfield  {title} {\emph {\bibinfo {title} {Simulation of rare events in
  quantum error correction},\ }}\href@noop {} {\bibfield  {journal} {\bibinfo
  {journal} {Physical Review A}\ }\textbf {\bibinfo {volume} {88}},\ \bibinfo
  {pages} {062308} (\bibinfo {year} {2013})}\BibitemShut {NoStop}%
\bibitem [{\citenamefont {Kivlichan}\ \emph {et~al.}(2020)\citenamefont
  {Kivlichan}, \citenamefont {Gidney}, \citenamefont {Berry}, \citenamefont
  {Wiebe}, \citenamefont {McClean}, \citenamefont {Sun}, \citenamefont {Jiang},
  \citenamefont {Rubin}, \citenamefont {Fowler}, \citenamefont {Aspuru-Guzik}
  \emph {et~al.}}]{kivlichan2020improved}%
  \BibitemOpen
  \bibfield  {author} {\bibinfo {author} {\bibfnamefont {I.~D.}\ \bibnamefont
  {Kivlichan}}, \bibinfo {author} {\bibfnamefont {C.}~\bibnamefont {Gidney}},
  \bibinfo {author} {\bibfnamefont {D.~W.}\ \bibnamefont {Berry}}, \bibinfo
  {author} {\bibfnamefont {N.}~\bibnamefont {Wiebe}}, \bibinfo {author}
  {\bibfnamefont {J.}~\bibnamefont {McClean}}, \bibinfo {author} {\bibfnamefont
  {W.}~\bibnamefont {Sun}}, \bibinfo {author} {\bibfnamefont {Z.}~\bibnamefont
  {Jiang}}, \bibinfo {author} {\bibfnamefont {N.}~\bibnamefont {Rubin}},
  \bibinfo {author} {\bibfnamefont {A.}~\bibnamefont {Fowler}}, \bibinfo
  {author} {\bibfnamefont {A.}~\bibnamefont {Aspuru-Guzik}},  \emph {et~al.},\
  }\bibfield  {title} {\emph {\bibinfo {title} {Improved fault-tolerant quantum
  simulation of condensed-phase correlated electrons via trotterization},\
  }}\href@noop {} {\bibfield  {journal} {\bibinfo  {journal} {Quantum}\
  }\textbf {\bibinfo {volume} {4}},\ \bibinfo {pages} {296} (\bibinfo {year}
  {2020})}\BibitemShut {NoStop}%
\bibitem [{\citenamefont {von Burg}\ \emph {et~al.}(2020)\citenamefont {von
  Burg}, \citenamefont {Low}, \citenamefont {H{\"a}ner}, \citenamefont
  {Steiger}, \citenamefont {Reiher}, \citenamefont {Roetteler},\ and\
  \citenamefont {Troyer}}]{von2020quantum}%
  \BibitemOpen
  \bibfield  {author} {\bibinfo {author} {\bibfnamefont {V.}~\bibnamefont {von
  Burg}}, \bibinfo {author} {\bibfnamefont {G.~H.}\ \bibnamefont {Low}},
  \bibinfo {author} {\bibfnamefont {T.}~\bibnamefont {H{\"a}ner}}, \bibinfo
  {author} {\bibfnamefont {D.~S.}\ \bibnamefont {Steiger}}, \bibinfo {author}
  {\bibfnamefont {M.}~\bibnamefont {Reiher}}, \bibinfo {author} {\bibfnamefont
  {M.}~\bibnamefont {Roetteler}}, \ and\ \bibinfo {author} {\bibfnamefont
  {M.}~\bibnamefont {Troyer}},\ }\bibfield  {title} {\emph {\bibinfo {title}
  {Quantum computing enhanced computational catalysis},\ }}\href@noop {}
  {\bibfield  {journal} {\bibinfo  {journal} {arXiv preprint arXiv:2007.14460}\
  } (\bibinfo {year} {2020})}\BibitemShut {NoStop}%
\bibitem [{\citenamefont {Su}\ \emph {et~al.}(2021)\citenamefont {Su},
  \citenamefont {Berry}, \citenamefont {Wiebe}, \citenamefont {Rubin},\ and\
  \citenamefont {Babbush}}]{su2021faulttolerant}%
  \BibitemOpen
  \bibfield  {author} {\bibinfo {author} {\bibfnamefont {Y.}~\bibnamefont
  {Su}}, \bibinfo {author} {\bibfnamefont {D.~W.}\ \bibnamefont {Berry}},
  \bibinfo {author} {\bibfnamefont {N.}~\bibnamefont {Wiebe}}, \bibinfo
  {author} {\bibfnamefont {N.}~\bibnamefont {Rubin}}, \ and\ \bibinfo {author}
  {\bibfnamefont {R.}~\bibnamefont {Babbush}},\ }\bibfield  {title} {\emph
  {\bibinfo {title} {Fault-tolerant quantum simulations of chemistry in first
  quantization},\ }}\href {https://arxiv.org/abs/2105.12767} {\  (\bibinfo
  {year} {2021})},\ \Eprint {http://arxiv.org/abs/2105.12767} {arXiv:2105.12767
  [quant-ph]} \BibitemShut {NoStop}%
\bibitem [{\citenamefont {Krishna}\ and\ \citenamefont
  {Poulin}(2020)}]{krishna2020topological}%
  \BibitemOpen
  \bibfield  {author} {\bibinfo {author} {\bibfnamefont {A.}~\bibnamefont
  {Krishna}}\ and\ \bibinfo {author} {\bibfnamefont {D.}~\bibnamefont
  {Poulin}},\ }\bibfield  {title} {\emph {\bibinfo {title} {Topological
  wormholes: Nonlocal defects on the toric code},\ }}\href@noop {} {\bibfield
  {journal} {\bibinfo  {journal} {Physical Review Research}\ }\textbf {\bibinfo
  {volume} {2}},\ \bibinfo {pages} {023116} (\bibinfo {year}
  {2020})}\BibitemShut {NoStop}%
\bibitem [{Note10()}]{Note10}%
  \BibitemOpen
  \bibinfo {note} {For instance, logical $\protect \overline {X}$ and $\protect
  \overline {Z}$ operators of a toric code or planar code are traceable, but
  the logical $\protect \overline {Y}$ is not, due to the unavoidable
  self-intersection of string operators (see, for example, Fig.~\ref
  {figLogicalBlockConcatenation}).}\BibitemShut {Stop}%
\bibitem [{\citenamefont {Gottesman}(2010)}]{gottesman2010introduction}%
  \BibitemOpen
  \bibfield  {author} {\bibinfo {author} {\bibfnamefont {D.}~\bibnamefont
  {Gottesman}},\ }in\ \href@noop {} {\emph {\bibinfo {booktitle} {Quantum
  information science and its contributions to mathematics, Proceedings of
  Symposia in Applied Mathematics}}},\ Vol.~\bibinfo {volume} {68}\ (\bibinfo
  {year} {2010})\ pp.\ \bibinfo {pages} {13--58}\BibitemShut {NoStop}%
\bibitem [{\citenamefont {Aaronson}\ and\ \citenamefont
  {Gottesman}(2004)}]{aaronson2004improved}%
  \BibitemOpen
  \bibfield  {author} {\bibinfo {author} {\bibfnamefont {S.}~\bibnamefont
  {Aaronson}}\ and\ \bibinfo {author} {\bibfnamefont {D.}~\bibnamefont
  {Gottesman}},\ }\bibfield  {title} {\emph {\bibinfo {title} {Improved
  simulation of stabilizer circuits},\ }}\href@noop {} {\bibfield  {journal}
  {\bibinfo  {journal} {Physical Review A}\ }\textbf {\bibinfo {volume} {70}},\
  \bibinfo {pages} {052328} (\bibinfo {year} {2004})}\BibitemShut {NoStop}%
\bibitem [{\citenamefont {Pastawski}\ \emph {et~al.}(2015)\citenamefont
  {Pastawski}, \citenamefont {Yoshida}, \citenamefont {Harlow},\ and\
  \citenamefont {Preskill}}]{pastawski2015holographic}%
  \BibitemOpen
  \bibfield  {author} {\bibinfo {author} {\bibfnamefont {F.}~\bibnamefont
  {Pastawski}}, \bibinfo {author} {\bibfnamefont {B.}~\bibnamefont {Yoshida}},
  \bibinfo {author} {\bibfnamefont {D.}~\bibnamefont {Harlow}}, \ and\ \bibinfo
  {author} {\bibfnamefont {J.}~\bibnamefont {Preskill}},\ }\bibfield  {title}
  {\emph {\bibinfo {title} {Holographic quantum error-correcting codes: Toy
  models for the bulk/boundary correspondence},\ }}\href@noop {} {\bibfield
  {journal} {\bibinfo  {journal} {Journal of High Energy Physics}\ }\textbf
  {\bibinfo {volume} {2015}},\ \bibinfo {pages} {1} (\bibinfo {year}
  {2015})}\BibitemShut {NoStop}%
\bibitem [{\citenamefont {Cao}\ and\ \citenamefont
  {Lackey}(2021)}]{cao2021quantum}%
  \BibitemOpen
  \bibfield  {author} {\bibinfo {author} {\bibfnamefont {C.}~\bibnamefont
  {Cao}}\ and\ \bibinfo {author} {\bibfnamefont {B.}~\bibnamefont {Lackey}},\
  }\bibfield  {title} {\emph {\bibinfo {title} {Quantum lego: Building quantum
  error correction codes from tensor networks},\ }}\href@noop {} {\bibfield
  {journal} {\bibinfo  {journal} {arXiv preprint arXiv:2109.08158}\ } (\bibinfo
  {year} {2021})}\BibitemShut {NoStop}%
\bibitem [{\citenamefont {Farrelly}\ \emph {et~al.}(2021)\citenamefont
  {Farrelly}, \citenamefont {Tuckett},\ and\ \citenamefont
  {Stace}}]{farrelly2021local}%
  \BibitemOpen
  \bibfield  {author} {\bibinfo {author} {\bibfnamefont {T.}~\bibnamefont
  {Farrelly}}, \bibinfo {author} {\bibfnamefont {D.~K.}\ \bibnamefont
  {Tuckett}}, \ and\ \bibinfo {author} {\bibfnamefont {T.~M.}\ \bibnamefont
  {Stace}},\ }\bibfield  {title} {\emph {\bibinfo {title} {Local tensor-network
  codes},\ }}\href@noop {} {\bibfield  {journal} {\bibinfo  {journal} {arXiv
  preprint arXiv:2109.11996}\ } (\bibinfo {year} {2021})}\BibitemShut {NoStop}%
\bibitem [{\citenamefont {PsiQuantum}(2022)}]{bombin2023unifying}%
\BibitemOpen
\bibfield  {author} {\bibinfo {author} {\bibfnamefont {H.}\ \bibnamefont
		{Bombin}}, \bibinfo {author} {\bibfnamefont {D.}\ \bibnamefont
		{Litinski}}, \bibinfo {author} {\bibfnamefont {N.}\ \bibnamefont
		{Nickerson}}, \bibinfo {author} {\bibfnamefont {F.}\ \bibnamefont
		{Pastawski}}, \ and\ \bibinfo {author} {\bibfnamefont {S.}\ \bibnamefont
		{Roberts}},\  }\bibfield  {title} {\emph {\bibinfo {title} {Unifying flavors of fault tolerance with the ZX calculus},\ }}\href{https://arxiv.org/abs/2303.08829} {\bibfield  {journal} {\bibinfo  {journal} {arXiv preprint arXiv:2303.08829}\ } (\bibinfo {year} {2023})}\BibitemShut {NoStop}%
\bibitem [{\citenamefont {Li}(2015)}]{li2015magic}%
  \BibitemOpen
  \bibfield  {author} {\bibinfo {author} {\bibfnamefont {Y.}~\bibnamefont
  {Li}},\ }\bibfield  {title} {\emph {\bibinfo {title} {A magic state’s
  fidelity can be superior to the operations that created it},\ }}\href@noop {}
  {\bibfield  {journal} {\bibinfo  {journal} {New Journal of Physics}\ }\textbf
  {\bibinfo {volume} {17}},\ \bibinfo {pages} {023037} (\bibinfo {year}
  {2015})}\BibitemShut {NoStop}%
\bibitem [{\citenamefont {Bombín}\ \emph {et~al.}(2022)\citenamefont
  {Bombín}, \citenamefont {Pant}, \citenamefont {Roberts},\ and\ \citenamefont
  {Seetharam}}]{bombin2022fault}%
  \BibitemOpen
  \bibfield  {author} {\bibinfo {author} {\bibfnamefont {H.}~\bibnamefont
  {Bombín}}, \bibinfo {author} {\bibfnamefont {M.}~\bibnamefont {Pant}},
  \bibinfo {author} {\bibfnamefont {S.}~\bibnamefont {Roberts}}, \ and\
  \bibinfo {author} {\bibfnamefont {K.~I.}\ \bibnamefont {Seetharam}},\
  }\bibfield  {title} {\emph {\bibinfo {title} {Fault-tolerant post-selection
  for low overhead magic state preparation},\ }}\href
  {https://arxiv.org/abs/2212.00813} {\bibfield  {journal} {\bibinfo  {journal}
  {arXiv preprint arXiv:2212.00813}\ } (\bibinfo {year} {2022})}\BibitemShut
  {NoStop}%
\bibitem [{\citenamefont {Poulin}(2005)}]{poulin2005stabilizer}%
  \BibitemOpen
  \bibfield  {author} {\bibinfo {author} {\bibfnamefont {D.}~\bibnamefont
  {Poulin}},\ }\bibfield  {title} {\emph {\bibinfo {title} {Stabilizer
  formalism for operator quantum error correction},\ }}\href@noop {} {\bibfield
   {journal} {\bibinfo  {journal} {Physical review letters}\ }\textbf {\bibinfo
  {volume} {95}},\ \bibinfo {pages} {230504} (\bibinfo {year}
  {2005})}\BibitemShut {NoStop}%
\bibitem [{\citenamefont {Bacon}(2006)}]{bacon2006operator}%
  \BibitemOpen
  \bibfield  {author} {\bibinfo {author} {\bibfnamefont {D.}~\bibnamefont
  {Bacon}},\ }\bibfield  {title} {\emph {\bibinfo {title} {Operator quantum
  error-correcting subsystems for self-correcting quantum memories},\
  }}\href@noop {} {\bibfield  {journal} {\bibinfo  {journal} {Physical Review
  A}\ }\textbf {\bibinfo {volume} {73}},\ \bibinfo {pages} {012340} (\bibinfo
  {year} {2006})}\BibitemShut {NoStop}%
\bibitem [{\citenamefont {PsiQuantum}(2022)}]{modularDecoding}%
\BibitemOpen
\bibfield  {author} {\bibinfo {author} {\bibfnamefont {H.}\ \bibnamefont
		{Bombin}}, \bibinfo {author} {\bibfnamefont {C.}\ \bibnamefont
		{Dawson}}, \bibinfo {author} {\bibfnamefont {Y.}\ \bibnamefont
		{Liu}}, \bibinfo {author} {\bibfnamefont {N.}\ \bibnamefont
		{Nickerson}}, \bibinfo {author} {\bibfnamefont {F.}\ \bibnamefont
		{Pastawski}}, \ and\ \bibinfo {author} {\bibfnamefont {S.}\ \bibnamefont
		{Roberts}},\  }\bibfield  {title} {\emph {\bibinfo {title} {Modular decoding: parallelizable real-time decoding for quantum computers},\ }}\href{https://arxiv.org/abs/2303.04846} {\bibfield  {journal} {\bibinfo  {journal} {arXiv preprint arXiv:2303.04846}\ } (\bibinfo {year} {2023})}\BibitemShut {NoStop}%
\bibitem [{\citenamefont {Bombin}\ \emph
  {et~al.}(2012{\natexlab{b}})\citenamefont {Bombin}, \citenamefont
  {Duclos-Cianci},\ and\ \citenamefont {Poulin}}]{bombin2012universal}%
  \BibitemOpen
  \bibfield  {author} {\bibinfo {author} {\bibfnamefont {H.}~\bibnamefont
  {Bombin}}, \bibinfo {author} {\bibfnamefont {G.}~\bibnamefont
  {Duclos-Cianci}}, \ and\ \bibinfo {author} {\bibfnamefont {D.}~\bibnamefont
  {Poulin}},\ }\bibfield  {title} {\emph {\bibinfo {title} {Universal
  topological phase of two-dimensional stabilizer codes},\ }}\href@noop {}
  {\bibfield  {journal} {\bibinfo  {journal} {New Journal of Physics}\ }\textbf
  {\bibinfo {volume} {14}},\ \bibinfo {pages} {073048} (\bibinfo {year}
  {2012}{\natexlab{b}})}\BibitemShut {NoStop}%
\bibitem [{\citenamefont {Scruby}\ and\ \citenamefont
  {Browne}(2020)}]{scruby2020hierarchy}%
  \BibitemOpen
  \bibfield  {author} {\bibinfo {author} {\bibfnamefont {T.}~\bibnamefont
  {Scruby}}\ and\ \bibinfo {author} {\bibfnamefont {D.}~\bibnamefont
  {Browne}},\ }\bibfield  {title} {\emph {\bibinfo {title} {A hierarchy of
  anyon models realised by twists in stacked surface codes},\ }}\href@noop {}
  {\bibfield  {journal} {\bibinfo  {journal} {Quantum}\ }\textbf {\bibinfo
  {volume} {4}},\ \bibinfo {pages} {251} (\bibinfo {year} {2020})}\BibitemShut
  {NoStop}%
\bibitem [{\citenamefont {Gaiotto}\ \emph {et~al.}(2015)\citenamefont
  {Gaiotto}, \citenamefont {Kapustin}, \citenamefont {Seiberg},\ and\
  \citenamefont {Willett}}]{gaiotto2015generalized}%
  \BibitemOpen
  \bibfield  {author} {\bibinfo {author} {\bibfnamefont {D.}~\bibnamefont
  {Gaiotto}}, \bibinfo {author} {\bibfnamefont {A.}~\bibnamefont {Kapustin}},
  \bibinfo {author} {\bibfnamefont {N.}~\bibnamefont {Seiberg}}, \ and\
  \bibinfo {author} {\bibfnamefont {B.}~\bibnamefont {Willett}},\ }\bibfield
  {title} {\emph {\bibinfo {title} {Generalized global symmetries},\
  }}\href@noop {} {\bibfield  {journal} {\bibinfo  {journal} {Journal of High
  Energy Physics}\ }\textbf {\bibinfo {volume} {2015}},\ \bibinfo {pages} {172}
  (\bibinfo {year} {2015})}\BibitemShut {NoStop}%
\bibitem [{\citenamefont {Kapustin}\ and\ \citenamefont
  {Thorngren}(2017)}]{kapustin2017higher}%
  \BibitemOpen
  \bibfield  {author} {\bibinfo {author} {\bibfnamefont {A.}~\bibnamefont
  {Kapustin}}\ and\ \bibinfo {author} {\bibfnamefont {R.}~\bibnamefont
  {Thorngren}},\ }in\ \href@noop {} {\emph {\bibinfo {booktitle} {Algebra,
  Geometry, and Physics in the 21st Century}}}\ (\bibinfo  {publisher}
  {Springer},\ \bibinfo {year} {2017})\ pp.\ \bibinfo {pages}
  {177--202}\BibitemShut {NoStop}%
\bibitem [{\citenamefont {Roberts}\ and\ \citenamefont
  {Bartlett}(2020)}]{roberts2020symmetry}%
  \BibitemOpen
  \bibfield  {author} {\bibinfo {author} {\bibfnamefont {S.}~\bibnamefont
  {Roberts}}\ and\ \bibinfo {author} {\bibfnamefont {S.~D.}\ \bibnamefont
  {Bartlett}},\ }\bibfield  {title} {\emph {\bibinfo {title}
  {Symmetry-protected self-correcting quantum memories},\ }}\href@noop {}
  {\bibfield  {journal} {\bibinfo  {journal} {Physical Review X}\ }\textbf
  {\bibinfo {volume} {10}},\ \bibinfo {pages} {031041} (\bibinfo {year}
  {2020})}\BibitemShut {NoStop}%
\bibitem [{\citenamefont {Farrelly}\ \emph {et~al.}(2020)\citenamefont
  {Farrelly}, \citenamefont {Harris}, \citenamefont {McMahon},\ and\
  \citenamefont {Stace}}]{farrelly2020parallel}%
  \BibitemOpen
  \bibfield  {author} {\bibinfo {author} {\bibfnamefont {T.}~\bibnamefont
  {Farrelly}}, \bibinfo {author} {\bibfnamefont {R.~J.}\ \bibnamefont
  {Harris}}, \bibinfo {author} {\bibfnamefont {N.~A.}\ \bibnamefont {McMahon}},
  \ and\ \bibinfo {author} {\bibfnamefont {T.~M.}\ \bibnamefont {Stace}},\
  }\bibfield  {title} {\emph {\bibinfo {title} {Parallel decoding of multiple
  logical qubits in tensor-network codes},\ }}\href@noop {} {\bibfield
  {journal} {\bibinfo  {journal} {arXiv preprint arXiv:2012.07317}\ } (\bibinfo
  {year} {2020})}\BibitemShut {NoStop}%
\end{thebibliography}

%

\end{document}